\documentclass[11pt, a4paper, logo, copyright]{deepseek}

\pdftrailerid{redacted}

\makeatletter
\renewcommand\bibentry[1]{\nocite{#1}{\frenchspacing\@nameuse{BR@r@#1\@extra@b@citeb}}}
\makeatother

\usepackage{CJKutf8}
\usepackage{newunicodechar}

\usepackage{amsmath}
\usepackage{amsfonts}
\usepackage{mathtools}
\usepackage{bm}
\usepackage{nicefrac}

\usepackage{float}
\usepackage{placeins}
\usepackage{wrapfig}
\usepackage{subcaption}
\usepackage{capt-of}
\usepackage[rightcaption]{sidecap}
\usepackage{array}
\usepackage{multirow}
\usepackage{makecell}
\usepackage[flushleft]{threeparttable}
\usepackage{arydshln}
\usepackage{siunitx}
\usepackage{tabularray}
\UseTblrLibrary{booktabs}

\renewcommand\tabularxcolumn[1]{m{#1}}
\newcolumntype{C}{>{\centering\arraybackslash}X}
\newcolumntype{L}{>{\raggedright\arraybackslash}X}

\usepackage{algorithm}
\usepackage{algorithmicx}
\usepackage{algpseudocode}
\usepackage{listings}
\setlist[itemize]{
  leftmargin=1.4em,
  labelwidth=0.8em,
  labelsep=0.45em,
  labelindent=0pt,
  topsep=0pt,
  partopsep=0pt,
  itemsep=0.18em plus 0.03em,
  parsep=0pt
}
\setlist[itemize,2]{
  leftmargin=1.55em,
  label={\textbullet},
  topsep=0pt,
  partopsep=0pt,
  itemsep=0.1em plus 0.02em,
  parsep=0pt
}
\setlist[enumerate]{
  leftmargin=1.65em,
  labelwidth=1.0em,
  labelsep=0.45em,
  labelindent=0pt,
  topsep=0pt,
  partopsep=0pt,
  itemsep=0pt,
  parsep=0pt
}

\usepackage{gdm-colors}
\usepackage[most]{tcolorbox}
\usepackage{soul}
\usepackage{xcolor}

\definecolor{thinkcolor}{RGB}{227,196,144}
\definecolor{observecolor}{RGB}{153,201,227}
\definecolor{explorecolor}{RGB}{178,217,200}
\definecolor{taomindlink}{RGB}{158,74,18}
\definecolor{taomindcite}{RGB}{24,62,112}
\definecolor{taomindurl}{RGB}{48,78,88}

\usepackage{tikz}
\usetikzlibrary{shapes.geometric, arrows.meta, positioning}

\usepackage{multicol}
\usepackage{calc}
\usepackage[bottom]{footmisc}

\usepackage{fontawesome}

\hypersetup{
    colorlinks=true,
    linkcolor=taomindlink,
    citecolor=taomindcite,
    urlcolor=taomindurl,
    filecolor=taomindlink,
    pdfborder={0 0 0},
    breaklinks=true
}
\usepackage[nameinlink]{cleveref}
\crefname{table}{Table}{Tables}
\Crefname{table}{Table}{Tables}
\crefname{figure}{Figure}{Figures}
\Crefname{figure}{Figure}{Figures}
\crefname{section}{Section}{Sections}
\Crefname{section}{Section}{Sections}
\crefname{equation}{Equation}{Equations}
\Crefname{equation}{Equation}{Equations}

\usepackage[numbers, sort&compress, square]{natbib}

\usepackage[normalem]{ulem}
\usepackage{dashrule}
\usepackage{blindtext}
\usepackage{tablefootnote}
\usepackage{xspace}

\newcounter{caseexample}[section]

\newcounter{promptexample}[section]

\newcommand{\strongemph}[1]{\textbf{\emph{#1}}}
\newcommand{\ourmethod}{ShopX\xspace}
\newcommand{\framework}{ShopX\xspace}

\xspaceaddexceptions{-}

\newcommand{\tightparagraph}[1]{\vspace{0.1ex}\noindent\textbf{#1}\hspace{0.5em}}

\newcommand{\titlelogo}{\hspace{0.08em}\raisebox{-0.3\height}{\includegraphics[height=1.5em,trim=150 100 145 95,clip]{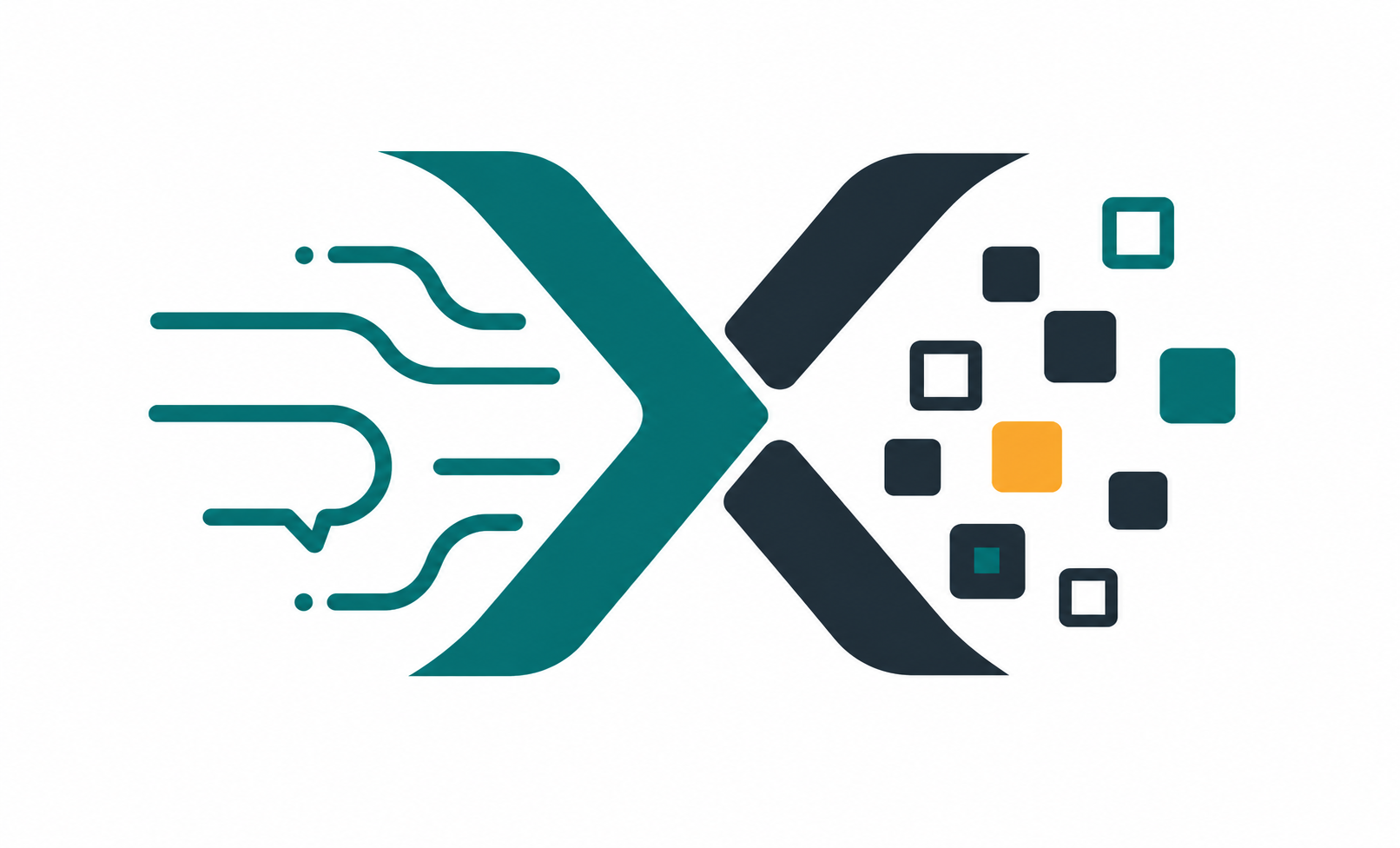}}}
\title{\centering \titlelogo~\ourmethod: A Foundation Model for Intent-to-Item \\ Fulfillment in Agentic Shopping}

\author{\ourmethod Team}
\renewcommand{\today}{}

\makeatletter
\renewcommand{\abscontent}{
    \noindent
    \parbox{\dimexpr\linewidth}{\absfont \theabstract}
    \@ifundefined{@keywords}{}{
        \vskip0.6em \noindent \keywordsfont Keywords: \@keywords}
}
\renewcommand{\maketitle}{\bgroup\setlength{\parindent}{0pt}
    \vspace*{0pt}
    \begin{adjustwidth}{0pt}{0pt}
        \begin{flushleft}
            {
                {\raggedright \titlefont \@title\par}%
                \vskip12pt
                {\raggedright \@author\par}
                \vskip18pt
            }%
        \end{flushleft}
    \end{adjustwidth}
    \egroup
    {%
        {\abscontent}
    }%
    \thispagestyle{firststyle}
}
\makeatother

\begin{document}
\begin{CJK*}{UTF8}{gbsn}

\begin{abstract}
\setlength{\parskip}{0.25\baselineskip}
\setlength{\parindent}{0pt}

The wave of AI-native applications is moving shopping beyond page- and feed-based browsing toward intent-driven experiences orchestrated by LLM agents. A common design wraps an LLM around existing search and recommendation pipelines, forcing complex intents through low-bandwidth retrieval or ranking interfaces and leaving a gap between language understanding and item-space fulfillment. Generative recommendation gives LLMs a direct item-space interface through semantic IDs (SIDs), but existing models mainly generate candidates for retrieval rather than translate flexible intents into item-space outcomes.

We propose \textbf{\ourmethod} to address this bottleneck by unifying intent understanding, execution planning, and flexible SID-native item-space operations into a single foundation model. We deploy \ourmethod in agentic shopping workflows through a model-native item-fulfillment framework with a serving harness that defines a model-facing action protocol and exposes support surfaces for context access, catalog grounding, and state management. Within this framework, \ourmethod plans and composes SID-based item-space operations such as SID beam-search retrieval, listwise ranking, or product bundling. This model-centric design reduces lossy hand-offs between agent orchestration and item-space execution.

To build \ourmethod, we design semantically recoverable, LLM-operable SIDs and a training recipe that equips a general LLM for flexible multi-turn item-space fulfillment while retaining the knowledge and instruction-following abilities needed by a shopping agent. We evaluate the \ourmethod framework against tool-mediated agentic systems on single- and multi-turn fulfillment tasks derived from anonymized Taobao production logs, showing that model-native fulfillment improves overall framework behavior, especially on complex or ambiguous requests.
We further assess \ourmethod through a fine-grained capability breakdown, yielding practices for SID design, continued pre-training, and post-training strategies.

\end{abstract}

\maketitle

\noindent\begin{minipage}{\textwidth}
\captionsetup{font=small,skip=4pt,hypcap=false}
\makebox[\textwidth][c]{%
\begin{minipage}[c]{0.61\textwidth}
  \centering
  \includegraphics[width=\linewidth,height=0.57\textwidth]{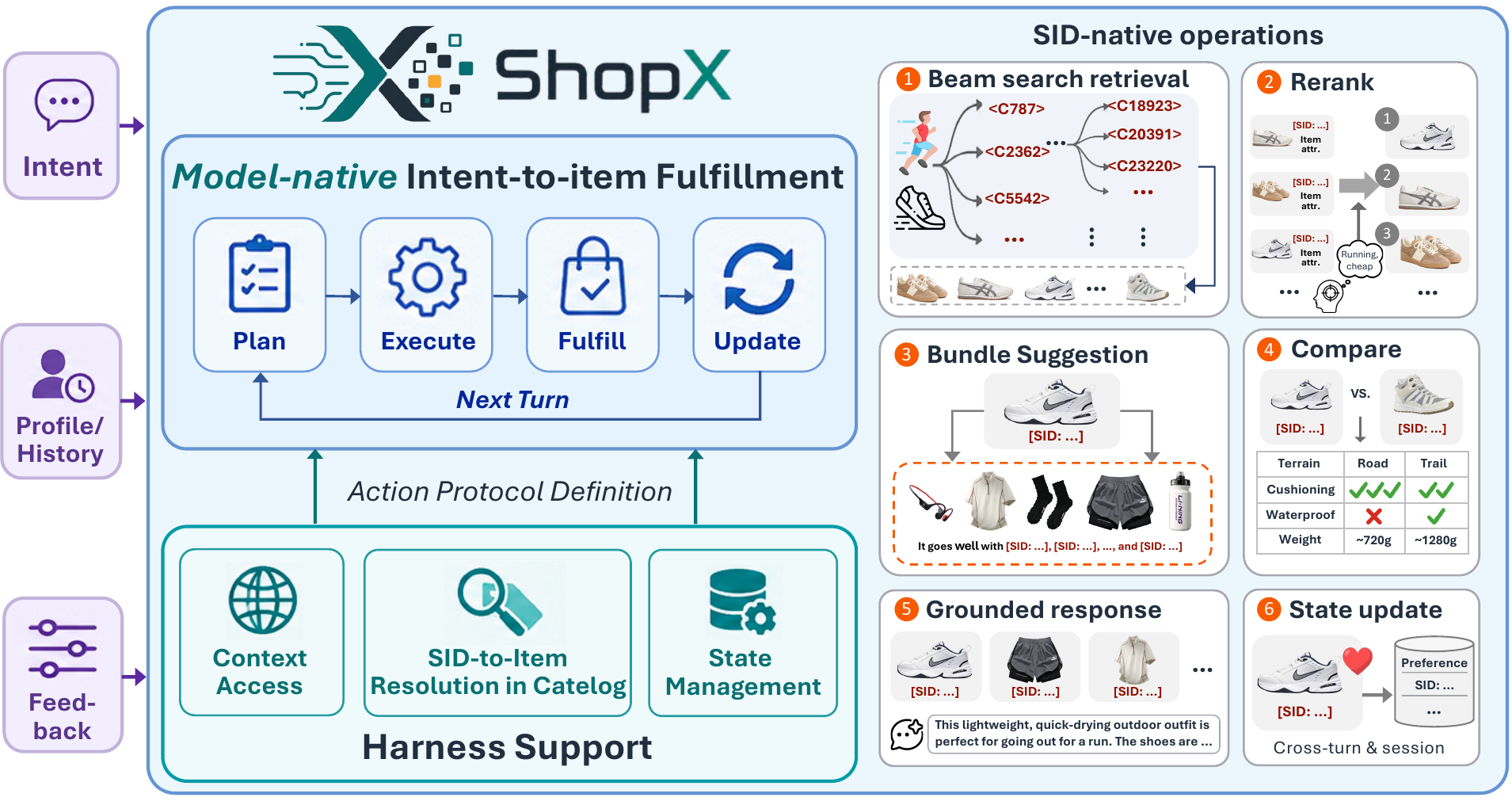}
\end{minipage}\hspace{0.008\textwidth}
\begin{minipage}[c]{0.39\textwidth}
  \centering
\includegraphics[width=\linewidth]{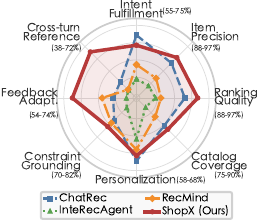}%
  
\end{minipage}%
}
\captionof{figure}{
(Left) \ourmethod is a SID-aware foundation model that serves as the fulfillment core, planning SID-native item operations through a harness for context, catalog grounding, and state support.
(Right) On Taobao-derived complex or ambiguous shopping tasks, the \ourmethod framework improves system-level fulfillment over LLM agents mediated by external tools.
}
\label{fig:graphical_abstract}
\end{minipage}

\setcounter{tocdepth}{2} %
\begingroup
\hypersetup{linkcolor=black} %
\color{black}
\renewcommand{\baselinestretch}{1.06}\normalsize
\setlength{\parskip}{5.2pt plus 0.8pt}
\tableofcontents %
\endgroup
\clearpage

\section{Introduction}
\label{sec:introduction}

Frontier LLMs and agentic applications are reshaping product entry points: users increasingly express intent in natural language and expect AI systems to reason over context, coordinate capabilities, and execute workflows~\citep{gpt5,claude4,gemini3,qwen3,codex,claude_code,openclaw}.
E-commerce follows this shift, moving from fixed catalog-browsing surfaces toward AI-native experiences with personal context, item-grounded summaries or comparisons, feedback-driven refinement, and catalog-grounded responses.
In this setting, recommendation is no longer limited to ranked lists or feeds; it must support \strongemph{intent-to-item fulfillment}: converting flexible shopping intents into catalog-grounded outcomes such as item lists, summaries, comparisons, bundles, or follow-up actions.
Recent AI shopping and search products such as Amazon Rufus, ChatGPT Shopping, and Qwen-Taobao assistants reflect this shift~\citep{amazon_rufus,chatgpt_shopping,chatgpt_instant_checkout,qwen_taobao,google_ai_mode_shopping,taobao_ai_search,xiaohongshu_diandian}.

\begin{figure}[!t]
\centering
\includegraphics[width=1.0\textwidth]{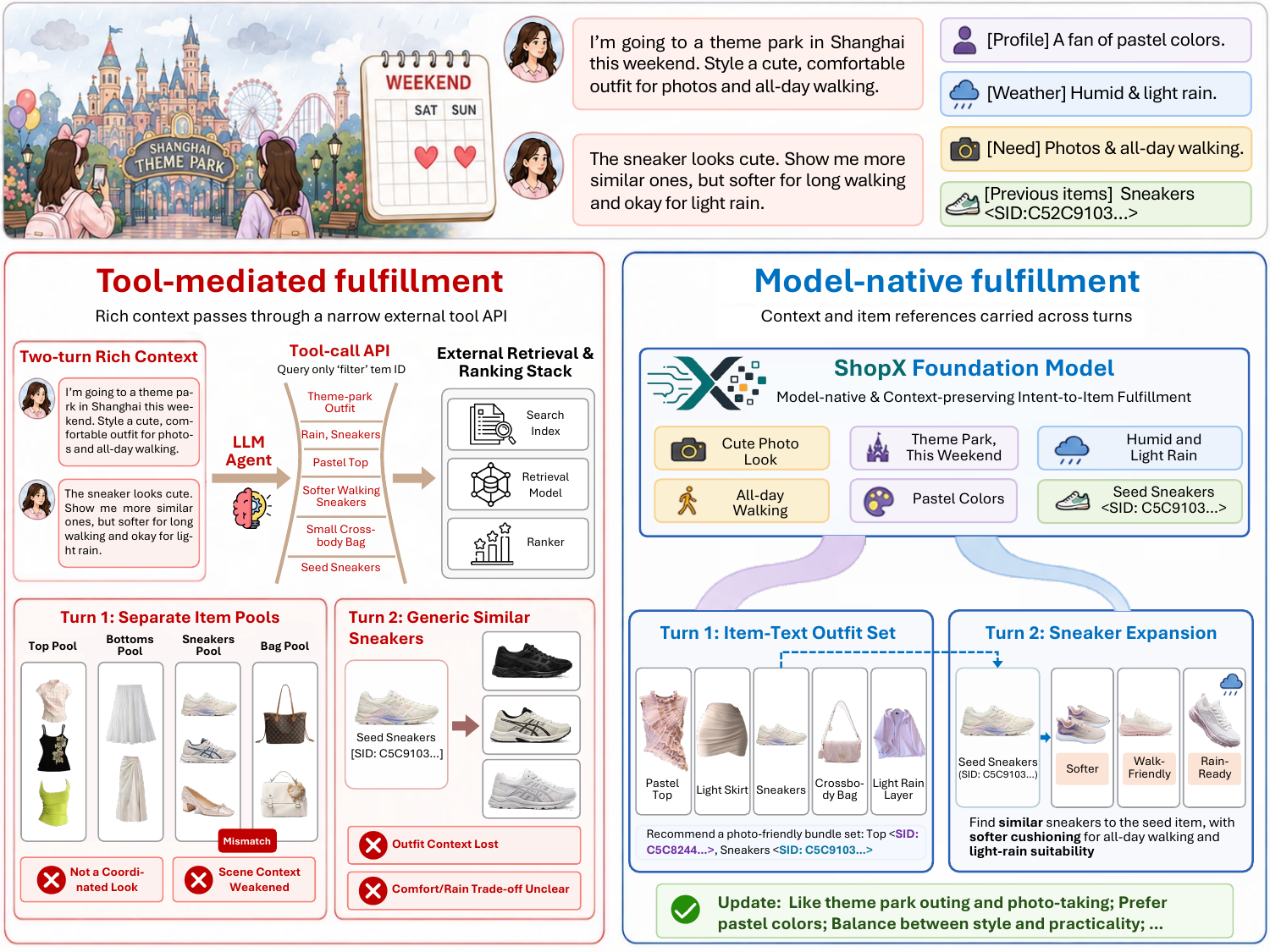}
\caption{
\textbf{Tool-mediated versus model-native fulfillment.}
In a two-turn shopping scenario, tool-mediated fulfillment compresses rich context into external retrieval and ranking calls, while \ourmethod keeps scenario context, item references, and feedback within a model-native item-space interface.
The example highlights two interface losses: intent details can be dropped as an LLM query maps to a rigid tool API, and prior item state can detach during refinement.
}
\label{fig:teaser}
\end{figure}

A common implementation wraps an LLM agent around existing search and recommendation pipelines, separating the workflow into two roles:
\begin{itemize}
    \item \textbf{Agent-side orchestration.}
    The LLM handles intent understanding, query rewriting, profile summarization, dialogue, planning, explanation, and tool orchestration.
    \item \textbf{Tool-side item-space execution.}
    External item-space services receive compact tool inputs and return candidates, scores, rankings, or item lists.
\end{itemize}
Representative systems include Chat-REC, InteRecAgent, RecMind, RecAI, RA-Rec, and industrial systems such as RecGPT~\citep{llm_for_rec_survey,llm_agents_rec_search_survey,chat_rec,interecagent,recmind,recai,ra_rec,recgpt,recgpt_v2}.
We refer to this family as \strongemph{tool-mediated systems}: LLM-agent systems in which the model plans and issues tool calls while external retrieval, ranking, or recommendation services perform item-space execution.
This separation brings LLMs' language priors and reasoning into recommendation, but the boundary remains low-bandwidth: the model must translate rich context into compact tool inputs rather than operate directly on items.
Fine-grained signals such as multi-turn preferences, scenario constraints, cross-item compatibility, and user feedback may therefore be compressed or fragmented before item-space execution, since retrieval and ranking services often expose narrow, fixed interfaces.
\autoref{fig:teaser} illustrates this loss in a two-turn shopping scenario: rich intent is squeezed into compact tool calls, and follow-up feedback may no longer stay tied to the prior items it is meant to refine.
This tension motivates a more direct question: \strongemph{how can rich shopping intent be grounded into item-space operations with fewer lossy tool hand-offs?}

To reduce such interface loss, we need an item-space interface that models can operate over more directly.
Generative recommendation provides an important substrate: items are represented as discrete semantic IDs (SIDs), and recommendation is formulated as autoregressive SID generation~\citep{tiger}.
Subsequent work further refines SID construction and item tokenization for accuracy, scalability, and deployment~\citep{lc_rec,letter,etegrec,forge}.
The OneRec series further shows that SID generation can consolidate conventional retrieval--ranking cascades into end-to-end frameworks and extend to recommendation-specific reasoning~\citep{onerec,onerec_v2,openonerec,onerec_think}.
However, existing generative recommenders remain optimized around a narrower contract: given a user history or additional reasoning trace, generate SID candidates.
This objective does not train the model to carry the full fulfillment state of an AI-native shopping session, including action choice, item selection or bundling, grounded explanation, clarification, and feedback or preference updates.
Therefore, even if SID generation reduces the hand-off inside conventional retrieval--ranking pipelines, existing generative recommender systems still leave the broader language-to-item fulfillment hand-off to an external~orchestrator.

To address this interface-loss problem, we propose \strongemph{\ourmethod, a foundation model for intent-to-item fulfillment in agentic shopping.}
\ourmethod retains the knowledge, instruction-following, and interaction abilities needed by an agentic shopping model while being trained to understand e-commerce domain knowledge, operate over SIDs, and translate flexible shopping intents into catalog-grounded outcomes.
Rather than acting as a conversational LLM wrapped around a recommender, or as a SID-based recommender limited to next-item prediction, \ourmethod serves as the central fulfillment layer that decides how the current context should be grounded in the item space.
Depending on the request and interaction state, \ourmethod can produce several kinds of fulfillment outcomes:
\begin{itemize}
    \item \textbf{Item-space operations.}
    It can retrieve items through SID beam-search retrieval, rerank and select items using both SID tokens and textual item information, expand from seed items, and compose complementary items into task-specific sets or bundles.
    \item \textbf{Item-grounded responses.}
    It can directly generate item-grounded summaries, comparisons, and text--SID interleaved responses, or answer user questions directly.
    \item \textbf{Interaction and state updates.}
    It can ask clarification questions, incorporate feedback, or emit context, preference-summary, and preference-memory update signals.
\end{itemize}
We deploy \ourmethod through the \textbf{\ourmethod serving framework}, a model-native fulfillment framework that pairs the central model with a lightweight serving harness.
The harness defines a model-facing action protocol and exposes support surfaces for context access, catalog grounding, and state management, while \ourmethod fills the protocol by selecting and composing model-native fulfillment actions rather than following a fixed retrieval--ranking--response chain.
This model-first design reduces the interface loss between LLM-agent orchestration and recommendation execution.
It is instantiated through three design choices:
\begin{enumerate}
    \item \textbf{Model--harness boundary.}
    \ourmethod carries fulfillment decisions, while the serving harness defines the action protocol through which the model plans, executes, fulfills, and emits updates, and provides the support surfaces through which it requests context, obtains catalog-grounded evidence, and writes updates back to serving state.
    \item \textbf{SIDs for semantic recoverability and LLM operability.}
    We study SIDs that preserve recoverable item semantics while remaining operable by an autoregressive foundation model, which serve as a coherent bridge between the language and item space.
    \item \textbf{Training recipe for item-space fulfillment.}
    Starting from a general LLM, we train \ourmethod to acquire SID-native item-space fulfillment abilities while preserving broad instruction-following and language capabilities.
    The recipe combines SID token alignment, domain continued pre-training, fulfillment SFT, specialist teacher training, and MOPD with task-specific outcome rewards, consolidating item-space specialization without turning the model into a narrow SID predictor.
\end{enumerate}

We evaluate the \ourmethod framework and training recipe at three levels.
First, at the framework level, we compare complete agentic fulfillment configurations under a shared serving abstraction: the \ourmethod framework combines the central model with its serving harness, while tool-mediated baselines combine an agent model with external retrieval and ranking tools.
The benchmark uses single- and multi-turn tasks derived from anonymized Taobao production logs, user-context signals, and catalog snapshots, and evaluates fulfillment quality, grounding, personalization, and feedback adaptation.
Second, capability breakdown examines the model abilities that support these outcomes, including instruction following and knowledge preservation, shopping semantics, context evidence extraction, SID semantic recovery and item mapping, and item-space operation quality.
Finally, ablations on SID design, continued pre-training mixtures, and post-training strategies trace how the SID representation, domain-CPT mixture, and joint OPD--RL post-training recipe shape the final model behavior.

\section{\ourmethod: Framework and Serving Patterns}
\label{sec:shopx}
\label{sec:method}

\begin{figure}[t]
\centering
\includegraphics[width=\textwidth]{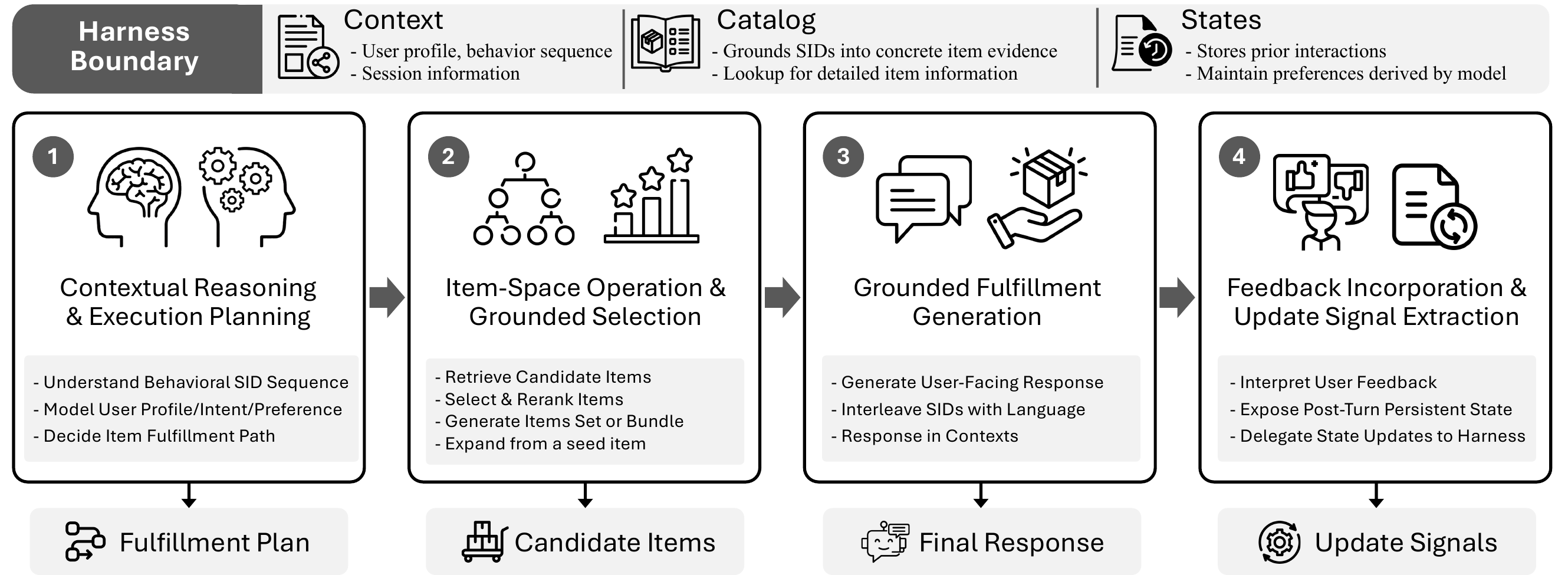}
\caption{
\textbf{The model-native serving framework of \ourmethod.} The \ourmethod model serves as the fulfillment core of the serving framework, enabling contextual intent understanding, fulfillment path planning, SID-native item-space operation, grounded fulfillment generation, feedback incorporation and update-signal extraction. Around this core, the surrounding harness defines a model-facing action protocol and exposes support surfaces for context access, catalog grounding, and state management, which bridge personalization context with the model-native item space and maintain evolving interaction state across turns. 
} 
\label{fig:harness_workflow}
\end{figure}

The \ourmethod serving framework is built around the \ourmethod model and a lightweight harness. The harness defines a model-facing action protocol and support surfaces for context access, catalog grounding, and state management; \autoref{fig:harness_workflow} illustrates this boundary. This separation lets \ourmethod own fulfillment decisions over the item space, while the harness manages the context, catalog evidence, and cross-turn state needed to serve those decisions.

\subsection{Serving-Time Responsibilities of \ourmethod}

We now focus on the model side of the boundary in \autoref{fig:harness_workflow}. The harness plays two distinct roles. First, it defines a model-facing action space with four action slots: \textit{Plan}, \textit{Execute}, \textit{Fulfill}, and \textit{Update}. These slots specify the kinds of decisions and outputs the model can produce at serving time (through prompts). \ourmethod fills these slots by inspecting the current request, deciding what evidence is needed, determining how the intent should be grounded in the item space, carrying out SID-native item-space operations, producing the user-facing response, and emitting update signals for subsequent turns. Second, the harness exposes three support surfaces that the model can use while filling these action slots: \textit{Context} provides profile and history behavior; \textit{Catalog} validates SIDs and grounds them in item evidence; and \textit{State} exposes prior turns, intermediate candidate sets, preference summaries, and write-back of model-emitted updates. The four responsibilities below define the model-side action slots, along with their inputs and outputs.

\begin{enumerate}
    \item \textbf{\textit{Plan}: Contextual Intent Understanding and Execution Planning.}
    \ourmethod interprets the assembled context and decides which fulfillment path is appropriate for the current turn.
    \begin{itemize}
        \item \textbf{Input:} current natural-language request, explicit constraints, and any model-requested \textit{Context} or \textit{State}, including session history, user profile, behavioral SID sequences, previous recommendations, preference memory, and feedback signals.
        \item \textbf{Output:} a fulfillment plan that specifies which evidence is needed and whether to answer directly, perform SID beam-search retrieval, rerank or compare candidate items, expand from seed items, compose a bundle, ask for clarification, or emit state-update signals.
    \end{itemize}

    \item \textbf{\textit{Execute}: SID-Native Item-Space Operations.}
    \ourmethod operates over the SID-defined item space. When generated SIDs need to be tied to concrete products, the model uses \textit{Catalog} to obtain item evidence and then continues selection, reranking, or comparison over catalog-grounded candidates.
    \begin{itemize}
        \item \textbf{Input:} fulfillment plan, interpreted intent, profile signals, constraints, optional seed items or previous candidates from \textit{State}, and any item evidence returned by \textit{Catalog}.
        \item \textbf{Output:} candidate SIDs or structured candidate groups produced through SID beam-search retrieval, seed expansion, or bundle-oriented generation; after \textit{Catalog} grounding, a selected, reranked, compared, or bundled set of catalog-grounded items, optionally with comparison rationales, rejection signals, or bundle structure for response generation.
    \end{itemize}

    \item \textbf{\textit{Fulfill}: Grounded Fulfillment Generation.}
    \ourmethod turns the executed item-space result into the user-facing response, including grounded item presentation and text--SID interleaving.
    \begin{itemize}
        \item \textbf{Input:} fulfillment plan, selected items, resolved \textit{Catalog} evidence, assembled \textit{Context} or \textit{State}, and user-facing constraints.
        \item \textbf{Output:} a recommendation list, product comparison, grounded summary, bundle presentation, text--SID interleaved answer, direct contextual response, or clarification question.
    \end{itemize}

    \item \textbf{\textit{Update}: Feedback Incorporation and Update-Signal Extraction.}
    \ourmethod exposes what should persist after the current turn, while the harness remains responsible for writing the updates into the serving state.
    \begin{itemize}
        \item \textbf{Input:} current request, response, selected items, explicit user feedback, and prior \textit{State}.
        \item \textbf{Output:} preference summaries, preference-memory updates, and task-context updates that \textit{State} can persist and make available in later context assembly.
    \end{itemize}
\end{enumerate}

\newcommand{\applicationpatternflow}[1]{%
\par\vspace{0.55em}
\noindent\makebox[\linewidth][c]{%
\begin{tikzpicture}[
    font=\sffamily\scriptsize,
    >=Stealth,
    harness/.style={draw=thinkcolor!85!black,fill=thinkcolor!18,rounded corners=2pt,align=center,text width=2.02cm,minimum height=0.72cm,inner sep=2.8pt},
    model/.style={draw=observecolor!85!black,fill=observecolor!22,rounded corners=2pt,align=center,text width=2.08cm,minimum height=0.72cm,inner sep=2.8pt},
    userinput/.style={draw=explorecolor!80!black,fill=explorecolor!20,rounded corners=2pt,align=center,text width=1.78cm,minimum height=0.62cm,inner sep=2.4pt},
    call/.style={->,draw=gray!70,line width=0.45pt,shorten >=2pt,shorten <=2pt},
    inputcall/.style={->,draw=explorecolor!70!black,densely dashed,line width=0.45pt,shorten >=2pt,shorten <=2pt}
]
#1
\end{tikzpicture}%
}
\par\vspace{0.25em}
}

\subsection{Serving Patterns in Agentic Shopping}

Rather than enumerating isolated task types, we group recurring agentic shopping uses into three serving patterns that instantiate the action protocol defined above. The first two patterns describe \textit{single-turn fulfillment}, while the third shows how the same protocol becomes \textit{stateful across turns} by reading and writing \textit{State}. In the diagrams, green nodes denote user-provided input, yellow nodes denote harness-side support surfaces, and blue nodes denote action slots filled by \ourmethod.

\begin{enumerate}
    \item \textbf{Intent-to-Item Fulfillment.}
    This pattern covers explicit shopping requests, search-like recommendations, grounded comparisons, explanations, and clarification turns that map current intent to catalog-grounded item evidence.
	    \applicationpatternflow{
	        \node[userinput,text width=2.6cm] (q) {\textbf{User}\\request, constraints};
	        \node[model,text width=2.6cm,right=0.24cm of q] (d) {\textbf{Plan}\\intent \& evidence};
	        \node[model,text width=2.6cm,right=0.24cm of d] (x) {\textbf{Execute}\\retrieval \& rerank};
	        \node[harness,text width=1.75cm,right=0.28cm of x] (g) {\textbf{Catalog}\\item evidence};
	        \node[model,text width=2.5cm,right=0.28cm of g] (o) {\textbf{Fulfill}\\grounded response};
	        \draw[inputcall] (q) -- (d);
	        \foreach \a/\b in {d/x,x/g,g/o} {\draw[call] (\a) -- (\b);}
	    }

    \item \textbf{Context-Augmented Personalization.}
    This pattern applies when the latest user message is vague, context-driven, or underspecified, and \ourmethod must decide whether profile or behavior history is needed before item-space fulfillment.
	    \applicationpatternflow{
	        \node[userinput,text width=2.1cm] (q) {\textbf{User}\\vague / context-rich request};
	        \node[model,text width=1.8cm,right=0.24cm of q] (d) {\textbf{Plan}\\fetch context, then execute};
	        \node[harness,right=0.24cm of d] (h) {\textbf{Context}\\profile, history behavior, \ldots};
	        \node[model,text width=2.8cm,right=0.24cm of h] (e) {\textbf{Execute}\\multi-intent retrieval, bundle generation, \ldots};
	        \node[harness,text width=1.75cm,right=0.24cm of e] (g) {\textbf{Catalog}\\item evidence};
	        \node[model,text width=2.2cm,right=0.24cm of g] (f) {\textbf{Fulfill}\\personalized response};
	        \draw[inputcall] (q) -- (d);
	        \foreach \a/\b in {d/h,h/e,e/g,g/f} {\draw[call] (\a) -- (\b);}
	    }

    \Needspace{4\baselineskip}
    \item \textbf{Stateful Multi-Turn Fulfillment.}
    This pattern covers compound requests, product bundling, and follow-up refinement. \ourmethod reads prior turns, preference memory, or intermediate candidates from \textit{State}, and writes update signals that condition future turns.
	    \applicationpatternflow{
	        \node[userinput,text width=1.8cm] (q) {\textbf{User}\\follow-up request};
	        \node[model,text width=2.1cm,right=0.24cm of q] (d) {\textbf{Plan}\\fetch state, then execute};
	        \node[harness,right=0.24cm of d] (s) {\textbf{State}\\memory, preference, \ldots};
	        \node[model,right=0.24cm of s] (x) {\textbf{Execute}\\revise, expand, explain, \ldots};
	        \node[harness,text width=1.75cm,right=0.24cm of x] (g) {\textbf{Catalog}\\item evidence};
	        \node[model,right=0.24cm of g] (o) {\textbf{Fulfill}\\reply \& updates};
	        \draw[inputcall] (q) -- (d);
	        \foreach \a/\b in {d/s,s/x,x/g,g/o} {\draw[call] (\a) -- (\b);}
	        \draw[call,densely dashed] (o.south) .. controls +(0,-0.35) and +(0,-0.35) .. node[below,font=\sffamily\tiny,text=gray!75] {write-back / later turn} (s.south);
	    }
\end{enumerate}
\vspace{-1.5em}
\noindent Appendix~\ref{app:case_studies_and_traces} collects concrete case-style materials for these patterns, including detailed serving-time traces that show how the model fills the action slots and how the harness grounds or carries state across the trajectory.

\section{Evaluation Protocol and Benchmarks}
\label{sec:evaluation_protocol}

The serving patterns in \S\ref{sec:shopx} require an accompanying evaluation suite.
This section defines the protocols used to compare complete serving frameworks and diagnose the model capabilities underlying these outcomes.
We organize the evaluation around two complementary views:
\begin{itemize}
    \item \textbf{Framework-level evaluation (\S\ref{sec:framework_level_evaluation_protocol}).}
    This is a framework-level comparison of model-native and tool-mediated fulfillment paradigms under a shared serving abstraction.
    The protocol uses original user queries from anonymized Taobao production logs, together with profile context and catalog snapshots, to test whether each system can carry rich shopping context into catalog-grounded outcomes.
    \item \textbf{Capability-breakdown evaluation (\S\ref{sec:capability_breakdown_evaluation_protocol}).}
    This protocol breaks down the model-side abilities behind those outcomes, using shopping-domain diagnostics together with public general-capability benchmarks that can be related back to framework-level behavior.
\end{itemize}

\subsection{Framework-Level Evaluation}
\label{sec:framework_level_evaluation_protocol}

\begin{figure}[t]
\centering
\includegraphics[width=\textwidth]{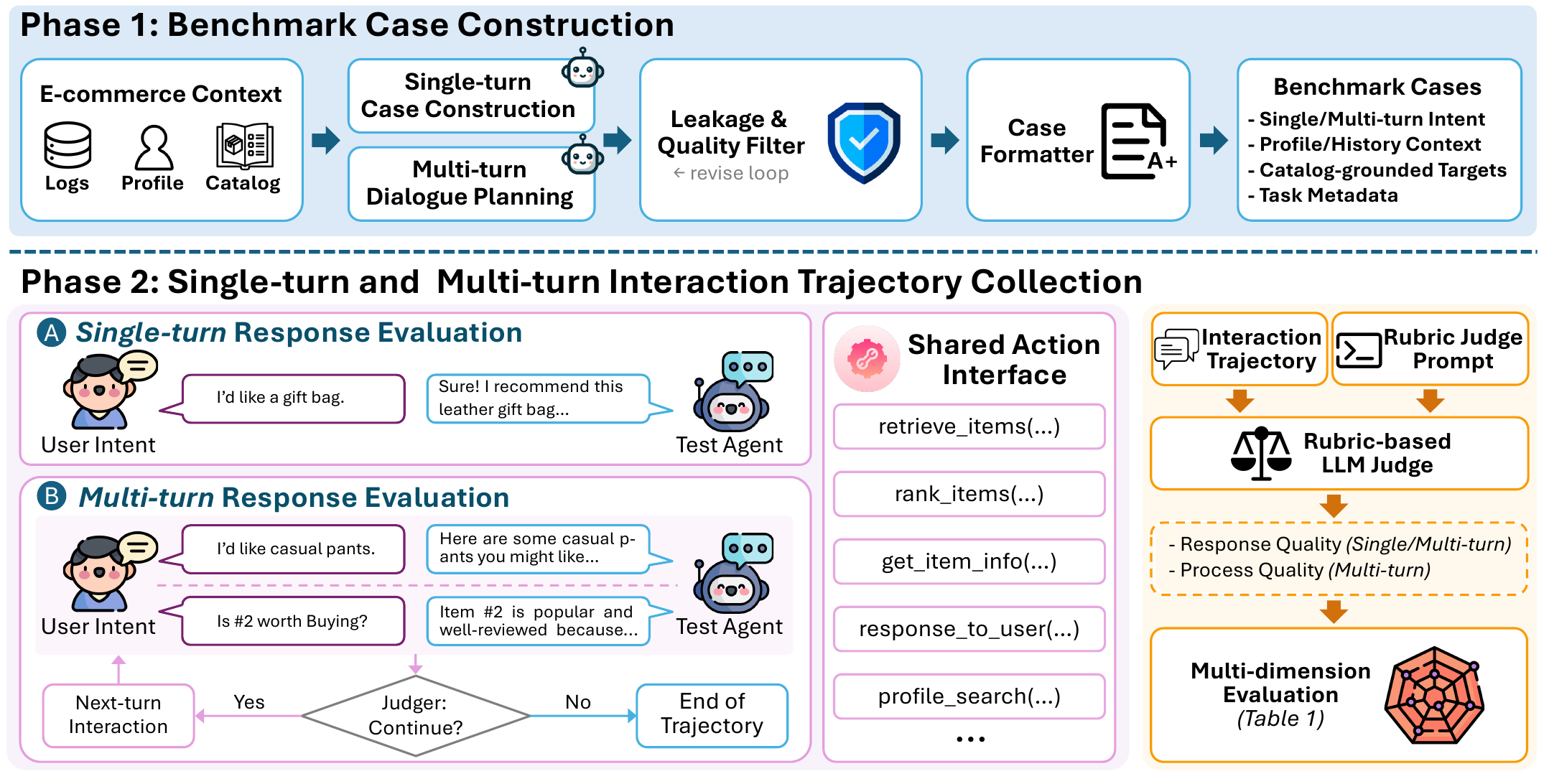}
\caption{
\textbf{Framework-level evaluation.}
Complete \ourmethod and tool-mediated serving frameworks are compared on shared tasks built from original user queries, context, catalog snapshots, and output specifications derived from anonymized Taobao production logs.
The overview metrics summarize intent fulfillment, item precision, ranking quality, category coverage, personalization, constraint grounding, feedback adaptation, and cross-turn reference.
}
\label{fig:evaluation_framework}
\end{figure}

Framework-level evaluation compares complete fulfillment frameworks under a shared task context, catalog-grounding scope, and output contract.
We first define the compared paradigms and item universe, then describe the two-phase case-construction and interaction-collection protocol in \autoref{fig:evaluation_framework}.

\paragraph{Compared fulfillment paradigms.}
The benchmark is designed to evaluate the entire path from a user's shopping intent to catalog-grounded fulfillment outcomes.
The evaluation pipeline fixes the user request, profile context, behavior history, catalog snapshot, and output format for every evaluated system.
Within this fixed context, we compare two fulfillment paradigms:
\begin{enumerate}
    \item \emph{Model-native fulfillment.} \ourmethod performs contextual reasoning, SID-native item-space operations, grounded fulfillment generation, and feedback or state updates through a model-native interface.
    This covers retrieval and selection, comparisons or bundles, text--SID interleaved responses, and multi-turn follow-up refinement.
    \item \emph{Tool-mediated fulfillment.} Following prior LLM-agent recommender designs~\citep{chat_rec,interecagent,recmind}, we equip all baseline agents with a shared LLM-external Taobao item-search tool, backed by a mature retrieval and ranking stack.
    The tool takes keyword queries and optional price filters, returns catalog item metadata in search-engine order, and the baseline LLM may then compress, rerank, or summarize the returned candidates before producing the final user-visible items.
\end{enumerate}

\paragraph{Catalog grounding scope.}
All catalog-grounded evaluations are anchored to a large Taobao item snapshot containing approximately 1.2B items, covering a major portion of the production item pool.
This snapshot defines the evaluation item universe for catalog resolution, item evidence lookup, and fulfillment judging.

\paragraph{Benchmark case construction.}
Phase~1 in \autoref{fig:evaluation_framework}, \emph{Benchmark Case Construction}, constructs controlled cases from anonymized Taobao interaction logs, profile and behavior signals, and catalog snapshots through single-turn case creation and multi-turn dialogue planning.
The seed queries are real Taobao user queries in their original form; we do not rewrite, paraphrase, or synthetically expand them for framework-level evaluation.
Each benchmark case records the original query, shopping intent, profile/history context, catalog-grounded target information, and task metadata, then passes leakage and quality filters before a human validity audit.
The resulting framework-level benchmark contains 279 single-turn cases and 80 multi-turn trajectories.
The single-turn split is stratified by intent type (\emph{precise-category}, \emph{scene-based}, \emph{multi-category}, \emph{broad semantic}, and \emph{minimal-query}) and difficulty tier (\emph{robust}, \emph{adversarial}, and \emph{semantic}).
The multi-turn split is grouped by seed type (\emph{vague}, \emph{constrained}, \emph{clear}, \emph{multi-need}, and \emph{ambiguous}), so trajectory-level scoring can test clarification, constraint persistence, follow-up refinement, and ambiguity resolution in addition to direct fulfillment.
Appendix~\ref{app:framework_benchmark_composition} reports the full benchmark~composition.

\paragraph{Interaction trajectory collection.}
Phase~2 in \autoref{fig:evaluation_framework}, \emph{Single-turn and Multi-turn Interaction Trajectory Collection}, rolls each constructed case into a single-turn response or multi-turn interaction trace for evaluation.
For multi-turn cases, Claude Sonnet 4.6 acts as a fixed shopper-side simulator, producing user turns from the profile information, seed query, and the agent's previous response.
Appendix~\ref{app:multi_turn_rollout_protocol} gives the rollout cap and stopping rule.
Across systems, the rollout fixes the case context, shared action interface, and final output contract: response text with at most five user-visible items.
Intermediate retrieval, ranking, item-information lookup, profile search, and candidate selection are left to each framework, so the comparison targets the resulting catalog-grounded trajectory rather than a common internal pipeline.

\paragraph{Rubrics-based LLM judge scoring.}
Following recent LLM-as-judge and rubric-based evaluation practice~\citep{g_eval,llm_as_judge_mtbench,prometheus2,rubrics_as_rewards}, rubric-based components use task-specific judge prompts rather than a single free-form preference question.
All framework-level rubric scoring uses Claude Sonnet 4.6 as a fixed judge, separate from the multi-turn user simulator.
The rubric judge prompts cover response quality for both single- and multi-turn outputs and process quality for multi-turn trajectories.
For each recorded response or trajectory, the judge receives the user intent, profile/history context, returned items, catalog evidence, and model response, then emits structured dimension scores after invalid-response checks for missing items, unsupported claims, or mismatched categories.
For reporting, we map the single-turn and multi-turn scores into eight 0--100 overview metrics, using the fixed metric compositions summarized in \autoref{tab:framework_overview_metrics}.
\begin{table}[t]
\centering
\caption{\textbf{Framework overview metrics.}
Eight 0--100 metrics used to summarize framework-level fulfillment behavior from single-turn responses and multi-turn trajectories.}
\label{tab:framework_overview_metrics}
\scriptsize
\begingroup
\renewcommand\tabularxcolumn[1]{p{#1}}
\setlength{\tabcolsep}{3pt}
\renewcommand{\arraystretch}{1.18}
\begin{tabularx}{\linewidth}{@{}>{\centering\arraybackslash}p{0.045\linewidth}>{\raggedright\arraybackslash}p{0.18\linewidth}>{\raggedright\arraybackslash}p{0.26\linewidth}>{\raggedright\arraybackslash}X@{}}
\toprule
\textbf{No.} & \textbf{Metric} & \textbf{Score source / composition} & \textbf{What it measures} \\
\midrule
1 &
Intent Fulfillment &
Mean of single-turn intent served rate and multi-turn goal achievement &
Successful conversion of the user's shopping intent into a satisfactory catalog-grounded outcome. \\
\midrule[0.15pt]
2 &
Item Precision &
Single-turn category precision &
Match between the served item set and the annotated shopping category set. \\
\midrule[0.15pt]
3 &
Ranking Quality &
Single-turn NDCG@5 over the displayed item list &
Ordering quality of displayed items under item-level relevance labels. \\
\midrule[0.15pt]
4 &
Category Coverage &
Mean of single-turn and multi-turn category coverage &
Coverage of required category facets for multi-facet, bundle-style, or complementary-item requests. \\
\midrule[0.15pt]
5 &
Personalization &
Mean of single-turn and multi-turn profile alignment &
Use of profile and behavior evidence in item selection, ordering, and recommendation rationale. \\
\midrule[0.15pt]
6 &
Constraint Grounding &
Mean of single-turn attribute accuracy and multi-turn constraint-grounding score &
Faithful handling of hard constraints, accumulated preferences, clarification needs, and catalog-evidence limits. \\
\midrule[0.15pt]
7 &
Feedback Adaptation &
Multi-turn feedback-response score &
Ability to incorporate user feedback, revise recommendations, preserve unaffected preferences, and avoid regressions. \\
\midrule[0.15pt]
8 &
Cross-turn Reference &
Multi-turn reference-resolution score &
Resolution of references to previously discussed items and correct application of the requested operation. \\
\bottomrule
\end{tabularx}
\endgroup
\end{table}

These metrics support the framework-level results and capability-breakdown analyses in \S\ref{sec:end_to_end_system_evaluation} and \S\ref{sec:capability_diagnostics_workflow_transfer}, as well as the radar visualization in \autoref{fig:graphical_abstract}.
Appendix~\ref{app:framework_overview_metric_protocol} gives the corresponding judge dimensions and aggregation details.

\subsection{Capability Breakdown Evaluation}
\label{sec:capability_breakdown_evaluation_protocol}

\begin{figure}[!t]
\centering
\includegraphics[width=\textwidth,keepaspectratio]{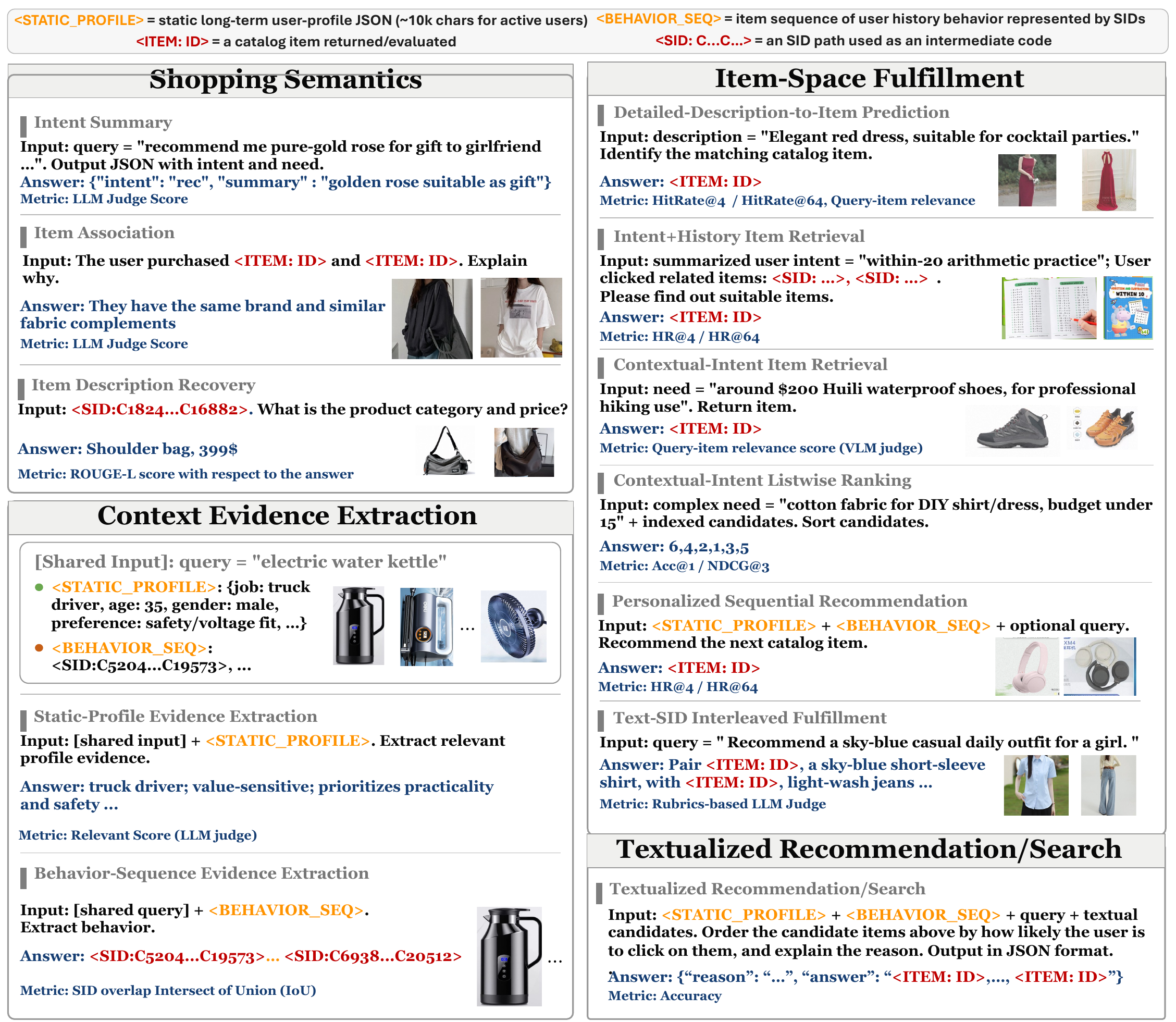}
\caption{
\textbf{Shopping capability diagnostics.}
Representative input--output patterns and primary metric types for the four shopping diagnostic blocks used in \autoref{tab:capability_diagnostics_workflow_transfer}: shopping semantics, context evidence extraction, textualized recommendation/search, and item-space fulfillment.
}
\label{fig:ecommerce_task_examples}
\end{figure}

Capability breakdown evaluation diagnoses which model abilities account for framework-level outcomes.
We split the suite into two scopes: general abilities that remain necessary inside agentic shopping, and shopping-specific diagnostics.
The resulting tasks, metrics, and results are reported in \S\ref{sec:capability_diagnostics_workflow_transfer}; \S\ref{app:capability_metric_protocol} gives detailed metric definitions and rubric-based judge details.

\paragraph{General capabilities for agentic shopping.}
Agentic shopping still depends on core foundation-model abilities: following user instructions and output contracts, applying broad factual and domain knowledge, handling multilingual or culturally specific product language, and retaining basic reasoning, math, and coding robustness.
We therefore use public benchmarks not as an unrelated leaderboard, but to verify that shopping specialization keeps the model usable as the central agentic interface rather than narrowing it into a pure SID predictor.
MMLU-Pro~\citep{mmlu_pro}, CMMLU~\citep{cmmlu}, and IFEval~\citep{ifeval} are the primary checks for broad knowledge and instruction following, while BBH~\citep{bbh}, GPQA-Diamond~\citep{gpqa}, MATH-500~\citep{math}, GSM8K~\citep{gsm8k}, and MBPP+~\citep{mbpp,evalplus} provide additional probes for reasoning, math, and coding robustness.

\paragraph{Shopping capability diagnostics.}
The shopping diagnostics, illustrated in \autoref{fig:ecommerce_task_examples}, are anchored in Taobao-derived catalog, interaction, and profile signals.
When catalog grounding is required, the diagnostic cases and candidate items are sampled from the approximately 1.2B-item Taobao snapshot defined in \S\ref{sec:framework_level_evaluation_protocol}.
The suite probes four model-facing abilities:
\begin{itemize}
    \item \emph{Shopping semantics} tests whether the model can turn underspecified or noisy shopping language into a coherent purchase intent, reason about product relations such as substitutes and complements, and recover item semantics from SIDs.
    \item \emph{Context evidence extraction} tests whether it can select the user-profile facts and behavior-history signals that are actually relevant to the current request, rather than treating the full context as uniformly useful.
    \item \emph{Textualized recommendation/search} tests catalog-aware preference matching when candidate items are exposed only through text, separating semantic shopping judgment from SID generation and allowing general LLMs, including GPT and Gemini models without native catalog-item representations, to be evaluated on the same decision problems.
    \item \emph{Item-space fulfillment} tests whether the model can operate over catalog items after the intent is understood, including detailed-description-to-item resolution, next-item prediction, retrieval, ranking, and open-ended responses that interleave natural language with resolvable SIDs.
\end{itemize}
The Interleaved Text--SID Fulfillment rubric further checks whether generated SIDs are contextually appropriate, correctly grounded, and well integrated into the response; Appendix~\ref{app:text_sid_interleaved_fulfillment_rubric} gives the full rubric and penalty rules.

\section{Semantic IDs for Bridging Language and Item Space}
\label{sec:semantic_ids}

Semantic IDs make item spaces generable by sequence models.
We build on FORGE~\citep{forge}, a collaborative-heavy SID construction framework developed for industrial generative retrieval and deployed at full scale in Taobao's ``Guess You Like'' scenario.
FORGE shows that discrete item codes can serve as effective generation targets for candidate retrieval, and that SID quality can be diagnosed before costly downstream generative-retrieval training.
In \ourmethod, the SID serves a different role: it becomes a model-native item language for connecting natural-language intents, user context, product semantics, and catalog-grounded fulfillment actions.
This role shift leads to two design goals:
\begin{itemize}
    \item \textbf{Recoverability.}
    The representation and SID should preserve item semantics, so the model can recover or reason about category, attributes, style, function, and visual appearance rather than merely memorize opaque codes.
    \item \textbf{LLM operability.}
    The SID should be learnable and stable as an autoregressive generation target, with prefix structure that avoids high-entropy next-token choices and keeps related items in coherent neighborhoods.
\end{itemize}
We therefore redesign the item representation, SID structure, and validation protocol around these goals: \S\ref{sec:sid_item_representation} introduces semantic item representations, \S\ref{sec:sid_structure_intrinsics} constructs hybrid global/local SIDs, and \S\ref{sec:sid_codebook_validation} describes a lightweight protocol for comparing and selecting candidate SID versions before the full training recipe.

\begin{figure}[!tbp]
\centering
\includegraphics[width=\textwidth]{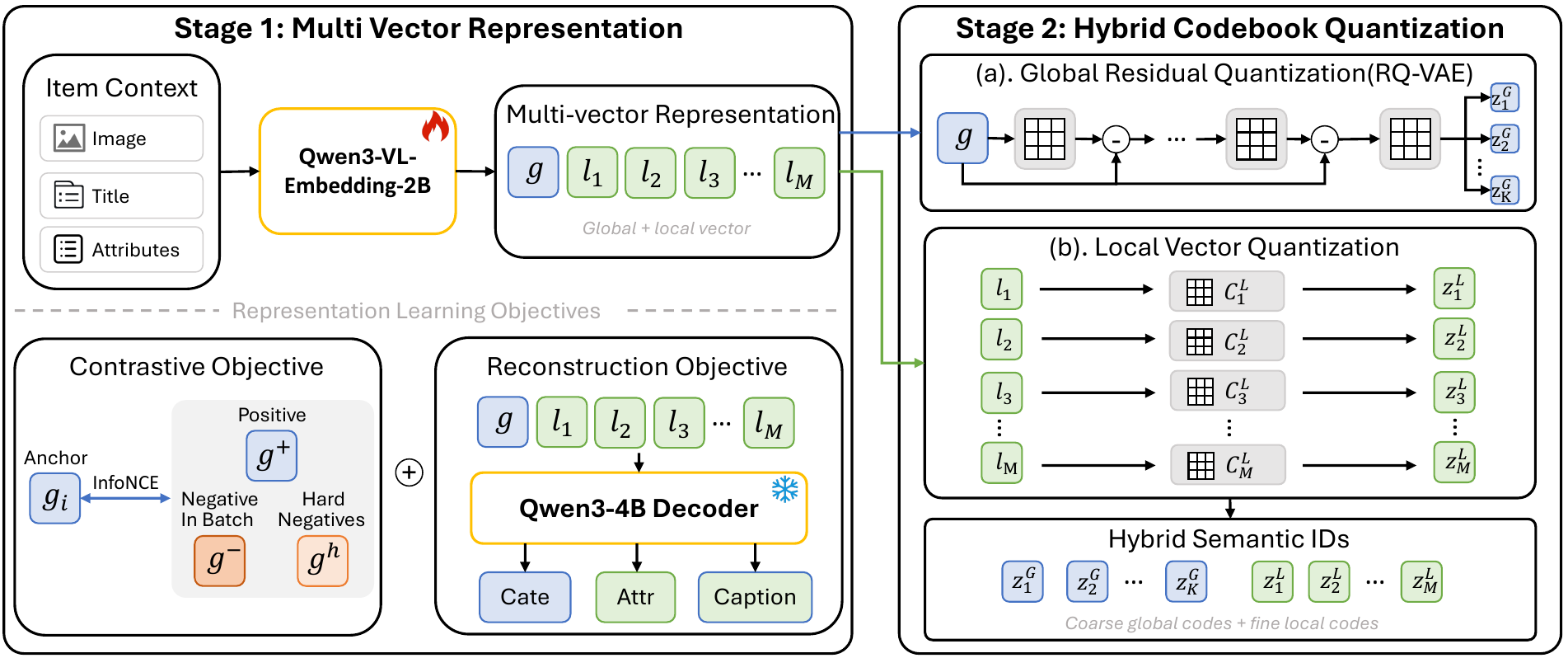}
\caption{
\textbf{Semantic ID architecture for bridging language and item space.}
Multimodal item content is encoded into a multi-vector representation with one global embedding and multiple local facet embeddings.
The global embedding is contrastively organized and residual-quantized into autoregressive prefix tokens, while local embeddings are trained with reconstruction objectives and vector-quantized into suffix tokens for additional recoverable semantic capacity.
}
\label{fig:sid}
\end{figure}

\subsection{Multimodal Item Representation Learning}
\label{sec:sid_item_representation}

\autoref{fig:sid} gives the overall design: multimodal item content is first encoded into continuous item representations, which are then discretized into global-prefix and local-suffix SIDs.
For each catalog item $i$, we denote its multimodal input as $x_i=(v_i,a_i,t_i)$, where $v_i$ denotes product images, $a_i$ structured attributes, and $t_i$ the item title.
Let $E_{\theta}$ be the multimodal item encoder that maps $x_i$ into continuous item representations.
Starting from the prior collaborative SID pipeline, we modify this representation stage along three axes: the encoder backbone, the training supervision, and the output representation structure.

\paragraph{Semantic backbone.}
We upgrade the representation backbone from a modular multimodal encoder to Qwen3-VL-Embedding-2B~\citep{qwen3vl}, used as the VLM embedding encoder.
The prior pipeline encodes modalities separately, using CLIP-style visual encoders and BERT-style text encoders~\citep{clip,bert}, before fusing features through a Q-Former-style bridge~\citep{blip2}.
Qwen3-VL-Embedding-2B instead processes the full item input $x_i=(v_i,a_i,t_i)$ directly, so image evidence, structured attributes, and title language cues can interact before SID construction.

\paragraph{Semantic supervision.}
Training supervision enters through the positive-pair relation used by contrastive learning.
Let $\mathcal{R}$ be a relation over items, and let $\mathcal{P}_{\mathcal{R}}(i)$ be the positive items for anchor $i$ under that relation.
Collaborative-product supervision instantiates this relation as $\mathcal{R}_{\mathrm{CP}}$, whose positives are behavior-derived pairs such as items frequently clicked or purchased by the same users.
These pairs are useful for retrieval, but they are not guaranteed to be semantically similar, making CP less ideal for grounding language or user intent to item semantics.
We instead use equivalent-product supervision $\mathcal{R}_{\mathrm{EP}}$, whose positives are different listings that correspond to the same underlying product.
For the equivalent-product setting, we also add same-category but non-equivalent items into each batch as hard negatives.
Rather than treating them as ordinary negatives, we use a soft-target InfoNCE objective~\citep{cpc,soft_target_infonce}.
For a mini-batch $\mathcal{B}$, let $\mathcal{H}(i)$ be the sampled hard negatives, $\mathcal{N}(i)$ the remaining in-batch negatives, and $\mathcal{C}(i)=\mathcal{P}_{\mathcal{R}}(i)\cup\mathcal{H}(i)\cup\mathcal{N}(i)$.
For normalized global embeddings $g_i$, we define the soft target $q_{i,k}$ by assigning unnormalized weights $1$ to $k\in\mathcal{P}_{\mathcal{R}}(i)$, $\alpha$ to $k\in\mathcal{H}(i)$, and $0$ to $k\in\mathcal{N}(i)$, followed by normalization over $\mathcal{C}(i)$.
The contrastive loss is
\begin{equation}
    \mathcal{L}_{\mathrm{CL}}(\mathcal{R})
    =
    -\frac{1}{|\mathcal{B}|}
    \sum_{i\in\mathcal{B}}
    \sum_{k\in\mathcal{C}(i)}
    q_{i,k}
    \log
    \frac{\exp(\operatorname{sim}(g_i,g_k)/\tau)}
    {\sum_{u\in\mathcal{C}(i)}\exp(\operatorname{sim}(g_i,g_u)/\tau)} .
\end{equation}
Here $\operatorname{sim}(\cdot,\cdot)$ is cosine similarity, $\tau$ is the temperature, and $\alpha\in(0,1)$ controls the hard-negative weight.
Together, equivalent-product supervision and the graded target make equivalent products the strongest positives, give same-category hard negatives a weaker semantic pull, and avoid collaborative priors that are not directly recoverable from item content.

\paragraph{Single- and multi-vector outputs.}
Single-vector representations compress all item evidence into one pooled embedding, which can lose fine-grained matching signals that are useful for item semantics~\citep{colbert,metaembed}.
To retain these signals while keeping a global representation for contrastive learning and prefix quantization, we compare two output structures: SV uses a single global vector, whereas MV augments the global vector with local vectors for finer semantic aspects such as attributes and visual details:
\[
    E_{\theta}^{\mathrm{SV}}(x_i)\coloneqq g_i,
    \qquad
    E_{\theta}^{\mathrm{MV}}(x_i)\coloneqq \left(g_i,\ell_{i,1},\ldots,\ell_{i,M}\right).
\]
Here $M$ is the number of local vectors.
The global embedding $g_i$ is used for contrastive learning and later global-prefix quantization.
The local embeddings $\ell_{i,m}$ are trained through caption and attribute reconstruction, which gives them a direct semantic prediction role and discourages collapse into redundant copies of the global embedding; $g_i$ is also supervised by category reconstruction.
For reconstruction training, let $D_{\psi}$ be a frozen LLM decoder and $A_{\eta}$ a trainable adapter that maps item representations into decoder-conditioning tokens.
Given a representation sequence $Z_i$, prompt $q_i$, and target text $y_i=(y_{i,1},\ldots,y_{i,|y_i|})$, the decoder-side next-token prediction loss is
\begin{equation}
    \mathcal{L}_{\mathrm{NTP}}(Z_i,q_i,y_i)
    =
    -
    \sum_{r=1}^{|y_i|}
    \log p_{\psi}\!\left(y_{i,r}\mid y_{i,<r}, q_i, A_{\eta}(Z_i)\right),
\end{equation}
where $p_{\psi}$ is the decoder next-token distribution; $\psi$ is frozen and gradients update the item encoder and adapter.
We use three prompt-specific reconstruction targets: category reconstruction from $Z_i^{\mathrm{cat}}\coloneqq(g_i)$, VLM-caption reconstruction from $Z_i^{\mathrm{cap}}\coloneqq(g_i,\ell_{i,1},\ldots,\ell_{i,M})$, and facet-attribute reconstruction from $Z_{i,m}^{\mathrm{attr}}\coloneqq(g_i,\ell_{i,m})$.
Let $\ell_i^{\mathrm{cat}}$, $\ell_i^{\mathrm{cap}}$, and $\ell_{i,m}^{\mathrm{attr}}$ denote the corresponding $\mathcal{L}_{\mathrm{NTP}}$ losses with their task-specific prompts and targets.
The reconstruction objective averages these losses as
\begin{equation}
    \mathcal{L}_{\mathrm{TR}}
    =
    \frac{1}{|\mathcal{B}|}\sum_{i\in\mathcal{B}}
    \left(\ell_i^{\mathrm{cat}}+\ell_i^{\mathrm{cap}}\right)
    +
    \frac{1}{|\mathcal{B}|M}
    \sum_{i\in\mathcal{B}}\sum_{m=1}^{M}
    \ell_{i,m}^{\mathrm{attr}}.
\end{equation}
The full representation objective is $\mathcal{L}_{\mathrm{rep}}=\mathcal{L}_{\mathrm{CL}}(\mathcal{R})+\lambda_{\mathrm{TR}}\mathcal{L}_{\mathrm{TR}}$, with $\lambda_{\mathrm{TR}}$ controlling reconstruction.
\autoref{tab:item_rep_intrinsics_ablation} ablates these representation choices before SID construction, with the corresponding results, metrics, and analysis discussed in \S\ref{sec:sid_design_ablation}.

\subsection{Hybrid Global--Local SID Tokenization}
\label{sec:sid_structure_intrinsics}

After item representations are learned, the SID tokenizer converts them into a discrete sequence that can be generated by an LLM.
We use $G_K$ to denote $K$ global-prefix tokens produced from the global embedding $g_i$, and $L_M$ to denote $M$ local-suffix tokens produced from the local embeddings $\{\ell_{i,m}\}_{m=1}^{M}$.
A global-only SID and a hybrid global--local SID are written as
\[
    G_K(i)\coloneqq\left(z^G_{i,1},\ldots,z^G_{i,K}\right),
    \qquad
    G_K(i)\Vert L_M(i)\coloneqq
    \left(z^G_{i,1},\ldots,z^G_{i,K},z^L_{i,1},\ldots,z^L_{i,M}\right).
\]
Here $\Vert$ denotes sequence concatenation.

\paragraph{Global RQ prefix.}
The global prefix is obtained by residual quantization of $g_i$, following vector-quantized autoencoders (VQ-VAE) and residual-quantized VAE (RQ-VAE)~\citep{vqvae,rqvae,irvq}.
We focus on the SID assignment rule used after codebook learning; standard RQ-VAE codebook-training details are omitted for brevity.
Let $\mathcal{C}^G_k=\{e^G_{k,c}\}_{c=1}^{V^G_k}$ be the codebook at global level $k$, where $V^G_k$ is the number of codewords at that level.
Starting from $r_{i,0}=g_i$, each prefix token is selected by
\begin{equation}
    z^G_{i,k}
    =
    \arg\min_{c\in [V^G_k]}
    \left\|r_{i,k-1}-e^G_{k,c}\right\|_2^2,
    \qquad
    r_{i,k}
    =
    r_{i,k-1}-e^G_{k,z^G_{i,k}}.
\end{equation}
Here $[V]$ denotes the index set $\{1,\ldots,V\}$.
The reconstructed global vector is $\hat{g}_i=\sum_{k=1}^{K} e^G_{k,z^G_{i,k}}$.
\emph{The operability of this prefix comes from the semantic geometry learned in \S\ref{sec:sid_item_representation}:} equivalent-product contrastive learning and same-category hard negatives organize the global space before residual quantization turns it into an autoregressive prefix path.

\paragraph{Local VQ suffix.}
For MV representations, the local embeddings are quantized separately with VQ-style codebooks instead of being folded into the same global residual chain.
Let $\mathcal{C}^L_m=\{e^L_{m,c}\}_{c=1}^{V^L_m}$ be the codebook for local facet $m$, where $V^L_m$ is the number of local codewords.
The suffix token for that facet is
\begin{equation}
    z^L_{i,m}
    =
    \arg\min_{c\in [V^L_m]}
    \left\|\ell_{i,m}-e^L_{m,c}\right\|_2^2.
\end{equation}
These local suffix tokens add recoverable semantic capacity for attributes, captions, and visual facets, rather than replacing the global prefix as the main routing signal.

\paragraph{Difference from global-only SIDs.}
Prior RQ-VAE-style SID construction applies residual quantization to one global item representation, so every SID token must serve both semantic routing and fine-grained semantic recovery.
Our hybrid design separates these roles: $G_K$ supports stable LLM generation, while $L_M$ preserves facet-level information for recoverability.
A similar separation between hierarchical codes and fine-grained item capacity also appears in the RQ-OPQ design of OneSearch~\citep{onesearch}.
Thus variants such as $G_3$, $G_6$, and $G_2+L_4$ are not merely different SID lengths; they assign different roles to prefix and suffix tokens.
Their concrete settings, intrinsic diagnostics, and codebook-validation results are discussed in \S\ref{sec:sid_design_ablation}.
The final \ourmethod recipe uses the hybrid $G_2+L_4$ form, with two global-prefix levels for LLM-operable autoregressive generation and four local-suffix levels for finer item evidence.

\subsection{Lightweight Validation of SID Versions}
\label{sec:sid_codebook_validation}

The representation and SID tokenization steps above define candidate SID versions, but construction alone does not determine whether a SID is suitable for LLM-based item grounding.
We evaluate candidate versions at three levels.
Representation intrinsics test whether the continuous item space is semantically recoverable and neighborhood-stable.
Global-prefix intrinsics test whether quantization preserves semantic structure, prefix stability, and healthy code usage.
Lightweight codebook validation then checks the model-facing question: after a small amount of alignment, can an LLM use the SID interface to map between item language and catalog items?
For each candidate codebook, we assign SIDs to the same fixed 25M-item mini catalog and train the same 4B LLM with a small validation recipe: a subset of SID-alignment data followed by a lightweight SFT pass for task formatting.
With the starting checkpoint, catalog, alignment budget, and SFT budget held fixed, this validation tests how easily each SID version can be learned and used by the model.

The validation centers on bidirectional language--item grounding.
In the forward direction, detailed item descriptions are decoded into SID candidates and resolved to catalog items, testing whether the SID can serve as a model-operable item target.
In the reverse direction, SID-conditioned item-description recovery asks the model to answer attribute- or facet-level questions about the represented item, testing semantic recoverability under SID conditioning.
Concrete metric definitions and decoding details are given in \S\ref{sec:sid_design_ablation} and \S\ref{sec:experimental_implementation_details}.

\section{Continued Pre-Training}
\label{sec:cpt}

\begin{figure}[!tbp]
\centering
\includegraphics[width=\textwidth]{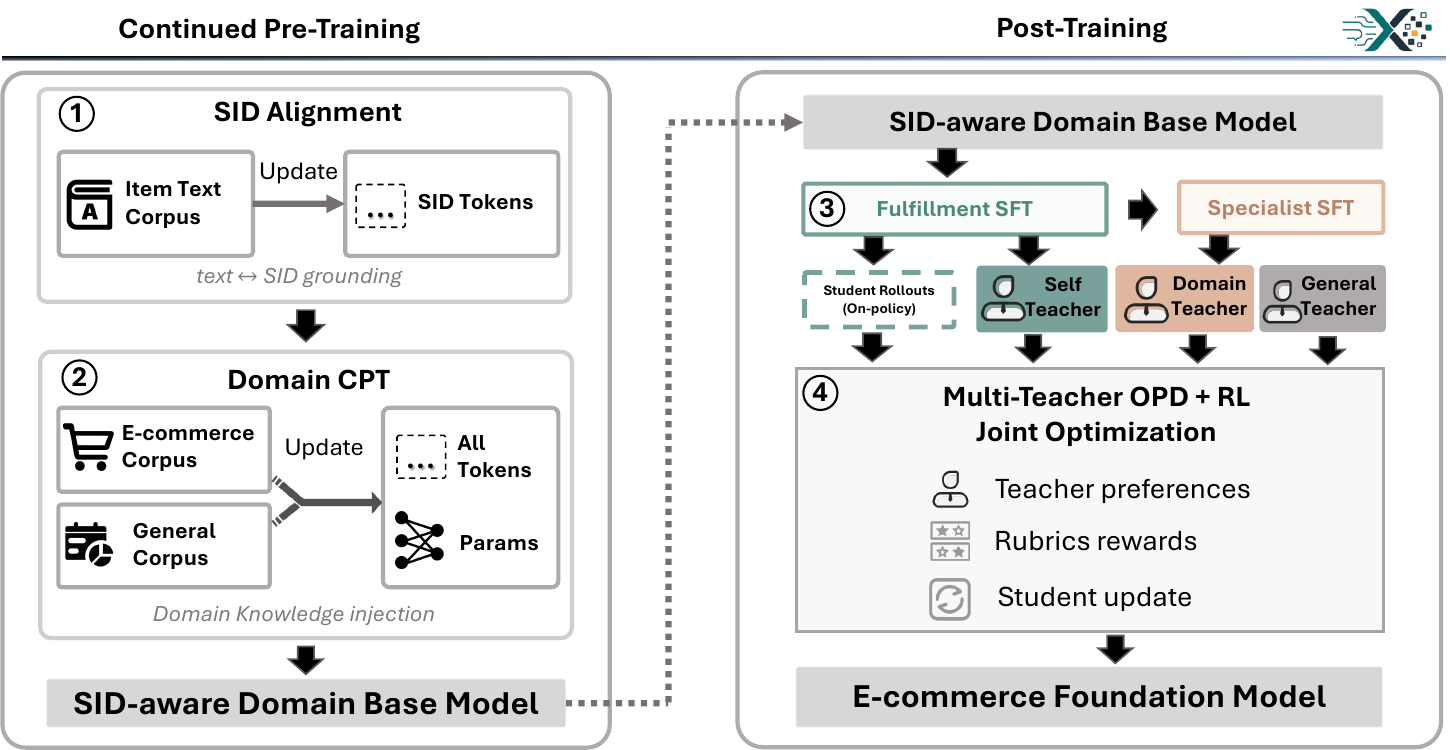}
\caption{
\textbf{Training pipeline for \ourmethod.}
The recipe starts from a general LLM and proceeds through SID token alignment, domain continued pre-training, fulfillment SFT, and joint OPD--RL post-training.
The final OPD--RL stage is expanded in \autoref{fig:task_routed_mopd}.
}
\label{fig:training_pipeline}
\end{figure}

After defining SIDs as a bridge between language and the item space in \S\ref{sec:semantic_ids}, we next describe how a general LLM is adapted toward SID-aware intent-to-item fulfillment.

\tightparagraph{Base model family.}
Unless otherwise stated, the reported 4B and 8B \ourmethod checkpoints are initialized from Qwen3-family backbones~\citep{qwen3} and use the same tokenizer, training recipe, and data mixtures; their main difference is model scale.

This section covers the first half of the training recipe, as summarized in \autoref{fig:training_pipeline}.
SID token alignment makes the SID vocabulary usable by the model, while domain continued pre-training adapts the model to catalog structure, item semantics, shopping intents, and behavior-grounded signals without eroding the general knowledge, reasoning, and instruction-following abilities needed for agentic shopping.
These stages establish the model's item language and shopping-domain grounding before task-level workflow learning begins.
The post-training stages that teach the model to execute the full fulfillment protocol are described separately in \S\ref{sec:post_training}.

\subsection{SID Token Alignment}

SID alignment establishes the bidirectional interface between natural-language item semantics and the SID token space before broader recommendation training.
At this stage, the data does not yet cover intent-to-item fulfillment examples.
Instead, we use catalog-derived item-to-SID and SID-to-item reconstruction data to teach the model that SIDs are model-operable item identifiers rather than opaque labels.
Because all original model parameters are frozen in this stage, the objective only needs to attach the SID token interface to the base LLM rather than rewrite its language or reasoning behavior.
The stage is therefore centered on broad item coverage and legal SID-token exposure rather than on downstream fulfillment supervision.
In the final recipe, this stage adds the vocabulary for our final SID configuration ($G_2+L_4$) and trains only the newly added token embeddings, keeping the original backbone frozen.
This embedding-only stage is trained for one epoch over 200B alignment tokens, using 1{,}024-token sequences, next-token pre-training loss on all tokens, and Adam with peak learning rate $1\times10^{-4}$.
Additional implementation details and the vocabulary size of our final SID are summarized in \S\ref{sec:experimental_implementation_details}.

\subsection{Domain Continued Pre-training: Data Mixture and Settings}

Domain continued pre-training prepares the model for catalog--SID grounding, intent- and behavior-conditioned item reasoning, user preference understanding, and grounded generation by mixing shopping-domain data with general replay.
Below, we describe the replay ratio, the resulting data mixture, and the training setup used for this stage.

\tightparagraph{Replay ratio selection.}
General replay prevents catastrophic forgetting of instruction following, broad knowledge, multilingual ability, and basic reasoning robustness during domain adaptation~\citep{llama3}.
The ratio between domain and replay data controls the specialization--preservation trade-off, so we select it through the ablation in \S\ref{sec:cpt_data_mixture_ablation}.
The final full-scale recipe uses approximately 70B shopping-domain tokens and 35B general-replay tokens, corresponding to a 2:1 domain-to-general token ratio.
The detailed sweep, source-level accounting, and resulting trade-off are analyzed in \S\ref{sec:cpt_data_mixture_ablation}.

\tightparagraph{Data mixture.}
\autoref{tab:cpt_data_distribution} summarizes the final CPT mixture, which contains 303.0\,M examples and 113.96\,B tokens.
The general replay portion accounts for 9.0\% of examples and 33.4\% of tokens, covering knowledge and reasoning, instruction following, and math and coding data to preserve broad model capabilities during domain adaptation.
For this CPT replay pool, we combine curated in-house data with filtered public corpora for web knowledge, reasoning, math, and code, including FineWeb~\citep{fineweb}, Step-3.5-Flash~\citep{step35_flash}, Nemotron-CC-Math~\citep{nemotron_cc_math}, and the general CPT data collected by OpenOneRec~\citep{openonerec}.
All candidate sources are deduplicated, length-filtered, and quality-filtered before constructing the replay mixture.
The shopping-domain portion accounts for 91.0\% of examples and 66.6\% of tokens.
We group shopping data by the capabilities that domain CPT is intended to inject.
Catalog--SID grounding data teaches the model to associate SID tokens with product titles, attributes, category structure, and long-form item descriptions.
Intent/behavior-to-item grounding data connects user queries, histories, profiles, and text-described candidates to item-space outcomes, including query--item mapping, sequential next-item prediction, intent-conditioned SID retrieval, profile-based recommendation, textualized recommendation/search, and interleaved text--SID generation.
User preference understanding data teaches the model to summarize long behavioral logs into structured shopping profiles.

\begin{table}[!t]
\centering
\scriptsize
\setlength{\tabcolsep}{2.5pt}
\resizebox{0.9\textwidth}{!}{
\begin{tabular}{llrrrrrr}
\toprule
\multirow{2}{*}{\textbf{Domain}} &
\multirow{2}{*}{\makecell[c]{\textbf{CPT} \textbf{Capability}}} &
\multicolumn{2}{c}{\textbf{Examples}} &
\multicolumn{3}{c}{\textbf{Tokens}} \\
\cmidrule(lr){3-4}\cmidrule(lr){5-7}
& &
\makecell[c]{\textbf{Count}} &
\makecell[c]{\textbf{Share}\\\textbf{(\%)}} &
\makecell[c]{\textbf{Total}} &
\makecell[c]{\textbf{Share}\\\textbf{(\%)}} &
\makecell[c]{\textbf{Avg.}\\} \\
\midrule
\multirow{4}{*}{General}
    & Knowledge \& Reasoning          & 20.0\,M  & 6.6  & 23.82\,B & 20.9 & 1,192  \\
    & Instruction Following           & 1.41\,M  & 0.5  & 2.89\,B  & 2.5  & 2,043  \\
    & Math \& Coding                  & 6.03\,M  & 2.0  & 11.41\,B & 10.0 & 1,892  \\
\cmidrule(lr){2-7}
    & \textit{Subtotal}               & \textit{27.4\,M} & \textit{9.0} & \textit{38.12\,B} & \textit{33.4} & \textit{1,390} \\
\midrule
\multirow{4}{*}{Shopping}
    & Catalog--SID Grounding          & 240.7\,M & 79.4 & 31.33\,B & 27.5 & 130    \\
    & Intent/Behavior-to-Item Grounding & 34.7\,M & 11.5 & 40.84\,B & 35.8 & 1,177  \\
    & User Preference Understanding   & 200.0\,K & 0.1  & 3.67\,B  & 3.2  & 18,365 \\
\cmidrule(lr){2-7}
    & \textit{Subtotal}               & \textit{275.6\,M} & \textit{91.0} & \textit{75.85\,B} & \textit{66.6} & \textit{275} \\
\midrule
    & \textbf{Total}                  & \textbf{303.0\,M} & \textbf{100.0} & \textbf{113.96\,B} & \textbf{100.0} & \textbf{376} \\
\bottomrule
\end{tabular}
}
\caption{
CPT data mixture used for domain continued pre-training, grouped by the capabilities injected during the stage. Percentages are computed over the full CPT mixture; average tokens are reported per example.
}
\label{tab:cpt_data_distribution}
\end{table}

\tightparagraph{Training settings.}
During domain CPT, we update the full model on the mixed training stream rather than restricting optimization to SID-specific parameters.
Starting from the aligned checkpoint, the final recipe trains for one epoch over the CPT mixture.
Training uses a 20{,}480-token context length.
Optimization uses Adam with peak learning rate $5\times10^{-5}$, a 1{,}000-step warmup, and a constant schedule after warmup.

\section{Post-Training}
\label{sec:post_training}

After SID token alignment and domain continued pre-training establish a SID-aware foundation model (\autoref{fig:training_pipeline}), post-training teaches the model to follow the \framework action protocol, perform SID-native item-space operations, generate catalog-grounded responses, and emit state-update signals for multi-turn interactions.
A key challenge is capability imbalance: strengthening SID prediction can regress general language and other shopping or workflow abilities, a form of the see-saw effect discussed in multi-skill post-training~\citep{mimo_v2_flash}. We address this with a recipe that assigns training examples to task families and combines Multi-teacher On-policy Distillation (MOPD)~\citep{mimo_v2_flash,deepseek_v4,nemotron_cascade_2} with task-specific reward optimization.

Our post-training pipeline consists of three stages.
\textbf{Stage~1: Supervised Fine-tuning (SFT).}
We train on a mixture of general instruction and SID-related data, enabling basic instruction-following across both domains.
\textbf{Stage~2: Specialist Preparation.}
We prepare task-specialized teacher policies used by MOPD in the OPD--RL stage.
\textbf{Stage~3: OPD--RL Joint Optimization.}
We route training examples into five task families with task-dependent combinations of OPD teacher supervision and reward feedback.

\subsection{Supervised Fine-tuning}
\label{sec:sft_post_training}

\begin{table}[!t]
\centering
\small
\setlength{\tabcolsep}{2.5pt}
{\renewcommand{\arraystretch}{1.06}
\resizebox{\textwidth}{!}{
\begin{tabular}{lllrrrrr}
\toprule
\multirow{2}{*}{\textbf{Domain}} &
\multirow{2}{*}{\makecell[c]{\textbf{SFT Capability}\\\textbf{Group}}} &
\multirow{2}{*}{\textbf{Covered Tasks}} &
\multicolumn{2}{c}{\textbf{Examples}} &
\multicolumn{3}{c}{\textbf{Tokens}} \\
\cmidrule(lr){4-5}\cmidrule(lr){6-8}
& & &
\makecell[c]{\textbf{Count}} &
\makecell[c]{\textbf{Share}\\\textbf{(\%)}} &
\makecell[c]{\textbf{Total}} &
\makecell[c]{\textbf{Share}\\\textbf{(\%)}} &
\makecell[c]{\textbf{Avg.}\\} \\
\midrule
\multirow{9}{*}{\makecell[c]{General}}
  & \makecell[l]{Knowledge \&\\Reasoning}
  & \makecell[l]{(1) Broad Knowledge QA;\\(2) General Reasoning;}
  & 2.32\,M & 40.8 & 2.57\,B & 25.3 & 1,104 \\
\cmidrule(lr){2-8}
  & \makecell[l]{Instruction\\Following}
  & \makecell[l]{(1) General Instruction Following;\\(2) Multi-Turn Dialogue;\\(3) Tool-Use and Search-Agent Instructions}
  & 1.88\,M & 33.1 & 2.65\,B & 26.1 & 1,407 \\
\cmidrule(lr){2-8}
  & \makecell[l]{Math \&\\Coding}
  & \makecell[l]{(1) Mathematical Reasoning;\\(2) Code Generation;\\(3) Code-Agent Tool Use}
  & 416.5\,K & 7.3 & 2.40\,B & 23.6 & 5,764 \\
\cmidrule(lr){2-8}
  & \textit{Subtotal} & & \textit{4.62\,M} & \textit{81.2} & \textit{7.62\,B} & \textit{74.9} & \textit{1,647} \\
\midrule
\multirow{9}{*}{\makecell[c]{Shopping}}
  & \makecell[l]{SID-Native\\ Fulfillment}
  & \makecell[l]{(1) SID--Text Alignment;\\(2) Textualized Item Ranking;\\(3) SID Retrieval/Ranking;\\(4) Interleaved Text--SID Recommendation}
  & 542.9\,K & 9.5 & 1.33\,B & 13.1 & 2,454 \\
\cmidrule(lr){2-8}
  & \makecell[l]{User/Context\\ Understanding}
  & \makecell[l]{(1) Profile/Behavior Summarization;\\(2) Context Evidence Extraction}
  & 229.0\,K & 4.0 & 1.07\,B & 10.5 & 4,658 \\
\cmidrule(lr){2-8}
  & \makecell[l]{Shopping\\ Dialogue}
  & \makecell[l]{(1) Multi-Turn Shopping Guide;\\(2) Item Comparison;\\(3) Preference and Constraint Refinement}
  & 300.4\,K & 5.3 & 148.7\,M & 1.5 & 495 \\
\cmidrule(lr){2-8}
  & \textit{Subtotal} & & \textit{1.07\,M} & \textit{18.8} & \textit{2.55\,B} & \textit{25.1} & \textit{2,375} \\
\midrule
\multicolumn{3}{l}{\textbf{Total}} & \textbf{5.70\,M} & \textbf{100.0} & \textbf{10.17\,B} & \textbf{100.0} & \textbf{1,784} \\
\bottomrule
\end{tabular}
}
}
\caption{
SFT data mixture grouped by supervised capability and representative covered tasks. Percentages are computed over the full SFT mixture; average tokens are reported per example.
}
\label{tab:sft_data_distribution}
\end{table}

SFT aligns the domain-adapted model with the task formats required by the \framework serving workflow.
The central design question is the task mixture: which capabilities to supervise and how to balance SID-generation tasks against broader shopping, workflow, and language tasks.

\tightparagraph{Supervision groups.}
Whereas CPT injects item-space grounding, SFT converts this grounding into supervised task formats and workflow behaviors.
\autoref{tab:sft_data_distribution} groups the data by capability and lists representative tasks.
The mixture is intentionally general-heavy: general data accounts for 81.2\% of examples and 74.9\% of tokens, while shopping data contributes 18.8\% of examples and 25.1\% of tokens across SID-native fulfillment, user/context understanding, and dialogue supervision.
The general portion is selected from the public and in-house general data pool with stricter quality filtering and instruction-format checks than CPT replay.
It retains high-quality Step-3.5-Flash examples~\citep{step35_flash}, adds dialogue-oriented public data such as Tulu~3~\citep{tulu3}, SmolTalk~\citep{smollm2}, UltraChat~\citep{ultrachat}, and COIG-CQIA~\citep{coig_cqia}, and complements them with curated in-house data.
Together, these sources support the general and multi-turn dialogue, tool-use and search-agent, and math, coding, and code-agent groups shown in \autoref{tab:sft_data_distribution}.
We do not treat these examples as shopping-domain supervision.
They are included to preserve action-format following, search-like decomposition, and procedural reasoning abilities when the model fills the \framework action protocol and uses harness-provided context, catalog, and state surfaces.

\tightparagraph{Mixture ratio and balancing.}
SID generation is a major capability to strengthen for model-native item fulfillment, but a SID-only mixture would over-specialize the model and weaken the broader task formats required by the serving workflow.
We therefore balance SID-native fulfillment, user/context understanding, dialogue, and general instruction data with capability-group weighting and per-task temperature sampling.
\autoref{tab:sft_data_distribution} reports the resulting data composition, and \S\ref{sec:post_training_ablation} compares this balanced mixture with SID-specialized variants.

\tightparagraph{Training settings.}
SFT uses full-parameter fine-tuning from the domain-CPT checkpoint with the Qwen3 chat template.
We train for one epoch.
Training uses an 81{,}920-token context length.
Optimization uses Adam with peak learning rate $5{\times}10^{-6}$, a 300-step warmup, and a constant schedule.

\subsection{Overview of OPD--RL Joint Training}
\label{sec:opd_rl_routing}

SFT provides broad task-format coverage but leaves three gaps:
\textbf{(1)}~SID prediction requires deeper item-token specialization than a balanced SFT mixture can deliver without over-fitting to SID-only data;
\textbf{(2)}~continued exposure to e-commerce-domain CPT and SFT data can weaken general instruction-following and open-domain language understanding unless explicit preservation signals are introduced;
\textbf{(3)}~tasks such as listwise ranking and text--SID interleaved recommendation require outcome-level RL feedback beyond imitation learning.
These gaps do not admit a single uniform RL objective: SID prediction rewards are sparse in the large item space, while aggressive SID-only optimization creates the same specialization--preservation trade-off discussed above.
We therefore use a joint OPD--RL stage: multi-teacher on-policy distillation (MOPD) provides dense token-level guidance from specialist teachers to strengthen SID prediction while recovering general ability, and reinforcement learning with outcome rewards optimizes behavior on tasks where reliable automatic evaluation is available.
Both signals are optimized in a unified on-policy training run implemented with ROLL~\citep{roll}, a reinforcement-learning scaling library for large-scale LLM~training.

\tightparagraph{Task routing.}
We route training examples into five task families with task-dependent optimization signals (\autoref{tab:opd_rl_routing}).
This routing differs from the supervised data grouping in \S~\ref{sec:sft_post_training}: SFT groups examples by capability coverage, whereas OPD--RL routing groups training examples by the signal used for policy updates.
The following subsections define the OPD and reward components, followed by the joint optimization objective and training details.

\begin{table}[h]
\centering
\small
\setlength{\tabcolsep}{3.5pt}
\resizebox{\textwidth}{!}{
\begin{tabular}{lrcccl}
\toprule
\textbf{Task Family} &
\makecell[c]{\textbf{Consumed}\\\textbf{Examples}} &
\makecell[c]{\textbf{Sampling}\\\textbf{Share}} &
\textbf{Rollout} &
\textbf{OPD Teacher} &
\textbf{Reward} \\
\midrule
General
  & 10{,}240
  & 20.0\%
  & Sampled
  & General Teacher
  & General judge \\
SID Prediction
  & 10{,}240
  & 20.0\%
  & Beam search
  & SID Prediction Teacher
  & --- \\
Ranking
  & 10{,}240
  & 20.0\%
  & Sampled
  & ---
  & NDCG reward \\
Interleave
  & 10{,}240
  & 20.0\%
  & Sampled
  & ---
  & Interleaved Fulfillment Reward \\
Other
  & 10{,}240
  & 20.0\%
  & Sampled
  & Self Teacher (frozen)
  & --- \\
\midrule
\textbf{Total}
  & \textbf{51{,}200}
  & \textbf{100.0\%}
  & ---
  & ---
  & --- \\
\bottomrule
\end{tabular}
}
\caption{
Task routing in the OPD--RL stage.
The final run samples five task families uniformly, using 200 update steps with rollout batch size 256, for a total of 51{,}200 consumed prompts.
Each task family receives a subset of the available training signals: OPD teacher supervision, reward feedback, or both.
}
\label{tab:opd_rl_routing}
\end{table}

\begin{figure}[t]
\centering
\includegraphics[width=\textwidth]{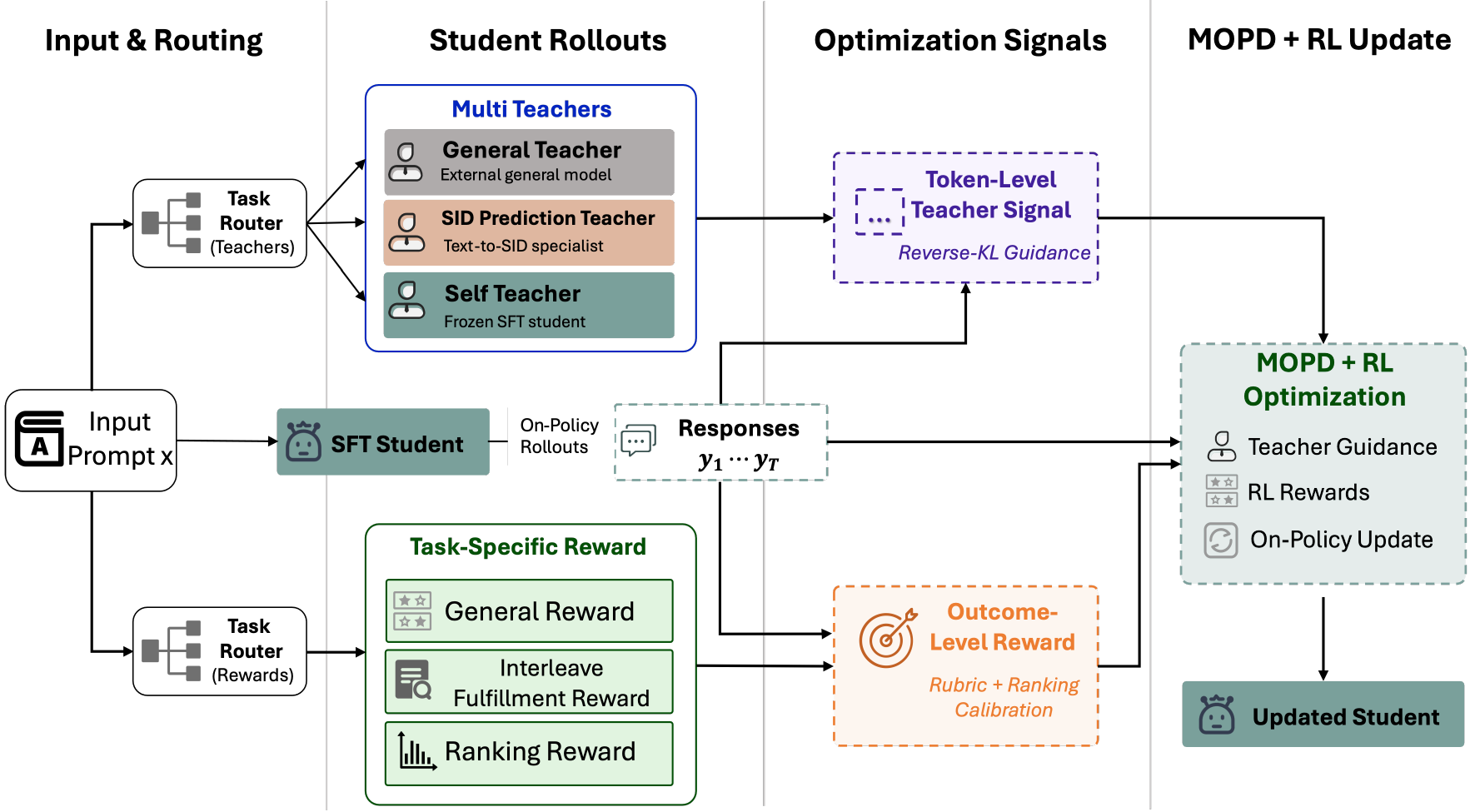}
\caption{
\textbf{Task routing in the OPD--RL stage.}
The OPD--RL stage assigns training examples to five task families with different training signals.
\emph{General} tasks use both the General Teacher and a general response judge reward; \emph{SID Prediction} and \emph{Other} tasks use OPD-only supervision from the SID Prediction Teacher and frozen Self Teacher; \emph{Ranking} and \emph{Interleave} tasks use reward-only optimization with ranking and text--SID fulfillment rewards.
}
\label{fig:task_routed_mopd}
\end{figure}

\subsection{Multi-teacher On-policy Distillation}
\label{sec:mopd_objective}

Our MOPD design follows recent multi-skill and multi-domain on-policy distillation recipes~\citep{mimo_v2_flash,nemotron_cascade_2}, but routes teacher supervision by the task families needed for \ourmethod.

\tightparagraph{Specialist Preparation.}
\label{sec:specialist_training}
MOPD is applied to the task families with OPD supervision: \emph{General}, \emph{SID Prediction}, and \emph{Other}. We prepare one teacher policy for each of these families:
\begin{itemize}[leftmargin=1.5em,itemsep=2pt,topsep=2pt]
\item \textbf{General Teacher.}
We use the pre-alignment base checkpoint as the General Teacher, which anchors open-domain language understanding after SID alignment, domain continued pre-training, and fulfillment-oriented SFT.
\item \textbf{SID Prediction Teacher.}
We further fine-tune the post-SFT student on SID prediction data, including detailed-description-to-item retrieval, intent-and-history retrieval, and contextual-intent retrieval, to obtain a teacher with stronger pure-SID generation.
The main teacher-training sources include 4.97M detailed-description-to-item retrieval samples and 4.92M intent-plus-history retrieval samples; additional source details are provided in Appendix~\ref{app:post_training_ablation_configs}.
\item \textbf{Self Teacher.}
For the \emph{Other} task family, including Item Description Recovery and Textualized Recommendation/Search, we use the frozen post-SFT model as the Self Teacher to preserve SFT-acquired task behavior.
\end{itemize}

\tightparagraph{MOPD Objective.}
Let $\pi_{\theta}$ denote the student policy, and let $\phi(k)$ denote the teacher selected for task family $k \in \mathcal{K}_{\mathrm{OPD}}$, where
\begin{equation}
\mathcal{K}_{\mathrm{OPD}}
=
\{\text{General},\text{SID Prediction},\text{Other}\}.
\end{equation}
Given a prompt $x$, the student policy generates a group of $G$ trajectories $\mathcal{G}(x)=\{\tau_1,\tau_2,\ldots,\tau_G\}$, where each trajectory $\tau_g=(y_{g,1},y_{g,2},\ldots,y_{g,T_g})$ is a complete response sequence.
For \emph{General} and \emph{Other} tasks, trajectories are sampled from the student policy using standard stochastic decoding.
For \emph{SID Prediction} tasks, the rollout group is constructed with beam search to match the pure-SID generation setting used at evaluation time.

For a task family $k$ with OPD supervision, the selected teacher
$\pi_{\phi(k)}$ provides dense token-level guidance by minimizing the
on-policy reverse KL from the student policy to the teacher policy.
For one rollout $\tau_g$, this trajectory-level objective is
\begin{equation}
\mathcal{L}^{\phi(k)}_{\mathrm{RKL}}(\theta)
=
-
\mathbb{E}_{\tau_g\sim\pi_{\theta}(\cdot|x)}
\left[
\sum_{t=1}^{T_g}
\log
\frac{
\pi_{\phi(k)}(y_{g,t}|x,y_{g,<t})
}{
\pi_{\theta}(y_{g,t}|x,y_{g,<t})
}
\right].
\label{eq:reverse_kl}
\end{equation}

In implementation, we use the local teacher--student log-probability ratio as a stop-gradient token-level preference weight:
\begin{equation}
\hat{A}^{\phi(k)}_{\mathrm{Teacher},g,t}
=
\mathrm{sg}\!\left[
\log
\frac{
\pi_{\phi(k)}(y_{g,t}|x,y_{g,<t})
}{
\pi_{\theta}(y_{g,t}|x,y_{g,<t})
}
\right],
\label{eq:teacher_advantage}
\end{equation}
where $\mathrm{sg}[\cdot]$ denotes the stop-gradient operator. This means that
the log-ratio is treated as a fixed teacher-derived weight, and gradients are taken only through the student log-probability in the policy-gradient surrogate.
The MOPD loss over tasks with OPD supervision is
\begin{equation}
\mathcal{L}_{\mathrm{MOPD}}(\theta)
=
-
\mathbb{E}_{(x,k)\sim\mathcal{D}_{\mathrm{OPD}},\;\{\tau_g\}_{g=1}^{G}\sim\pi_{\theta}(\cdot|x)}
\left[
\frac{1}{G}
\sum_{g=1}^{G}
\sum_{t=1}^{T_g}
\hat{A}^{\phi(k)}_{\mathrm{Teacher},g,t}
\log
\pi_{\theta}(y_{g,t}|x,y_{g,<t})
\right].
\label{eq:mopd}
\end{equation}
Here $\mathcal{D}_{\mathrm{OPD}}$ is the routed training distribution over prompts and task families that receive teacher supervision.
This objective only covers the task families that have teacher supervision.
Reward-only task families are handled by the reward objective defined in the next subsection.

\subsection{Reward Design}
\label{sec:reward_design}

Reward feedback is used for task families where reliable outcome-level signals are available: \emph{General}, \emph{Ranking}, and \emph{Interleave}. For \emph{SID Prediction}, we avoid sparse exact-hit rewards and rely on the SID Prediction Teacher instead; Appendix~\ref{app:sparse_sid_rewards} discusses the SID-reward attempts that led to this choice. For \emph{Other} tasks, we use the frozen Self Teacher to preserve SFT-acquired behavior without introducing an additional reward.

\tightparagraph{General response judge reward.}
For general tasks, we adopt an LLM-as-a-Judge framework to assess the quality of model responses. The judge takes the generated response and the corresponding reference answer as input, removes hidden reasoning spans when present, and assigns an integer score ranging from 0 to 10 according to the semantic evaluation rubrics. The resulting score is subsequently normalized to the range $[0,1]$ and used as the response-level~reward.
The detailed rubric used for the LLM judge is provided in Appendix~\ref{app:runtime_reward_prompts}.

\tightparagraph{Interleaved Fulfillment Reward.}
For outputs that interleave natural language with SID tokens, such as text--SID interleaved recommendation responses, we introduce the \textit{Interleaved Fulfillment Reward}. The reward first extracts the visible answer region after hidden reasoning spans, then parses all valid SID tokens from the response and the reference answer. The parsed SIDs are resolved through the product catalog index to obtain item metadata such as title, price, brand, and seller information.
The reward combines three sources of supervision:
\begin{itemize}[leftmargin=1.5em,itemsep=2pt,topsep=2pt]
\item \textbf{Rule-based validity and penalty signals.}
We apply rule-based checks and penalties for missing SIDs, unresolved response SIDs, duplicated SIDs, overlong answers, insufficient coverage when multiple reference SIDs are available, over-used SIDs within recent batches, fabricated item identifiers, and duplicated product listings. Reference SIDs that cannot be resolved can be masked according to the runtime configuration.

\item \textbf{Hierarchical SID matching reward.}
For valid predicted and reference SIDs, we compute an optimal matching between the two SID sets. Each matched pair receives partial credit according to the deepest matched SID prefix, with deeper hierarchy matches receiving higher credit.

\item \textbf{Catalog-grounded judge reward.}
We expand predicted and reference SIDs into catalog metadata and pass them, together with the original task and model response, to an LLM judge. The judge evaluates multiple dimensions, including aesthetic fit, alignment with the reference items, satisfaction of the user intent, response text quality, and consistency between the stated rationale and the recommended items. It also flags fabricated item identifiers and duplicated product listings.
The detailed rubric is in Appendix~\ref{app:runtime_reward_prompts}.
\end{itemize}
The response-level reward is computed by combining the normalized catalog-grounded judge dimensions with the hierarchical SID matching score, followed by multiplicative rule-based penalties. Implementation details and aggregation weights are provided in Appendix~\ref{app:interleaved_fulfillment_reward_details}.

\tightparagraph{Ranking reward.}
For listwise ranking tasks, the model is provided with a set of candidate items and is required to output a ranked permutation according to their relevance to the given user intent. We first apply a hard validity check: after extracting the visible answer, the output must be a comma-separated list that forms a complete permutation of the ground-truth item set, with no missing, duplicated, or extraneous items. Invalid outputs receive zero reward. Valid permutations are scored with an NDCG-based reward against the reference ranking.

\subsection{Joint OPD--RL Objective}
\label{sec:joint_opd_rl_objective}

Following the task routing defined in \S~\ref{sec:opd_rl_routing}, the OPD--RL stage optimizes a single student policy over the five task families:
\begin{equation}
\mathcal{K}
=
\{
\text{General},
\text{SID Prediction},
\text{Ranking},
\text{Interleave},
\text{Other}
\}.
\end{equation}
Let $\mathcal{K}_{\mathrm{OPD}}=\{\text{General},\text{SID Prediction},\text{Other}\}$ denote the task families with teacher supervision, and let $\mathcal{K}_{\mathrm{R}}=\{\text{General},\text{Ranking},\text{Interleave}\}$ denote the task families with reward feedback.
For a sampled training example, let $x$ denote the prompt and $k$ its task family.
The student produces a rollout group $\{\tau_g\}_{g=1}^{G}$, where $g$ indexes a response in the group and $\tau_g=(y_{g,1},\ldots,y_{g,T_g})$ is the generated token sequence.
For tasks with reward feedback, the rollout group obtains response-level rewards $\{r_1,\ldots,r_G\}$ from the corresponding reward function $\psi(k)$, where $r_g$ is the reward assigned to trajectory $\tau_g$.
We follow GRPO~\citep{deepseek_r1} to compute the group-normalized reward advantage:
\begin{equation}
\hat{A}^{\psi(k)}_{\mathrm{Reward},g}
=
\frac{
r_g-\mathrm{mean}(r_1,\ldots,r_G)
}{
\mathrm{std}(r_1,\ldots,r_G)
}.
\label{eq:reward_advantage}
\end{equation}
For task families without reward feedback, this reward advantage is set to zero.

For trajectory $\tau_g$ from task family $k$, the training signal at token $t$ is
\begin{equation}
\hat{A}^{k}_{g,t}
=
\mathbf{1}[k\in\mathcal{K}_{\mathrm{OPD}}]\,
\hat{A}^{\phi(k)}_{\mathrm{Teacher},g,t}
+
\alpha_k\,
\mathbf{1}[k\in\mathcal{K}_{\mathrm{R}}]\,
\hat{A}^{\psi(k)}_{\mathrm{Reward},g},
\label{eq:joint_advantage}
\end{equation}
where $\mathbf{1}[\cdot]$ is the indicator function, $\hat{A}^{\phi(k)}_{\mathrm{Teacher},g,t}$ is the token-level MOPD weight defined in \autoref{eq:teacher_advantage}, and $\alpha_k$ controls the reward weight for each task family with reward feedback.
This matches the task-family signal assignment in \autoref{tab:opd_rl_routing}.

The resulting joint objective is
\begin{equation}
\mathcal{L}_{\mathrm{Joint}}(\theta)
=
-
\mathbb{E}_{(x,k)\sim\mathcal{D},\;\{\tau_g\}_{g=1}^{G}\sim\pi_{\theta}(\cdot|x)}
\left[
\frac{1}{G}
\sum_{g=1}^{G}
\sum_{t=1}^{T_g}
\hat{A}^{k}_{g,t}
\log
\pi_{\theta}(y_{g,t}|x,y_{g,<t})
\right].
\label{eq:joint_objective}
\end{equation}
Here $\mathcal{D}$ is the routed OPD--RL training distribution over all five task families.

\tightparagraph{Training settings.}
The final OPD--RL run follows the task-family allocation in \autoref{tab:opd_rl_routing}.
We train for 200 update steps with a rollout batch of 256 prompts and $G=16$ responses per prompt.
The actor is optimized with Adam using a constant learning rate of $5\times10^{-7}$ after 50 warmup steps.
SID Prediction rollouts use six-level SID generation with beam search.

\section{Experiments}
\label{sec:experiments}

We first evaluate the complete shopping system, then trace its behavior back to the design choices that shape it.
Shopping and system-level evaluations follow the Taobao-derived protocols in \S\ref{sec:evaluation_protocol}, while public benchmarks evaluate the general abilities that remain necessary for agentic shopping. This section is organized as follows:
\begin{itemize}
    \item \textbf{Implementation details} (\S\ref{sec:experimental_implementation_details}) summarize the shared catalog, decoding, SID resolution, and reporting settings used across experiments.
    \item \textbf{Framework-level results} (\S\ref{sec:end_to_end_system_evaluation}) compare the \ourmethod serving framework with tool-mediated LLM-agent baselines under the same evaluation protocol.
    This establishes the complete-system effect of model-native fulfillment.
    \item \textbf{Capability breakdown} (\S\ref{sec:capability_diagnostics_workflow_transfer}) evaluates the model under a fixed harness, linking general abilities such as instruction following and broad knowledge, shopping semantics, context evidence extraction, SID semantic recovery, item-space retrieval, ranking, and grounded generation to end-to-end fulfillment outcomes.
    \item \textbf{SID design} (\S\ref{sec:sid_design_ablation}) isolates how identifier construction affects semantic recoverability and LLM operability, beyond retrieval-style candidate-generation accuracy alone.
    \item \textbf{Continued pre-training} (\S\ref{sec:cpt_data_mixture_ablation}) studies how domain data and general replay balance item-space grounding against the general abilities required by the agentic interface.
    \item \textbf{Post-training} (\S\ref{sec:post_training_ablation}) compares supervised specialization, task-family-specific teacher distillation, and RL rewards for consolidating SID-native fulfillment while preserving other shopping, workflow, and general capabilities.
\end{itemize}

\subsection{Experimental Implementation Details}
\label{sec:experimental_implementation_details}

Unless otherwise stated, full-scale \ourmethod training and evaluation use the same Taobao item snapshot introduced in \S\ref{sec:framework_level_evaluation_protocol}, containing approximately 1.2B items.
This snapshot defines the item universe for catalog grounding, item-evidence lookup, SID resolution, and fulfillment judging.
The SID beam search for item retrieval tasks in the capability breakdown experiments uses a progressive beam-size schedule \texttt{[8, 32, 64, 64, \ldots]} across SID levels and returns at most 64 generated SIDs per query.
For framework-level evaluation in \S\ref{sec:framework_level_evaluation_protocol}, SID beam search uses beam size 30 for our final six-level SID configuration ($G_2+L_4$).
We do not use constrained decoding or illegal-SID filtering, so invalid SID strings are counted as failed resolutions.

The $G_2+L_4$ hybrid SID adds one token set for each codebook level, for a total of $6\times8192$ new SID tokens.
Training and evaluation data are strictly deduplicated at the user level.
All reported bounded metrics are placed on a 0--100 scale unless otherwise noted.
Detailed row-level metric definitions and reporting details are provided in Appendix~\ref{app:capability_metric_protocol}.

\subsection{Framework-Level Results}
\label{sec:end_to_end_system_evaluation}

\begin{table}[t]
\centering
\setlength{\hfuzz}{2pt}
\scriptsize
\setlength{\tabcolsep}{0.7pt}
\renewcommand{\arraystretch}{1.10}
\begin{tabularx}{\linewidth}{@{}
>{\raggedright\arraybackslash}p{0.145\linewidth}
>{\raggedright\arraybackslash}p{0.14\linewidth}
>{\raggedright\arraybackslash}p{0.095\linewidth}
*{8}{>{\centering\arraybackslash}X}
@{}}
\toprule
\multirow{2}{*}{\textbf{System}} &
\multirow{2}{*}{\textbf{Fulfillment Path}} &
\multirow{2}{*}{\textbf{Model}} &
\multicolumn{8}{c}{\textbf{Framework-Level Overview Metrics}} \\
\cmidrule(lr){4-11}
& & &
\makecell{\textbf{Intent}\\\textbf{Ful.}} &
\makecell{\textbf{Item}\\\textbf{Prec.}} &
\makecell{\textbf{Rank.}\\\textbf{Qual.}} &
\makecell{\textbf{Cat.}\\\textbf{Cov.}} &
\makecell{\textbf{Pers.}} &
\makecell{\textbf{Constr.}\\\textbf{Ground.}} &
\makecell{\textbf{Feed.}\\\textbf{Adapt.}} &
\makecell{\textbf{Cross}\\\textbf{turn}\\\textbf{Ref.}} \\
\midrule
\multirow{3}{*}{\framework{}} &
\multirow{3}{*}{\makecell[l]{Model-native\\item-space ops.}} &
\ourmethod-4B &
65.6 & 91.1 & 92.0 & \textbf{84.7} & 65.7 & 74.7 & 66.5 & 60.9 \\
& &
\ourmethod-8B &
\underline{69.2} & \underline{95.0} & \underline{95.5} & \underline{83.9} & 65.9 & \underline{76.4} & \textbf{71.5} & \underline{68.3} \\
& &
\makecell[l]{\ourmethod-30B\\A3B} &
68.9 & \textbf{95.8} & \textbf{95.9} & 83.8 & \textbf{69.5} & \textbf{77.6} & \underline{68.5} & \textbf{80.0} \\
\midrule
InteRecAgent~\cite{interecagent} &
\multirow{3}{=}{Retrieval \& ranking tools} &
Qwen3-8B &
59.5 & 89.8 & 90.0 & 77.1 & 63.4 & 71.8 & 55.9 & 41.8 \\
Chat-REC~\cite{chat_rec} &
&
Qwen3-8B &
\textbf{72.2} & 94.0 & 93.9 & 82.8 & \underline{66.6} & 74.8 & 59.8 & 47.5 \\
RecMind~\cite{recmind} &
&
Qwen3-8B &
63.6 & 91.2 & 90.7 & 80.0 & 65.0 & 75.6 & 62.9 & 42.1 \\
\bottomrule
\end{tabularx}
\caption{
\textbf{Framework-level evaluation results.}
Complete serving configurations are compared using the eight overview metrics defined in \S\ref{sec:framework_level_evaluation_protocol}.
All numbers are on a 0--100 scale; bold and underlined values mark the best and second-best systems in each metric column.
\ourmethod-30B-A3B is evaluated with our final SID ($G_2+L_4$) only.
}
\label{tab:main_workflow_evaluation}
\end{table}

The framework-level results instantiate the protocol in \S\ref{sec:framework_level_evaluation_protocol}.
\autoref{tab:main_workflow_evaluation} compares complete system configurations rather than isolated recommenders, reporting the same eight framework-level overview metrics used by the radar profile in \autoref{fig:graphical_abstract}.
We reproduce several tool-mediated LLM-agent baselines, including InteRecAgent-style~\citep{interecagent}, Chat-REC-style~\citep{chat_rec}, and RecMind-style~\citep{recmind} agents, while replacing their original retrieval or ranking components with the same LLM-external search and ranking services exposed by our evaluation scaffold.
This keeps the comparison focused on the fulfillment paradigm rather than on differences between external tool implementations.

\autoref{tab:main_workflow_evaluation} shows that \ourmethod improves overall framework-level performance over tool-mediated baselines equipped with a strong Taobao item-search backend.
As defined in \S\ref{sec:framework_level_evaluation_protocol}, these baselines may further use the LLM to filter the retrieved items into a smaller shortlist or rerank them before response generation.
These baselines remain competitive on first-pass fulfillment and personalization, with Chat-REC even leading the table on first-pass intent fulfillment.
The main \ourmethod-8B gains instead appear on item precision, ranking quality, constraint grounding, feedback adaptation, and cross-turn reference, with the clearest separation on the two stateful multi-turn axes.
The \ourmethod-30B-A3B row shows that scaling \ourmethod with our final SID further improves item precision, ranking quality, personalization, constraint grounding, and especially cross-turn reference, which rises from 68.3 to 80.0.
Intent fulfillment and feedback adaptation, however, do not improve over the 8B model.

The trace-level pattern behind these gains is the interface-loss problem introduced in \S\ref{sec:introduction}.
In tool-mediated systems, the LLM may understand the request, but still has to hand off item-space execution through compact keyword queries, filters, or candidate-list operations.
This boundary is not a problem for many direct requests, which is why the baselines remain strong on first-pass matching.
It becomes more costly when constraints, item references, and feedback must stay attached to the same catalog evidence across turns.
\ourmethod keeps candidate generation, grounded selection, response generation, and state updates within a model-facing item-space workflow, reducing the need to reconstruct the shopping state after each external tool call.
Appendix~\ref{app:serving_trace_examples} provides qualitative case-level comparisons that illustrate this pattern in both single-turn and multi-turn settings, including examples where tool-mediated baselines retrieve plausible products initially but lose category, constraint, or reference state during subsequent fulfillment steps.

\ourmethod is therefore not uniformly best on every overview axis.
Its advantage is concentrated where the fulfillment state must remain coherent across planning, item-space execution, evidence grounding, and multi-turn updates.
This supports the central claim of the report: model-native fulfillment reduces the lossy hand-offs that arise when language understanding and item-space execution are separated by tool interfaces.

\subsection{Capability Breakdown Results}
\label{sec:capability_diagnostics_workflow_transfer}

\begin{table}[p]
\centering
\fontsize{7.2pt}{8.0pt}\selectfont
\definecolor{capGeneralBg}{RGB}{248,251,255}
\definecolor{capEcommerceBg}{RGB}{248,252,248}
\definecolor{capSystemBg}{RGB}{255,250,244}
\newcommand{\capmetrichang}{\hspace{0.1em}}
\newcommand{\capmetric}[1]{\\[-0.35ex]\capmetrichang{\tiny\itshape\textcolor{black!58}{#1}}}
\newcommand{\capsetting}[1]{\\[-0.35ex]\capmetrichang{\tiny\itshape\textcolor{black!52}{#1}}}
\newcommand{\capmetricrow}[1]{\capmetrichang{\tiny\itshape\textcolor{black!58}{#1}}}
\newcommand{\pairscore}[2]{#1\kern0.08em/\kern0.08em#2}
\newcommand{\notapp}{/}
\begin{tblr}{
  width=\textwidth,
  colspec={
    Q[c,m,wd=0.105\textwidth]
    Q[l,m,wd=0.145\textwidth]
    Q[l,m,wd=0.245\textwidth]
    Q[c,m,wd=0.081\textwidth]
    Q[c,m,wd=0.081\textwidth]
    Q[c,m,wd=0.081\textwidth]
    Q[c,m,wd=0.081\textwidth]
    Q[c,m,wd=0.081\textwidth]
    Q[c,m,wd=0.081\textwidth]
  },
  colsep=0.45pt,
  leftsep=0pt,
  rightsep=0pt,
  rowsep=0.9pt,
  row{3-10}={bg=capGeneralBg},
  row{11-25}={bg=capEcommerceBg},
  row{26-33}={bg=capSystemBg},
  hline{7,8,14,16,18}={2-9}{0.2pt,black!35},
  cell{1}{1}={r=2}{l,m},
  cell{1}{2}={r=2}{l,m},
  cell{1}{3}={r=2}{l,m},
  cell{1}{4}={c=2}{c,m},
  cell{1}{6}={c=2}{c,m},
  cell{1}{8}={c=2}{c,m},
  cell{3}{1}={r=8}{c,m},
  cell{3}{2}={r=4}{l,m},
  cell{7}{2}={r=1}{l,m},
  cell{8}{2}={r=3}{l,m},
  cell{11}{1}={r=15}{c,m},
  cell{11}{2}={r=3}{l,m},
  cell{14}{2}={r=2}{l,m},
  cell{16}{2}={r=2}{l,m},
  cell{18}{2}={r=8}{l,m},
  cell{26}{1}={r=8}{c,m},
  cell{26}{2}={r=8}{l,m},
  cell{26}{4}={r=8,c=2}{c,m},
  row{1-2}={font=\bfseries},
}
\toprule
Scope & Evaluation Axis & Task &
\makecell[c]{Qwen3} & &
\makecell[c]{ShopX\\w/ FORGE SID} & &
ShopX & \\
\cmidrule[lr]{4-5}\cmidrule[lr]{6-7}\cmidrule[lr]{8-9}
& & & 4B & 8B & 4B & 8B & 4B & 8B \\
\midrule
\makecell[c]{General\\Capabilities} &
\makecell[l]{Knowledge \&\\Reasoning} &
\makecell[l]{BBH CoT\capsetting{3-shot CoT; exact match}} & 69.0 & 73.2 & 70.4 & 75.0 & 71.3 & 74.1 \\
& & \makecell[l]{MMLU-Pro\capsetting{5-shot; exact match}} & 60.4 & 62.9 & 50.3 & 58.8 & 52.0 & 59.3 \\
& & \makecell[l]{GPQA-Diamond\capsetting{0-shot; exact match}} & 43.4 & 47.5 & 29.8 & 41.4 & 36.9 & 41.4 \\
& & \makecell[l]{CMMLU\capsetting{0-shot; accuracy}} & 66.8 & 74.2 & 69.1 & 71.8 & 70.4 & 75.8 \\
& \makecell[l]{Instruction\\Following} &
\makecell[l]{IFEval\capsetting{0-shot; strict prompt acc.}} & 79.1 & 81.9 & 73.8 & 75.2 & 72.3 & 78.4 \\
& \makecell[l]{Math \&\\Coding} &
\makecell[l]{MATH-500\capsetting{0-shot; exact match}} & 77.0 & 76.2 & 53.8 & 62.2 & 51.8 & 59.6 \\
& & \makecell[l]{GSM8K\capsetting{8-shot exemplars; exact match}} & 84.2 & 88.2 & 85.1 & 88.1 & 86.7 & 88.6 \\
& & \makecell[l]{MBPP+\capsetting{3-shot; pass@1}} & 73.8 & 75.1 & 77.0 & 81.2 & 76.5 & 83.1 \\
\midrule
\makecell[c]{Shopping\\Capabilities} &
\makecell[l]{Shopping\\Semantics} &
\makecell[l]{Intent Summary\capmetric{LLM-judge score}} & 72.5 & 76.6 & 79.0 & 80.9 & 78.0 & 79.0 \\
& & \makecell[l]{Item Association\capmetric{LLM-judge score}} & 26.9 & 28.3 & 46.3 & 46.9 & 48.5 & 49.7 \\
& & \makecell[l]{Item Description Recovery\capmetric{ROUGE}} & 9.9 & 10.1 & 26.0 & 25.1 & 31.7 & 33.2 \\
& \makecell[l]{Context Evidence\\Extraction} &
\makecell[l]{Static-Profile Evidence\capmetric{LLM-judge score}} & 45.6 & 52.9 & 75.1 & 77.7 & 75.0 & 76.9 \\
& & \makecell[l]{Behavior-Sequence Evidence\capmetric{Overlap IoU}} & 18.9 & 16.1 & 54.3 & 58.1 & 46.0 & 52.7 \\
& \makecell[l]{Textualized\\Rec./Search} &
\makecell[l]{Textualized Rec.\capmetric{Multiple Choice Question Score}} & 44.5 & 48.1 & 64.9 & 66.9 & 64.3 & 66.0 \\
& & \makecell[l]{Textualized Search\capmetric{Multiple Choice Question Score}} & 57.1 & 56.2 & 68.1 & 68.7 & 66.4 & 68.7 \\
& \makecell[l]{Item-Space\\Fulfillment} &
\makecell[l]{Detailed Desc. to Item\capmetric{Item HR@4/64}} & \notapp & \notapp & \pairscore{0.6}{3.8} & \pairscore{0.8}{5.2} & \pairscore{1.2}{3.2} & \pairscore{2.2}{5.3} \\
& & \capmetricrow{Query--Item Relevance (VLM judge)} & \notapp & \notapp & 32.8 & 34.2 & 43.8 & 46.5 \\
& & \makecell[l]{Personalized Seq. Rec.\capmetric{Item HR@4/64}} & \notapp & \notapp & \pairscore{2.9}{6.4} & \pairscore{2.8}{6.8} & \pairscore{6.2}{14.3} & \pairscore{7.1}{15.7} \\
& & \makecell[l]{Intent+History Item Ret.\capmetric{Item HR@4/64}} & \notapp & \notapp & \pairscore{6.5}{12.5} & \pairscore{5.6}{12.1} & \pairscore{12.7}{24.1} & \pairscore{13.1}{27.2} \\
& & \makecell[l]{Intent+History Item Rank.\capmetric{Acc@1/NDCG@3}} & \pairscore{34.2}{69.9} & \pairscore{34.9}{69.6} & \pairscore{45.3}{77.9} & \pairscore{43.2}{77.1} & \pairscore{44.8}{78.2} & \pairscore{46.5}{78.5} \\
& & \makecell[l]{Contextual-Intent Item Ret.\capmetric{Query--Item Relevance (VLM judge)}} & \notapp & \notapp & 57.8 & 58.7 & 57.7 & 59.4 \\
& & \makecell[l]{Contextual-Intent Item Rank.\capmetric{Acc@1/NDCG@3}} & \pairscore{45.4}{61.8} & \pairscore{48.9}{63.3} & \pairscore{48.3}{63.0} & \pairscore{50.0}{63.1} & \pairscore{42.5}{58.9} & \pairscore{46.6}{60.8} \\
& & \makecell[l]{Interleaved Text--SID\capmetric{5-dim detailed rubrics}} & \notapp & \notapp & 77.9 & 88.0 & 70.1 & 86.9 \\
\midrule
\makecell[c]{End-to-End\\System} &
\makecell[l]{Framework\\Overview} &
Intent Fulfillment & N/A & & 62.4 & 67.8 & 65.6 & 69.2 \\
& & Item Precision & & & 87.3 & 92.3 & 91.1 & 95.0 \\
& & Ranking Quality & & & 90.7 & 93.8 & 92.0 & 95.5 \\
& & Category Coverage & & & 85.1 & 89.7 & 84.7 & 83.9 \\
& & Personalization & & & 60.6 & 65.3 & 65.7 & 65.9 \\
& & Constraint Grounding & & & 71.8 & 79.0 & 74.7 & 76.4 \\
& & Feedback Adaptation & & & 55.3 & 62.0 & 66.5 & 71.5 \\
& & Cross-turn Reference & & & 38.0 & 57.8 & 60.9 & 68.3 \\
\bottomrule
\end{tblr}
\caption[Capability breakdown and framework overview]{
\textbf{Capability breakdown and framework overview.}
General, shopping, and framework-level metrics for the initial Qwen3 instruction-model references, ShopX with FORGE SID, and ShopX with our final SID ($G_2+L_4$).
The Qwen3 columns are direct-evaluation references on tasks that a general LLM can answer without SID-native item generation, indicating how well different general abilities are preserved after ShopX training.
The two ShopX variants use the same method family with different SID designs.
Detailed ShopX-30B-A3B diagnostics are reported in Appendix~\ref{app:shopx_30a3b_diagnostics}.
Scores are on a 0--100 scale, with compact setting or metric labels shown beneath task names where applicable.
}
\label{tab:capability_diagnostics_workflow_transfer}
\end{table}

The capability breakdown follows the diagnostic protocol in \S\ref{sec:capability_breakdown_evaluation_protocol}.
\autoref{tab:capability_diagnostics_workflow_transfer} compares the initial Qwen3 instruction-model references, ShopX with FORGE SID, and ShopX with our final SID ($G_2+L_4$) at both 4B and 8B.
The Qwen3 columns are evaluated only on tasks that do not require SID-native item generation; the framework rows are included so capability changes can be read together with end-to-end behavior.
The main observations are:
\begin{itemize}[leftmargin=*,itemsep=1pt,topsep=1pt,parsep=0pt,partopsep=0pt]
    \item General abilities are not preserved uniformly after shopping-domain training.
    The largest drops are on the harder knowledge and math rows: at 8B, GPQA-Diamond falls from 47.5 to 41.4 and MATH-500 from 76.2 to 59.6.
    The rest of the general suite is less affected.
    For shopping fulfillment, the degraded GPQA/MATH skills are less central than core shopping-interface skills, but the gap still shows the cost of domain specialization.
    \item Text-conditioned ranking is a case where SID knowledge is not always necessary.
    When the candidate items already come with sufficient textual evidence, the Qwen3 references perform strongly on ranking rows.
    The contextual-intent ranking row also shows a potential downside of domain specialization: ShopX with $G_2+L_4$ SID is below the Qwen3 8B reference, consistent with scenario-rich queries depending more on generalization and world knowledge than on SID-native retrieval.
    \item Scaling from 4B to 8B helps most model families and task groups.
    The pattern is broad rather than tied to one metric family, although a few diagnostics are not monotonic.
    \item Our final SID mainly improves item-space retrieval, especially when user profile or behavior history is part of the query.
    Compared with FORGE SID at 8B, Personalized Sequential Recommendation rises from 2.8/6.8 to 7.1/15.7 HR@4/64, and Intent+History Item Retrieval rises from 5.6/12.1 to 13.1/27.2.
    Detailed-Description-to-Item retrieval also becomes more relevant to the query: HR@4 improves from 0.8 to 2.2, and VLM-judged query--item relevance rises from 34.2 to 46.5.
    This matches the SID results in \S\ref{sec:sid_design_ablation}: the $G_2+L_4$ code is more LLM-operable at the prefix and more semantically specific in the suffix.
    \item The longer $G_2+L_4$ code is not free.
    Behavior-Sequence Evidence drops from 58.1 to 52.7 at 8B relative to FORGE SID, suggesting that exact extraction over longer SID strings is harder even when downstream retrieval improves.
    The Interleaved Text--SID row shows the same generation-side cost: the 4B model is more penalty-prone during free-form SID generation, and the 8B model remains slightly below the FORGE-SID 8B variant because the six-level code produces more unresolved SID paths during free-form response generation.
    \item These diagnostic shifts line up with the framework results.
    ShopX with $G_2+L_4$ SID improves the most stateful and item-facing framework axes over the FORGE-SID 8B variant, matching the stronger shopping capabilities and more stable general abilities above.
\end{itemize}
Row-level metric definitions and reporting details are given in Appendix~\ref{app:capability_metric_protocol}; the Interleaved Text--SID Fulfillment row uses the rubric in Appendix~\ref{app:text_sid_interleaved_fulfillment_rubric}.
For a broader view of how general-purpose models perform on textualized shopping tasks, see Appendix~\ref{app:general_llm_rec}.

\subsection{Semantic ID Design Ablation}
\label{sec:sid_design_ablation}

We next evaluate whether the SID design realizes the two goals from \S\ref{sec:semantic_ids}: semantic recoverability and LLM operability.

\autoref{tab:item_rep_intrinsics_ablation} ablates the continuous item representation before SID construction.
FORGE is the collaborative-heavy baseline; SV/MV denote single- and multi-vector representations; CP/EP denote collaborative-product and equivalent-product supervision; and CL/TR denote contrastive learning and text reconstruction.
Appendix~\ref{app:sid_design_ablation_metrics} gives the metric definitions, equivalent-product query pairs, candidate item set, and 25M-item mini catalog used by these diagnostics.
The main observations are:
\begin{itemize}[leftmargin=*,itemsep=2pt,topsep=2pt]
    \item (Row 1 vs. rows 2--5) For the language-to-item diagnostics studied here, the collaborative-heavy FORGE representation is noticeably behind the EP-supervised representations on equivalent-product retrieval and category purity, indicating weaker alignment with semantic item identity.
    \item (Rows 2--3) SV and MV have nearly identical global retrieval scores, so adding local vectors alone does not automatically improve the global item embedding.
    \item (Rows 4--5) Reconstruction also gives a direct check of what information the representation preserves.
    Row 4 reconstructs attributes from local embeddings and captions from all local embeddings, leaving the global embedding supervised only by contrastive learning.
    Row 5 uses the full objective from \S\ref{sec:sid_item_representation}: global+local embeddings reconstruct captions, while the global and local parts are separately tied to category and grouped-attribute reconstruction.
    Its higher caption recovery shows that this design encodes more recoverable multimodal information; cosine-similarity diagnostics show lower local-vector redundancy.
\end{itemize}

\autoref{tab:sid_design_ablation} fixes the final item representation and ablates SID construction.
Here $G_K$ denotes $K$ global-prefix tokens and $L_M$ denotes $M$ local-suffix tokens, with vocabulary size 8192 per codebook level.
The row-5 G+L configuration inherits the first two levels from the global-only $G_3$ prefix and appends four local VQ suffix tokens.
SID intrinsics measure global-prefix reconstruction, neighborhood quality, prefix stability, and path-load balance, while codebook validation uses the same fixed 4B LLM to measure bidirectional language--item grounding on the 25M-item mini catalog.
Avg. Item is the average number of catalog items attached to the beam-search top-1 SID, so lower values mean the generated code path is more specific.
Gini measures path-load imbalance across SID paths; lower Gini indicates more balanced path usage under the same codebook structure.
We interpret Gini together with Avg. Item: Gini describes the assignment distribution of the codebook, while Avg. Item describes the ambiguity of the model's generated top-1 SID path.
Bounded scores are reported on a 0--100 scale except Gini, which uses its native 0--1 scale; SNR remains in dB and Avg. Item remains an item count.
The SID results are:
\begin{itemize}[leftmargin=*,itemsep=2pt,topsep=2pt]
    \item (Rows 1--2) Row 1 inherits the FORGE representation mismatch above.
    In row 2, adding a FORGE-style contrastive auxiliary loss keeps quantized retrieval reasonable, but reconstruction fidelity and equivalent-product prefix consistency degrade, and Desc.$\rightarrow$Item remains weak.
    This points to the importance of a semantically stable early prefix for autoregressive SID generation.
    \item (Rows 3--4) Removing the auxiliary loss in row 3 restores prefix consistency and improves Desc.$\rightarrow$SID, but many generated paths still map to large item clusters.
    Extending the global code to $G_6$ lowers Gini and improves SID-level recovery, yet Avg. Item remains high; this explains why Desc.$\rightarrow$Item does not improve over row 3.
    \item (Row 5) The final $G_2+L_4$ design keeps an LLM-operable prefix and VQ suffixes, trading lower exact SID recovery for lower Gini, smaller buckets, and better Desc.$\rightarrow$Item and SID$\rightarrow$Desc.
\end{itemize}

Together, the representation and SID ablations show that the final design satisfies the two SID goals defined in \S\ref{sec:semantic_ids}.
The local suffix improves item specificity and semantic recoverability, while the global prefix remains operable for autoregressive generation.

\begin{table}[!t]
\centering
\footnotesize
\setlength{\tabcolsep}{4.0pt}
\begin{tabular*}{\textwidth}{@{\extracolsep{\fill}}lccc*{4}{S[table-format=2.1,detect-weight=true]}@{}}
\toprule
\multirow{2}{*}{\makecell[l]{\textbf{Item}\\\textbf{Rep.}}} &
\multirow{2}{*}{\textbf{Data}} &
\multirow{2}{*}{\textbf{Obj.}} &
\multirow{2}{*}{\textbf{Encoder}} &
\multicolumn{4}{c}{\textbf{Representation Intrinsics}} \\
\cmidrule(lr){5-8}
& & & &
\multicolumn{1}{c}{\makecell[c]{Lang. ROUGE}} &
\multicolumn{1}{c}{R@10} &
\multicolumn{1}{c}{NDCG@10} &
\multicolumn{1}{c}{\makecell[c]{Cat. Purity}} \\
\midrule

FORGE & CP & CL & CLIP-base & \multicolumn{1}{c}{--} & 86.8 & 78.4 & 91.1 \\
SV & EP & CL & Qwen3-VL-2B & \multicolumn{1}{c}{--} & 93.2 & 83.9 & 94.2 \\
MV & EP & CL & Qwen3-VL-2B & \multicolumn{1}{c}{--} & 93.1 & 83.7 & 93.5 \\
MV & EP & CL+TR\textit{(l)} & Qwen3-VL-2B & 25.6 & 92.9 & 82.9 & 94.5 \\
MV & EP & CL+TR\textit{(g+l)} & Qwen3-VL-2B & \bfseries 33.5 & \bfseries 93.5 & \bfseries 84.5 & \bfseries 95.1 \\

\bottomrule
\end{tabular*}
\caption{
\textbf{Representation intrinsic ablation.}
This table reports representation-side ablations before downstream codebook validation.
Configuration notation and metric groups are summarized in \S\ref{sec:sid_design_ablation}.
All bounded metric entries are reported on a 0--100 scale.
}
\label{tab:item_rep_intrinsics_ablation}
\end{table}

\begin{table}[!t]
\centering
\scriptsize
\setlength{\tabcolsep}{0.8pt}
\newcommand{\NA}{\multicolumn{1}{c}{--}}
\newcommand{\InheritedGlobalPrefix}{\multicolumn{4}{c}{\scriptsize inherited $G_3$ global-prefix intrinsics}}
\begin{tabular*}{\textwidth}{@{\extracolsep{\fill}}lcccS[table-format=1.2,detect-weight=true]*{6}{S[table-format=2.1,detect-weight=true]}S[table-format=3.2,detect-weight=true]*{3}{S[table-format=2.1,detect-weight=true]}@{}}
\toprule
\multirow{3}{*}{\makecell[l]{\textbf{SID}\\\textbf{Input}}} &
\multirow{3}{*}{\textbf{Data}} &
\multirow{3}{*}{\makecell[c]{\textbf{Aux.}\\\textbf{Loss}}} &
\multirow{3}{*}{\makecell[c]{\textbf{SID}\\\textbf{Structure}}} &
\multicolumn{5}{c}{\textbf{Global-Prefix Intrinsics}} &
\multicolumn{6}{c}{\textbf{Codebook Validation}} \\
\cmidrule(lr){5-9}
\cmidrule(lr){10-15}
& & & &
\multicolumn{1}{c}{\textbf{Recon.}} &
\multicolumn{1}{c}{\textbf{Quant.}} &
\multicolumn{3}{c}{\textbf{Structure}} &
\multicolumn{2}{c}{\textbf{Desc.$\rightarrow$SID}} &
\multicolumn{1}{c}{\makecell[c]{\textbf{Avg.}\\\textbf{Item}}} &
\multicolumn{2}{c}{\textbf{Desc.$\rightarrow$Item}} &
\multicolumn{1}{c}{\textbf{SID$\rightarrow$Desc.}} \\
\cmidrule(lr){5-5}
\cmidrule(lr){6-6}
\cmidrule(lr){7-9}
\cmidrule(lr){10-11}
\cmidrule(lr){12-12}
\cmidrule(lr){13-14}
\cmidrule(lr){15-15}
& & & &
\multicolumn{1}{c}{\makecell[c]{SNR\\(dB)}} &
\multicolumn{1}{c}{NDCG@10} &
\multicolumn{1}{c}{\makecell[c]{Cat.\\Pur.}} &
\multicolumn{1}{c}{\makecell[c]{Pref.\\Cons.@3}} &
\multicolumn{1}{c}{Gini} &
\multicolumn{1}{c}{HR@4} &
\multicolumn{1}{c}{HR@64} &
\multicolumn{1}{c}{\#Items} &
\multicolumn{1}{c}{HR@4} &
\multicolumn{1}{c}{HR@64} &
\multicolumn{1}{c}{\makecell[c]{ROUGE\\Score}} \\
\midrule

FORGE & CP & CL & G3
& 4.44 & 45.6 & 43.0 & 18.3 & 0.50
& 8.8 & 13.8 & 99.1 & 0.6 & 4.7 & 24.4 \\
\hdashline

G & EP & CL & G3
& 3.03 & 59.4 & 96.3 & 3.8 & 0.33
& 9.2 & 10.6 & 256.1 & 0.3 & 2.8 & 26.5 \\

G & EP & -- & G3
& 4.69 & 54.7 & 98.9 & 21.1 & 0.49
& 11.5 & 18.0 & 126.5 & 0.9 & 6.2 & 25.5 \\

G & EP & -- & G6
& 5.39 & 68.6 & 98.9 & 22.3 & 0.10
& 15.1 & 23.8 & 190.7 & 0.8 & 6.2 & 29.5 \\
\hdashline

G+L & EP & -- & G2+L4
& \InheritedGlobalPrefix & 0.06
&  3.1 &  3.5 & 13.8 & 2.3 & 11.3 & 31.5 \\

\bottomrule
\end{tabular*}
\caption{
\textbf{SID design ablation.}
SID variants built from the best representation in \autoref{tab:item_rep_intrinsics_ablation}.
G uses the global embedding, while G+L adds local facet embeddings.
Metrics are defined in \S\ref{sec:sid_design_ablation}; bounded scores use a 0--100 scale except Gini, which uses 0--1; SNR remains in dB and Avg. Item remains an item count.
}
\label{tab:sid_design_ablation}
\end{table}

\subsection{Domain Continued Pre-training Data Mixture}
\label{sec:cpt_data_mixture_ablation}

\S\ref{sec:cpt} defines the SID-alignment stage and the CPT data mixture.
Here we use the ablation only to answer two recipe questions: whether SID alignment is needed before CPT, and how much general replay to pair with the available shopping-domain CPT data.
Because a full 8B sweep is expensive, we use the dense 4B mixture grid to choose the replay ratio and apply the selected ratio in the final 4B and 8B recipes.

\autoref{tab:cpt_stage_ablations} first isolates the stage order.
Without CPT, the model preserves general benchmarks but remains weak on item retrieval.
CPT without SID alignment improves some shopping-understanding diagnostics, yet falls behind aligned CPT on item retrieval.
We therefore keep both stages: SID alignment makes the new tokens usable as generation targets, and CPT supplies the broader shopping-domain grounding.

\begin{table}[t]
\centering
\scriptsize
\setlength{\tabcolsep}{3.2pt}
\renewcommand{\arraystretch}{1.05}
\resizebox{\textwidth}{!}{
\begin{tabular}{ccccccccccc}
\toprule
\multicolumn{3}{c}{\textbf{CPT Configuration}} &
\multicolumn{2}{c}{\textbf{Item Retrieval}} &
\multicolumn{3}{c}{\textbf{Shopping Understanding}} &
\multicolumn{3}{c}{\textbf{General Capability}} \\
\cmidrule(lr){1-3}
\cmidrule(lr){4-5}
\cmidrule(lr){6-8}
\cmidrule(lr){9-11}
\makecell[c]{SID\\Alignment} &
\makecell[c]{Domain\\Tokens} &
\makecell[c]{General\\Tokens} &
\makecell[c]{Desc.\\$\rightarrow$ Item} &
\makecell[c]{Seq.\\Rec.} &
\makecell[c]{Intent\\Summary} &
\makecell[c]{Sequence\\Extract} &
\makecell[c]{Profile\\Extract} &
MMLU-Pro &
IFEval &
CMMLU \\
\midrule
\ding{51} & 0B & 0B
& 0.3 & 4.6 & 77.6 & 45.0 & 71.3 & \textbf{50.1} & \textbf{72.5} & \textbf{72.9} \\
\ding{55} & 70B & 70B
& 2.4 & 6.3 & \textbf{78.7} & 53.5 & \textbf{76.3} & 46.9 & 69.7 & 66.4 \\
\ding{51} & 70B & 35B
& \textbf{3.6} & 6.5 & 78.4 & 57.0 & 75.2 & 46.9 & 69.1 & 66.4 \\
\ding{51} & 70B & 70B
& 3.3 & \textbf{6.8} & 78.1 & \textbf{58.4} & 75.2 & 46.2 & 69.3 & 67.8 \\
\bottomrule
\end{tabular}
}
\caption{
\textbf{CPT stage ablations.}
Representative 4B configurations isolate the effects of SID alignment and domain CPT.
Bold marks the best score in each column; all metric entries are normalized to a 0--100 scale, with general benchmarks evaluated in non-thinking mode.
The full CPT grid in \autoref{fig:cpt_data_mixture_ablation} averages these three general columns together with GSM8K.
Under Item Retrieval, Desc.$\rightarrow$Item denotes Detailed-Description-to-Item HR@64 and Pers. Seq. Rec. denotes Personalized Sequential Recommendation HR@64.
}
\label{tab:cpt_stage_ablations}
\end{table}

\begin{figure}[H]
\centering
\includegraphics[width=\linewidth]{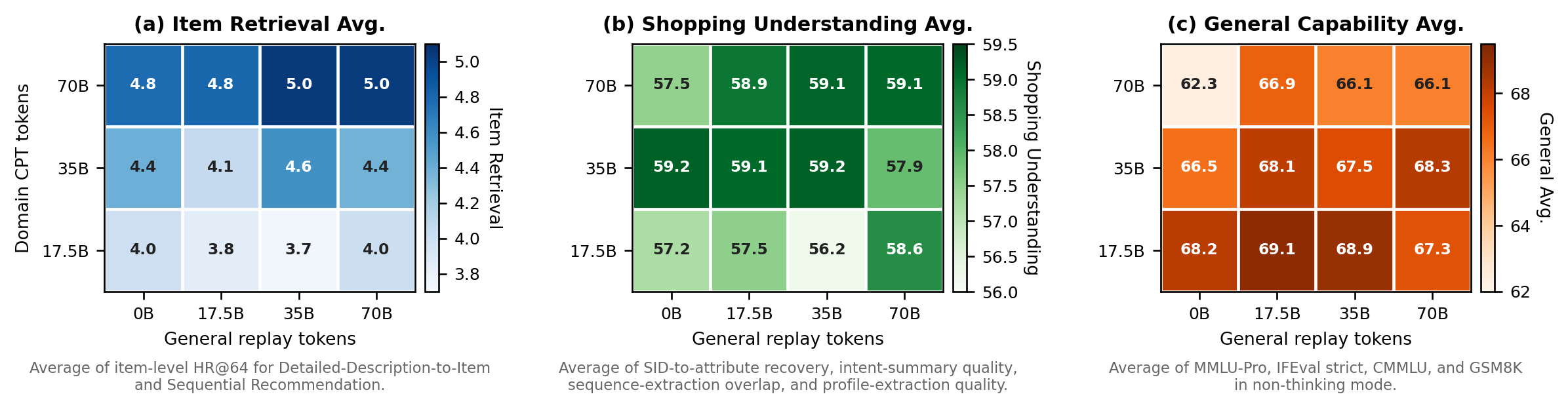}
\caption{
\textbf{CPT data-mixture grid.}
We sweep domain CPT tokens (rows) and general replay tokens (columns) on the 4B model.
Panels report averaged item retrieval, shopping understanding, and general capability scores; empty cells indicate configurations not run.
All scores are shown on the 0--100 scale.
}
\label{fig:cpt_data_mixture_ablation}
\end{figure}

With alignment enabled, we then choose the replay ratio within the evaluated token budget.
Under the 70B shopping-domain budget, 35B and 70B general replay are close: 35B replay is better for Detailed-Description-to-Item retrieval, while 70B replay is slightly better for sequential recommendation and sequence extraction.
The mixture grid in \autoref{fig:cpt_data_mixture_ablation} leads us to use 70B shopping-domain tokens with 35B general replay, a 2:1 domain-to-general ratio, for the final full-scale recipe.
This is a recipe choice within the available domain data and evaluated grid.

\subsection{Post-training Ablation}
\label{sec:post_training_ablation}

We conduct an ablation study on the proposed post-training pipeline described in \S\ref{sec:post_training} by progressively enabling each training component, including supervised fine-tuning (SFT), single-teacher on-policy distillation (OPD), multi-teacher on-policy distillation (MOPD), and RL rewards.
All variants in this ablation use the 8B model with the FORGE SID codebook.
\autoref{tab:post_training_ablation} summarizes the results.
Complementary comparisons among the teachers, SFT Student, and Final Model for ShopX with our final SID at 4B, 8B, and 30B-A3B scale are reported in \autoref{tab:rl_scaling_comparison}.
The analysis is organized into three stages, each examining the contribution of a specific component in the training recipe.

\begin{table}[!t]
\centering
\scriptsize
\setlength{\tabcolsep}{2.5pt}
\newcommand{\abltbd}{\textcolor{black!45}{TBD}}
\resizebox{\textwidth}{!}{
\begin{tabular}{@{}lcccccccccccc@{}}
\toprule
\multirow{2}{*}{\textbf{Variant}} &
\multicolumn{3}{c}{\textbf{MOPD Teachers}} &
\multicolumn{3}{c}{\textbf{Rewards}} &
\multicolumn{6}{c}{\textbf{Metrics}} \\
\cmidrule(lr){2-4}\cmidrule(lr){5-7}\cmidrule(lr){8-13}
&
\makecell[c]{General} &
\makecell[c]{SID} &
\makecell[c]{Self} &
\makecell[c]{LLM\\Judge} &
\makecell[c]{Ranking} &
\makecell[c]{IFR} &
\makecell[c]{CMMLU} &
\makecell[c]{IFEVal} &
\makecell[c]{Desc.\\$\rightarrow$Item} &
\makecell[c]{Prof.\\Ext.} &
\makecell[c]{Rank} &
\makecell[c]{Text--\\SID} \\
\midrule
General Teacher & -- & -- & -- & -- & -- & -- & 74.2 & 81.9 & 0.0 & 52.9 & 66.5 & 27.0 \\
SFT student & -- & -- & -- & -- & -- & -- & 70.2 & 72.5 & 3.7 & 77.6 & 63.9 & 69.5 \\
SID Prediction Teacher & -- & -- & -- & -- & -- & -- & 68.4 & 66.4 & 6.4 & 0.0 & 0.0 & 31.2 \\
\midrule
SID-only OPD & -- & \ding{51} & -- & -- & -- & -- & 66.7 & 70.4 & 5.4 & 0.8 & 0.0 & 35.9 \\
MOPD-2T & \ding{51} & \ding{51} & -- & -- & -- & -- & 74.5 & 77.5 & 5.35 & 4.4 & 57.4 & 77.0 \\
MOPD-3T & \ding{51} & \ding{51} & \ding{51} & -- & -- & -- & 73.0 & 76.9 & 5.12 & 78.1 & 63.1 & 75.3 \\
\midrule
+ LLM judge & \ding{51} & \ding{51} & \ding{51} & \ding{51} & -- & -- & 72.1 & 78.6 & 5.02 & 77.9 & 63.5 & 76.4 \\
+ Ranking & \ding{51} & \ding{51} & \ding{51} & -- & \ding{51} & -- & 73.0 & 75.6 & 5.00 & 77.6 & 67.3 & 77.0 \\
+ LLM judge\& Ranking & \ding{51} & \ding{51} & \ding{51} & \ding{51} & \ding{51} & -- & 73.7 & 76.0 & 5.11 & 77.4 & 68.6 & 78.2 \\
+ Full rewards & \ding{51} & \ding{51} & \ding{51} & \ding{51} & \ding{51} & \ding{51} & 71.8 & 75.2 & 5.2 & 77.7 & 70.1 & 88.0 \\
\bottomrule
\end{tabular}
}
\caption{
\textbf{Post-training ablation.}
All variants use the 8B model with the FORGE SID codebook.
We compare SFT, OPD/MOPD, and OPD--RL variants; checkmarks mark active task-family teachers and reward terms.
Desc.$\rightarrow$Item reports Detailed-Description-to-Item HR@64; Rank averages NDCG@3 over Intent+History and Contextual-Intent Item Ranking.
Text--SID reports the Interleaved Text--SID Fulfillment score.
}
\label{tab:post_training_ablation}
\end{table}

\tightparagraph{Stage 1: Effect of the SID prediction teacher.}
The first stage investigates whether specializing the model solely for SID prediction is sufficient to support the overall fulfillment capability.

\begin{itemize}[leftmargin=1.5em, itemsep=1pt, topsep=2pt, parsep=0pt, partopsep=0pt]
    \item \textbf{General Teacher.} Initialized from the pre-alignment Qwen3 checkpoint, it serves as the general-capability reference. It achieves the highest scores on CMMLU (74.2) and IFEval (81.9), but is unable to generate valid SIDs (Desc.$\rightarrow$Item HR@64 of 0.0).

    \item \textbf{SFT Student.} After supervised fine-tuning, it acquires broad shopping capabilities, achieving strong performance on Profile Extraction (77.6), Ranking (63.9), and Text\textendash SID (69.5). At the same time, its general capabilities remain reasonably preserved (CMMLU 70.2, IFEval 72.5).

    \item \textbf{SID Prediction Teacher.} Further specializing the SFT Student using only SID prediction data yields this teacher. This specialization substantially degrades the model's general capabilities (CMMLU 68.4, IFEval 66.4) and completely destroys its ability to perform Profile Extraction and Ranking, both of which collapse to zero.

    \item \textbf{SID-only OPD.} Applying single-teacher OPD transfers the SID expertise to the student, but fails to alleviate this specialization trade-off: Profile Extraction drops to 0.8 and Ranking remains unavailable.
\end{itemize}

\noindent\textbf{Takeaway.} These results indicate that optimizing exclusively for SID prediction produces a narrow specialist but severely compromises broader agentic-shopping capabilities.

\tightparagraph{Stage 2: Effect of multi-teacher distillation.}
We next examine whether MOPD can restore the lost general capabilities while preserving SID~specialization.

\begin{itemize}[leftmargin=1.5em, itemsep=1pt, topsep=2pt, parsep=0pt, partopsep=0pt]
    \item \textbf{MOPD-2T.} This variant incorporates both the General Teacher and the SID Prediction Teacher. Compared with SID-only OPD, introducing the General Teacher substantially recovers general capabilities, improving CMMLU from 66.7 to 74.5, IFEval from 70.4 to 77.5, and Ranking from 0.0 to 57.4. Nevertheless, Profile Extraction remains poor (4.4), since neither teacher provides supervision for this task family.

    \item \textbf{MOPD-3T.} To address this limitation, this variant further introduces the frozen \textbf{Self Teacher} (the post-SFT student), which supplies supervision for task families not covered by the other teachers. Consequently, Profile Extraction increases dramatically from 4.4 to 78.1, fully recovering the original SFT capability, while Ranking further improves to 63.1. This demonstrates that the Self Teacher effectively preserves behaviors already learned during SFT without interfering with SID specialization.
\end{itemize}

\noindent\textbf{Takeaway.} Overall, task-family-aware multi-teacher distillation reconciles SID specialization with general capability preservation, avoiding the single-teacher distillation trade-off.

\tightparagraph{Stage 3: Effect of RL rewards.}
Starting from \textbf{MOPD-3T}, we progressively introduce the proposed reward functions to evaluate their individual contributions.

\begin{itemize}[leftmargin=1.5em, itemsep=1pt, topsep=2pt, parsep=0pt, partopsep=0pt]
    \item \textbf{+ LLM Judge.} This variant introduces the response-quality reward, primarily improving instruction-following ability, as reflected by the increase of IFEval from 76.9 to 78.6, while leaving other metrics largely unchanged.

    \item \textbf{+ Ranking.} This variant adds the NDCG-based ranking reward, leading to a substantial improvement in Ranking (63.1 $\rightarrow$ 67.3) while maintaining comparable Text\textendash SID performance.

    \item \textbf{+ LLM Judge \& Ranking.} This variant combines the two rewards, producing the strongest Ranking performance (68.6) among all partial-reward variants while preserving high instruction-following capability. The two rewards exhibit complementary effects, with one improving response quality and the other directly optimizing ranking accuracy.

    \item \textbf{+ Full Rewards.} This variant further incorporates the proposed Interleaved Fulfillment Reward. This significantly boosts Text\textendash SID fulfillment from 78.2 to 88.0 and further improves Ranking to the best overall score (70.1), while only introducing a minor reduction in CMMLU (73.7 $\rightarrow$ 71.8) and IFEval (76.0 $\rightarrow$ 75.2).
\end{itemize}

\noindent\textbf{Takeaway.} These results suggest that each RL reward provides complementary task-specific optimization beyond MOPD, yielding consistent improvements with only marginal impact on general~capabilities.

\tightparagraph{Summary.}
The ablation study reveals a clear progression throughout the proposed post-training pipeline. SFT establishes broad task coverage but provides limited SID generation capability. Training solely for SID prediction sharpens SID specialization while sacrificing general and workflow capabilities. MOPD effectively resolves this conflict by leveraging task-family-aware teacher routing, simultaneously preserving SID specialization and recovering general behaviors. Finally, the proposed RL rewards further refine task-specific performance through complementary optimization objectives. As a result, the complete post-training pipeline achieves the strongest fulfillment-oriented performance while maintaining competitive general capabilities, demonstrating the effectiveness of progressively combining supervised alignment, multi-teacher distillation, and RL-reward optimization.

\section{Related Work}
\label{sec:related_work}

Existing LLM-augmented and generative recommender systems connect to \ourmethod in different roles: some define the tool-mediated serving paradigm we compare against, while others provide the generative item-space substrate that \ourmethod extends.
Below, we focus on how each family connects to our serving abstraction and workflow-level evaluation, rather than organizing the discussion as a conventional survey.

\tightparagraph{Tool-mediated LLM agents for recommendation.}
LLM-agent recommender systems commonly use the LLM for intent understanding, dialogue management, planning, memory, tool orchestration, critique, and response generation, while delegating item-space execution to external retrieval and ranking tools.
Representative systems include prompt-based conversational recommenders, tool-using recommendation agents, retrieval-augmented conversational recommenders, and industrial LLM-plus-retrieval pipelines~\citep{chat_rec,interecagent,recmind,recai,ra_rec,recgpt,recgpt_v2}.
In our terminology, they instantiate \strongemph{tool-mediated fulfillment}: rich context is compressed into tool plans, filters, queries, or candidate-set operations before external modules perform most item-space execution.
This is a strong and practical baseline family, but it exposes the interface-loss problem studied in this report: fine-grained constraints, personalized context, and multi-turn feedback must cross low-bandwidth tool boundaries before they can affect item-space fulfillment.
\ourmethod keeps the harness-side benefits of context assembly and workflow control, while training a SID-aware foundation model to carry contextual reasoning into SID-token retrieval, listwise selection, and grounded response generation.

Post-retrieval LLM reranking methods are a narrower neighbor of this family.
RankGPT and zero-shot listwise reranking show that prompted LLMs can improve ordering for retrieval tasks~\citep{rankgpt,lrl_rerank}, while recommendation-oriented systems such as LLM4Rerank and CARE adapt this idea to multi-objective recommendation reranking and conversational candidate selection~\citep{llm4rerank,care}.
These works are relevant to our listwise ranking/selection capability, but they assume that an external system has already produced the candidate set.
They therefore test post-retrieval ordering rather than the full path from context understanding to candidate retrieval, catalog grounding, response generation, and feedback handling.

\tightparagraph{Generative recommendation and semantic item identifiers.}
Generative recommendation provides a model-operable item-space interface by representing items as discrete tokens or Semantic IDs and formulating recommendation as autoregressive item generation~\citep{tiger,onerec,shi2025llada}.
The OneRec line shows that autoregressive item generation can consolidate conventional multi-stage cascaded recommendation pipelines into an end-to-end generative architecture~\citep{onerec,onerec_v2}; recent systems such as OpenOneRec and OneRec-Think further push this direction toward larger-scale data, reproducible training, foundation-model-style recommendation ability, and recommendation-specific reasoning~\citep{openonerec,onerec_think}.
Thus, instead of using SID generation mainly as a candidate-generation or pipeline-compression mechanism, we study how a SID-aware foundation model can serve as the central item-space fulfillment layer in agentic e-commerce.

SID construction has also become a core design problem.
Prior work improves identifier learning through collaborative-semantic alignment, content-collaboration fusion, learnable tokenizers, order-agnostic or LLM-oriented item tokens, end-to-end tokenizer optimization, transfer and multimodal variants, efficient long-SID decoding, and industrial-scale SID formation~\citep{tiger,lc_rec,content_based_collaborative_generation,letter,tokenrec,order_agnostic_identifier,etegrec,utgrec,bbqrec,long_semantic_ids,forge}.
These studies show that compactness, discrimination, collision control, codebook use, collaborative signal, and deployment efficiency all matter for SID quality.
Our setting adds a different requirement: SIDs must also be semantically recoverable and easy for a foundation model to operate on as an item language.
This motivates both our SID design ablation in \S~\ref{sec:sid_design_ablation} and the training recipe that integrates item--SID alignment, retrieval, ranking/selection, response generation, and preference-update extraction with broader language and reasoning abilities.

\section{Discussion and Future Work}
\label{sec:discussion_future_work}

\tightparagraph{Model-native fulfillment as a different serving point.}
The practical question is not whether a production system should use tools, retrieval indexes, ranking models, or serving-side state.
Industrial recommender systems will continue to use such infrastructure.
The question is where complex context is carried during fulfillment.
Tool-mediated agents place the LLM mainly before and after item-space operations.
Post-retrieval rerankers place the LLM after candidate generation.
Generative recommenders place item IDs inside the model output space, but are often evaluated mainly as item predictors or candidate generators.
\ourmethod targets a different serving point.
The serving harness defines the model-facing action protocol and exposes support surfaces for context access, catalog grounding, and state management, while the central model selects and composes SID-native fulfillment actions.
It can use SIDs not only as output labels, but as model-operable item representations across retrieval, ranking/selection, and grounded response generation.
This is why our main evaluation compares complete system configurations: the \ourmethod serving framework versus strong tool-mediated baselines backed by external retrieval and ranking tools.

\tightparagraph{Toward agentic training.}
The present training recipe does not yet include task-specific agentic reinforcement learning over full shopping trajectories.
In the current system, tool registration, tool understanding, and tool invocation rely largely on the native tool-use abilities of the Qwen3-family instruction models, further adapted through CPT and SFT data that include tool-calling examples.
This setup lets us evaluate model-native fulfillment under a fixed serving framework, but it does not directly optimize long-horizon agentic decisions such as when to retrieve, when to ask for clarification, how to revise item choices after feedback, or when to update serving-side state.
An important next step is trajectory-level reinforcement learning over multi-turn model--harness interactions.
The workflow-level evaluation dimensions in \S~\ref{sec:evaluation_protocol} provide a natural starting point for reward modeling, since they already measure task success, catalog grounding, feedback adaptation, personalization, and state-update quality.
Going further, a dynamic user-simulation agent could provide interactive reward feedback that better approximates real shopping sessions, enabling personalized agentic RL over evolving preferences and multi-turn fulfillment outcomes.

\section{Conclusion}
\label{sec:conclusion}

We introduced \ourmethod, a foundation model for intent-to-item fulfillment in agentic shopping. The report argues that AI-native shopping should translate rich intent, personalization context, and multi-turn feedback into catalog-grounded outcomes with fewer lossy tool hand-offs. \ourmethod is deployed through a model-native fulfillment framework in which a lightweight serving harness defines the model-facing action protocol and support surfaces, while the central model selects and composes SID-native fulfillment actions. Across SID design, continued pre-training, post-training, and system-level evaluation, the results support model-native fulfillment as a practical direction for complex and context-dependent shopping workflows.

\clearpage
\bibliographystyle{unsrtnat}
\nobibliography*
\bibliography{reference}

\clearpage
\appendix
\section*{Appendix}

\section{Author List}
\label{app:author_list}

\begin{multicols}{2}
\noindent
\textbf{Core Contributors} \\
Jiacheng Chen$^{*}$ \\
Tao Zhang$^{*}$ \\
Manxi Lin$^{*}$ \\
Dunxian Huang$^{*}$ \\
Teng Shi \\
Honghao Fu$^{1}$ \\
Mengyan Li \\
Xinming Zhang \\
Chenchi Zhang \\
Xuan Lu$^{2}$ \\
Xiaoxiong Du \\
Junjun Zheng$^{\dagger}$ \\
Xiangheng Kong \\
Yuning Jiang \\

\columnbreak

\noindent
\textbf{Contributors} \\
Haibin Chen \\
Shaolin Ye \\
Hao Chang \\
Xiaoqi Li \\
Shuwen Xiao \\
Yujin Yuan \\
Jingxuan Feng \\
Shaopan Xiong \\
Huimin Yi \\
Ju Huang \\
Qiu Shen$^{3}$ \\
Ying Chen \\
Dan Ou \\
Haihong Tang \\
Bo Zheng \\
\end{multicols}

\noindent
$^*$Equal contribution.\\
$^1$University of Queensland.\\
$^2$Shanghai Jiao Tong University.\\
$^3$Nanjing University.\\
$^\dagger$Corresponding contributor. Contact: \href{mailto:fangcheng.zjj@alibaba-inc.com}{\nolinkurl{fangcheng.zjj@alibaba-inc.com}}.

\section{Evaluation Details}
\label{app:eval_details}

\subsection{Framework-Level Benchmark Details}
\label{app:framework_benchmark_details}

\tightparagraph{Benchmark composition.}
\label{app:framework_benchmark_composition}

\autoref{tab:framework_benchmark_stats} gives the detailed composition of the framework-level benchmark described in \S\ref{sec:framework_level_evaluation_protocol}.
These counts are used to make the evaluation taxonomy explicit without interrupting the main protocol narrative.

\begin{table}[t]
\centering
\scriptsize
\setlength{\tabcolsep}{2.2pt}
\renewcommand{\arraystretch}{1.08}
\begin{tabularx}{\linewidth}{@{}
>{\raggedright\arraybackslash}p{0.18\linewidth}
>{\raggedright\arraybackslash}p{0.29\linewidth}
>{\centering\arraybackslash}p{0.13\linewidth}
>{\centering\arraybackslash}p{0.10\linewidth}
>{\centering\arraybackslash}p{0.10\linewidth}
>{\centering\arraybackslash}p{0.07\linewidth}
@{}}
\toprule
\textbf{Single-turn intent type} &
\textbf{Description} &
\textbf{Adversarial} &
\textbf{Robust} &
\textbf{Semantic} &
\textbf{Total} \\
\midrule
narrow\_category & Precise category requests, e.g., toothbrush heads & 65 & 0 & 0 & 65 \\
multi\_category & Multi-category or bundle-style needs & 36 & 29 & 0 & 65 \\
scene\_based & Scenario-driven requests, e.g., camping use & 0 & 27 & 38 & 65 \\
general\_semantic & Broad semantic needs, e.g., stylish and durable accessories & 0 & 23 & 36 & 59 \\
minimal & Minimal or underspecified queries & 0 & 25 & 0 & 25 \\
\midrule
\textbf{Single-turn total} & Full single-turn split & \textbf{101} & \textbf{104} & \textbf{74} & \textbf{279} \\
\bottomrule
\end{tabularx}

\vspace{0.55em}

\begin{minipage}[t]{0.49\linewidth}
\centering
\setlength{\tabcolsep}{2.2pt}
\begin{tabularx}{\linewidth}{@{}
>{\raggedright\arraybackslash}X
>{\centering\arraybackslash}p{0.18\linewidth}
>{\centering\arraybackslash}p{0.20\linewidth}
@{}}
\toprule
\textbf{Constraint/profile signal} & \textbf{Count} & \textbf{Coverage} \\
\midrule
User profile available & 279 & 100.0\% \\
Expected category annotated & 279 & 100.0\% \\
Soft constraints & 278 & 99.6\% \\
Hard constraints & 193 & 69.2\% \\
Target population specified & 63 & 22.6\% \\
Brand specified & 45 & 16.1\% \\
Negation expression & 35 & 12.5\% \\
Price budget & 27 & 9.7\% \\
\bottomrule
\end{tabularx}
\end{minipage}
\hfill
\begin{minipage}[t]{0.49\linewidth}
\centering
\setlength{\tabcolsep}{2.2pt}
\begin{tabularx}{\linewidth}{@{}
>{\raggedright\arraybackslash}p{0.25\linewidth}
>{\raggedright\arraybackslash}X
>{\centering\arraybackslash}p{0.15\linewidth}
@{}}
\toprule
\textbf{Multi-turn seed} & \textbf{Description} & \textbf{Count} \\
\midrule
vague & Vague query requiring clarification & 25 \\
constrained & Multi-constraint request requiring persistence & 20 \\
clear & Clear shopping intent & 15 \\
multi\_need & Multi-category need & 15 \\
ambiguous & Ambiguous query & 5 \\
\midrule
\textbf{Total} & Full multi-turn split & \textbf{80} \\
\bottomrule
\end{tabularx}
\end{minipage}
\caption{
\textbf{Framework-level benchmark composition.}
The single-turn split contains 279 cases, stratified by intent type and difficulty tier.
Coverage denotes the fraction of single-turn cases containing the corresponding profile, annotation, or constraint signal.
The multi-turn split contains 80 trajectories, grouped by seed type to cover clarification, persistent-constraint following, explicit fulfillment, multi-need requests, and ambiguity resolution.
}
\label{tab:framework_benchmark_stats}
\end{table}

\tightparagraph{Multi-turn rollout protocol.}
\label{app:multi_turn_rollout_protocol}

For multi-turn framework-level cases, Claude Sonnet 4.6 is used as a fixed shopper-side simulator.
At each turn, the simulator receives the user's profile information, the seed query, and the agent's previous response, then either produces the next user message or ends the interaction once it determines that the request has been satisfied.
Rollouts are capped at five user turns.
The simulator and rubric judge use separate prompts and roles; the judge only reads the stored trajectory after the interaction is complete.

\subsection{Framework-Level Overview Metric Protocol}
\label{app:framework_overview_metric_protocol}

The eight overview metrics in \autoref{tab:framework_overview_metrics} are derived from stored single-turn responses and multi-turn trajectories.
All judge calls use Claude Sonnet 4.6 as a fixed rubric judge.
The judge is separate from the multi-turn shopper simulator, and all rubric scores are normalized to a 0--100 scale before aggregation.

\tightparagraph{Single-turn judge dimensions.}
For each single-turn response, the judge receives the original query, annotated shopping intent, profile summary, response text, and the returned catalog items.
It emits four structured outputs.
\emph{Item relevance} labels each displayed item as \texttt{served}, \texttt{partially\_served}, or \texttt{not\_served}; these labels define intent served rate, category precision, and NDCG@5 using relevance gains 2, 1, and 0.
\emph{Attribute consistency} extracts explicit item-attribute claims from the response and marks each claim as consistent, conflicting, or not verifiable against catalog evidence; attribute accuracy is computed as consistent claims divided by consistent plus conflicting claims.
\emph{Profile utilization} is a 1--5 rubric score measuring whether item choice, ordering, or explanation reflects profile evidence not already stated in the query; the overview metric uses the average score divided by 5 and multiplied by 100, while the stricter score-$\geq$4 rate is kept only as a diagnostic.
\emph{Naturalness} is a 1--5 response-quality rubric and is reported in detailed single-turn diagnostics but is not one of the eight overview axes.
Single-turn category coverage measures how many annotated expected category facets are covered by the relevant returned items.

\tightparagraph{Multi-turn judge dimensions.}
For each multi-turn trajectory, the judge receives the full dialogue, profile context, final returned items, and catalog evidence.
The result-quality judge reports final goal achievement, per-item served labels, category coverage, and a 1--5 profile-match score.
The process-quality judge reports three trajectory dimensions used in the overview metrics.
The constraint-grounding dimension combines clarification behavior, information gain from clarification, preservation of accumulated constraints, candidate evidence for hard constraints, honesty when catalog evidence is missing, and turn efficiency.
The feedback-response dimension combines feedback understanding, effectiveness of the revised recommendations, preservation of unrelated preferences, correction speed, and absence of later regression; correction-style signals are merged into this dimension.
The cross-turn-reference dimension combines entity-reference accuracy and operation correctness when the user refers to a previously shown item.

\tightparagraph{Process-dimension normalization.}
The multi-turn process dimensions are weighted averages of normalized rubric fields:
\begin{itemize}
    \item \textbf{Constraint grounding:} clarification necessity and information gain (0.15 each), constraint preservation and candidate coverage (0.25 each), grounding honesty and turn efficiency (0.10 each).
    \item \textbf{Feedback response:} feedback accuracy and adjustment effectiveness (0.25 each), preference preservation (0.20), correction speed and no-regression behavior (0.15 each).
    \item \textbf{Cross-turn reference:} entity accuracy and requested-operation correctness (0.50 each).
\end{itemize}
Binary rubric fields are mapped to 0 or 100; 1--5 rubric fields are multiplied by 20; turn efficiency uses $\max(0,(5-t)/4)\times100$, where $t$ is the number of turns.

\tightparagraph{Overview aggregation.}
Let ST denote single-turn metrics and MT denote multi-turn metrics.
The eight framework-level axes are computed as:
\begin{align*}
\text{Intent Fulfillment} &= \operatorname{mean}(\text{ST intent served rate}, \text{MT goal achievement}), \\
\text{Item Precision} &= \text{ST category precision}, \\
\text{Ranking Quality} &= \text{ST NDCG@5}, \\
\text{Category Coverage} &= \operatorname{mean}(\text{ST category coverage}, \text{MT category coverage}), \\
\text{Personalization} &= \operatorname{mean}(\text{ST profile-utilization average}, \text{MT profile match}), \\
\text{Constraint Grounding} &= \operatorname{mean}(\text{ST attribute accuracy}, \text{MT constraint-grounding score}), \\
\text{Feedback Adaptation} &= \text{MT feedback-response score}, \\
\text{Cross-turn Reference} &= \text{MT cross-turn-reference score}.
\end{align*}
This aggregation is used both by the main framework table and by the radar visualization.

\subsection{Capability Diagnostic Metric Protocol}
\label{app:capability_metric_protocol}

Each task in \autoref{tab:capability_diagnostics_workflow_transfer} reports the primary metric shown in parentheses in the task column.
For target-item retrieval rows, the main diagnostic table reports item-level HR@4 and HR@64 together so short-list hit quality and broader candidate coverage can be read side by side.
For open-ended contextual-intent retrieval, where a broad query can admit multiple valid catalog items, the main table reports VLM-judged query--item relevance instead of strict hit rate.
We use the benchmark's standard primary metric for general capabilities and a fixed task-level metric for each remaining shopping diagnostic.
The main diagnostic table reports these primary metrics on a 0--100 scale.

\tightparagraph{Metric definitions.}
Accuracy and Acc@1 measure exact top-1 correctness after answer extraction or item-id resolution.
HR@K measures whether any annotated target item appears in the top-$K$ generated or resolved catalog items.
NDCG@K measures listwise ranking quality with graded item relevance, normalized by the ideal ordering.
Relevant-unit F1 computes set-level F1 over annotated profile facts or behavior SIDs selected as relevant to the current context.
Judge Score denotes the mean scalar score from Claude Sonnet 4.6 under the task-specific rubric after catalog-grounding checks.

\tightparagraph{General capabilities.}
BBH CoT, MMLU-Pro, GPQA-Diamond, CMMLU, MATH-500, and GSM8K Flex report accuracy under their standard answer-extraction protocols.
IFEval reports strict prompt-level instruction-following accuracy.
MBPP+ reports pass@1.

\tightparagraph{Shopping semantics.}
Intent Summary reports a judge-based summary-quality score.
Item Association reports Judge Score.
Item Description Recovery reports ROUGE.

\tightparagraph{Context evidence extraction.}
Static-Profile Evidence Extraction reports an LLM-judge relevance score, while Behavior-Sequence Evidence Extraction reports sequence-overlap IoU.

\tightparagraph{Textualized recommendation/search.}
Textualized Recommendation and Textualized Search report a Multiple Choice Question Score that gives partial credit when the selected hard negative is semantically close to the target.

\tightparagraph{Item-space fulfillment.}
Detailed-Description-to-Item retrieval reports item-level HR@4 and HR@64 in the main diagnostic table after resolving the intermediate SID generation output to catalog items.
For Detailed-Description-to-Item retrieval, Query--Item Relevance is reported as a supplementary sub-metric under the same task and is scored by a VLM judge.
After beam search, we take the top five predicted SIDs, resolve two catalog items for each SID, and provide the VLM judge with the input text, the target item metadata, and the same metadata for each candidate item.
The item metadata includes the main image, long title, short title, category, seller, brand, and price; the resulting relevance score is normalized to a 0--100 scale.
Personalized Sequential Recommendation and Intent+History Item Retrieval report HR@4 and HR@64 after resolving generated SIDs to catalog items.
Personalized Sequential Recommendation conditions on the static profile, the behavior sequence, and, for a subset of examples, an optional query.
Intent+History Item Retrieval uses the current intent or query together with selected relevant historical items rather than the full behavior sequence, while Contextual-Intent Item Retrieval uses underspecified, ambiguous, or scenario-rich intent queries without personal history evidence.
Because such contextual-intent queries often have multiple acceptable answers, Contextual-Intent Item Retrieval uses VLM-judged query--item relevance as its main-table metric; strict exact-target HR@K is treated only as an auxiliary diagnostic and omitted from the main table.
Intent+History Item Ranking and Contextual-Intent Item Ranking jointly report Acc@1 and NDCG@3.

\tightparagraph{End-to-end system outcomes.}
System outcomes reuse the framework-level overview metrics from \S\ref{sec:framework_level_evaluation_protocol}.
The main diagnostic table carries the same overview dimensions so model-facing capability changes can be related back to complete-system behavior under the fixed harness.

\subsection{SID Design Ablation Metrics}
\label{app:sid_design_ablation_metrics}

\autoref{tab:item_rep_intrinsics_ablation} and \autoref{tab:sid_design_ablation} use three groups of diagnostics.
All bounded scores are reported on a 0--100 scale in the main tables, except Gini, which is reported on its native 0--1 scale; SNR remains in dB and Avg. Item remains an item count.

\tightparagraph{Representation intrinsics.}
These diagnostics evaluate the continuous item representation before SID construction.
Lang. ROUGE is ROUGE-L between the decoder-reconstructed caption and the reference item caption generated by Qwen3-VL-235B-A3B.
R@10 and NDCG@10 use the global vector $g$ for nearest-neighbor retrieval over a fixed evaluation pool with 80{,}036 equivalent-product query pairs and 1{,}088{,}790 candidate items; the query item is excluded from the candidate pool, and the paired equivalent product is the positive item.
Cat. Purity uses the same catalog and KNN@100 neighbors, reporting the largest fraction of neighbors that share the same leaf-category path.

\tightparagraph{Global-prefix intrinsics.}
These diagnostics evaluate the discretized global-prefix codebook before lightweight validation.
SNR measures how well the quantized reconstruction $\hat{g}$ preserves the original global embedding $g$.
Quantized NDCG@10 and Cat. Purity repeat the retrieval and neighborhood-coherence tests above using $\hat{g}$.
Prefix Cons.@3 is computed on equivalent-product pairs as the pairwise exact-match rate of the first three global SID tokens.
Gini measures path-load imbalance in the SID assignment, i.e., whether catalog items are spread evenly across available SID paths or concentrated in a small subset of paths.
Lower Gini indicates more balanced path usage under the same codebook structure, while higher Gini indicates more concentrated assignments.
For G+L variants, global-prefix reconstruction, retrieval, category purity, and prefix consistency inherit the corresponding global-prefix diagnostics, while Gini and the lightweight codebook-validation metrics reflect the evaluated SID path.

\tightparagraph{Codebook validation.}
These diagnostics are model-facing evaluations after the lightweight validation protocol in \S\ref{sec:sid_codebook_validation}.
Desc.$\rightarrow$SID HR@K measures whether the gold SID appears among the top-$K$ SID candidates under the fixed beam-search setting; invalid SID strings are not filtered before scoring.
Desc.$\rightarrow$Item HR@K resolves generated SIDs against the 25M-item mini catalog and checks whether the target item is recovered.
When one generated SID maps to multiple items, resolved items follow a fixed random order for metric computation.
Avg. Item reports the average number of catalog items attached to the beam-search top-1 SID, measuring the ambiguity of the first generated code path; lower values indicate a more specific top-1 SID.
SID$\rightarrow$Desc. ROUGE evaluates SID-conditioned item-information recovery by comparing the model's facet-level answer against the corresponding reference item information.

\subsection{Interleaved Text--SID Fulfillment Rubric}
\label{app:text_sid_interleaved_fulfillment_rubric}

Interleaved Text--SID Fulfillment is a shopping item-space fulfillment diagnostic in which a model must produce a natural-language shopping recommendation while interleaving semantic item identifiers (SIDs).
Because there is no single canonical answer, we evaluate it with a rubric-based LLM judge under the same fixed judging setup used for the capability-breakdown evaluation.
For each case, the judge receives the user query, the generated response with hidden reasoning spans stripped, and the catalog records resolved from all predicted SIDs through IGraph lookup.
The judge scores the response along five dimensions on a 0--4 scale:
\begin{itemize}
    \item \textbf{Intent satisfaction} measures whether the recommended items and text address the user's scenario, constraints, and shopping goal.
    \item \textbf{Resolved item quality} measures whether predicted SIDs resolve to valid catalog items whose categories, attributes, and complementary roles match the query.
    \item \textbf{Interleaving validity} measures whether SIDs are correctly embedded alongside descriptive text, each with a clear role, and free of malformed, unresolved, or duplicated identifiers.
    \item \textbf{Grounded rationale} measures whether the stated recommendation reasons are supported by resolved item attributes rather than fabricated claims.
    \item \textbf{Response quality} measures fluency, informativeness, and whether the answer is actionable for shopping.
\end{itemize}
The five dimensions are weighted and normalized to a 0--100 composite score.
Three boolean penalty flags further discount the score: fabricated identifiers ($-20\%$), missing SIDs for explicitly recommended items ($-10\%$), and duplicate listings ($-5\%$).

\subsection{General Capability Evaluation Settings}
\label{app:general_capability_eval_settings}

The general-capability rows in \autoref{tab:capability_diagnostics_workflow_transfer} are evaluated with lm-evaluation-harness using a vLLM backend.
Unless otherwise specified by the benchmark task, we run inference in bfloat16 with tensor parallelism over 8 GPUs, apply the model chat template, disable thinking mode, and use greedy decoding with \texttt{temperature=0} and \texttt{do\_sample=false}.
Model maximum length is set between 16K and 32K tokens according to the run configuration, with GPU memory utilization in the 0.80--0.85 range.
For MBPP+, code execution is enabled through the harness unsafe-code confirmation flag.

\begin{table}[h]
\centering
\scriptsize
\setlength{\tabcolsep}{3pt}
\renewcommand{\arraystretch}{1.08}
\begin{tabularx}{\linewidth}{@{}>{\raggedright\arraybackslash}p{0.15\linewidth}>{\raggedright\arraybackslash}p{0.22\linewidth}>{\centering\arraybackslash}p{0.08\linewidth}>{\raggedright\arraybackslash}p{0.23\linewidth}>{\raggedright\arraybackslash}X@{}}
\toprule
\textbf{Benchmark} & \textbf{lm-eval task} & \textbf{Shots} & \textbf{Primary metric} & \textbf{Data and scope} \\
\midrule
BBH CoT & \texttt{bbh\_cot\_fewshot} & 3 & exact match after get-answer extraction & 27 BIG-Bench Hard subtasks; about 6.5K examples. \\
MMLU-Pro & \texttt{mmlu\_pro} & 5 & exact match after custom answer extraction & 14 subjects and about 12K 10-choice questions. \\
GPQA-Diamond & \makecell[l]{\texttt{gpqa\_diamond}\\\texttt{\_generative\_n\_shot}} & 0 & exact match after flexible option extraction & Diamond subset of GPQA with 198 graduate-level science questions. \\
CMMLU & \texttt{cmmlu} & 0 & multiple-choice accuracy by log likelihood & 67 Chinese subject areas; Parquet-converted CMMLU split for current \texttt{datasets} compatibility. \\
IFEval & \texttt{leaderboard\_ifeval} & 0 & prompt-level strict accuracy & 541 instruction-following prompts; loose and instruction-level variants are logged but not reported as the main score. \\
MATH-500 & \texttt{hendrycks\_math500} & 0 & exact match & 500 held-out MATH problems covering seven difficulty levels and multiple math branches. \\
GSM8K Flex & \texttt{gsm8k\_cot} & 8 & exact match after flexible numeric extraction & 1,319 grade-school math word problems; few-shot examples use the first-$n$ sampler. \\
MBPP+ & \texttt{mbpp\_plus} & 3 & pass@1 after code extraction and execution & 399 Python programming tasks from EvalPlus MBPP+. \\
\bottomrule
\end{tabularx}
\caption{\textbf{General-capability benchmark settings.}
This table expands the compact setting labels shown under the general-capability task names in \autoref{tab:capability_diagnostics_workflow_transfer}.}
\label{tab:general_capability_eval_settings}
\end{table}

\tightparagraph{Prompting, extraction, and generation details.}
BBH CoT uses each subtask's three chain-of-thought few-shot examples and stops generation at \texttt{</s>}, \texttt{Q:}, or \texttt{<|im\_end|>} with a 1024-token generation cap; answers are extracted from final-answer style spans and compared with case and punctuation normalization.
MMLU-Pro uses five validation-shot examples, a 10-option question format, a 2048-token cap, and stops at the next \texttt{Question:}; the reported score is the weighted average over subjects.
GPQA-Diamond is run as a 0-shot generative multiple-choice task with options A--D, a 4096-token cap, and flexible extraction of the final parenthesized option.
CMMLU is evaluated as a log-likelihood multiple-choice task with a Chinese answer prompt, using the highest-probability option as the prediction and averaging accuracy across subtasks.
IFEval directly uses the dataset prompt field and allows generation up to 1280 tokens without an early stop sequence; we report \texttt{prompt\_level\_strict\_acc}.
MATH-500 uses the \texttt{Problem: ... Answer:} format with an 8192-token cap and stops on the next problem marker or chat end token.
GSM8K uses eight few-shot exemplars, a 2048-token cap, and flexible extraction of the last numeric answer after stripping commas, dollar signs, and the \texttt{\#\#\#\#} answer marker.
MBPP+ follows the MBPP-style expert-Python prompt with three visible tests, a 2048-token cap, stops at \texttt{[DONE]} or the chat end token, and reports pass@1 after executing the extracted code against EvalPlus tests.

\section{Additional Results and Qualitative Examples}
\label{app:additional_results}
\label{app:case_studies_and_traces}

This section groups supplementary results that are easier to read as qualitative examples or focused auxiliary analyses.
We place fulfillment trace comparisons first because they directly support the framework-level results in \S\ref{sec:end_to_end_system_evaluation}, followed by generated-output cases and a broader textualized recommendation analysis for general LLMs.

\subsection{Qualitative Fulfillment Trace Comparisons}
\label{app:serving_trace_examples}

The case-level figures in this subsection are qualitative samples for the framework-level comparison in \S\ref{sec:end_to_end_system_evaluation}.
Each figure keeps the original Chinese request, dialogue, and catalog item titles as evidence, and adds an English intent summary for non-Chinese readers.
The upper trace panel shows how \ourmethod fills the \textit{Plan}, \textit{Execute}, \textit{Fulfill}, and \textit{Update} model-action slots from \S\ref{sec:shopx}, while distinguishing harness-side \textit{Context}, \textit{State}, and \textit{Catalog} support surfaces.
The lower panel compares \ourmethod with a tool-mediated baseline on the same case, including final catalog-grounded items and the observed failure mode.

The examples are intended as reading aids rather than an additional benchmark table.
Together they cover single-turn problem-to-product inference, hard constraint grounding, bundle-style category coverage, negated-style filtering, stateful refinement, active category preservation, and cross-turn item references.
For multi-turn cases, the shopper simulator may terminate a method's trajectory as soon as it judges the user's request to be satisfied; therefore, a multi-turn benchmark case can show only one \ourmethod turn when first-pass fulfillment is accepted.
Within each figure, score chips summarize the applicable judge outputs for that case; process-oriented chips are shown only when the corresponding multi-turn dimension is applicable, and the figure body gives the case-specific mechanism takeaway.

\newcommand{\qualcasefigure}[4]{%
\begin{figure}[t]
    \centering
    \includegraphics[width=\textwidth,height=0.78\textheight,keepaspectratio]{assets/qualitative_cases/#1}
    \caption[Qualitative case #2]{Qualitative fulfillment trace comparison, Case~#2. The case title, intent summary, method outputs, and mechanism takeaway are shown inside the figure.}
    \label{#4}
\end{figure}
}

\newcommand{\qualcasefigureearlystop}[4]{%
\begin{figure}[p]
    \centering
    \includegraphics[width=\textwidth,height=0.78\textheight,keepaspectratio]{assets/qualitative_cases/#1}
    \caption[Qualitative case #2]{Qualitative fulfillment trace comparison, Case~#2. Although this is a multi-turn benchmark case, the shopper simulator terminates the \ourmethod trajectory after the first response because it judges the request to be satisfied; the baseline trajectory continues until satisfaction or the turn limit.}
    \label{#4}
\end{figure}
}

\newcommand{\qualcasefigurepair}[8]{%
\begin{figure}[p]
    \centering
    \includegraphics[width=0.86\textwidth]{assets/qualitative_cases/#1}
    \vspace{0.7em}
    \includegraphics[width=0.86\textwidth]{assets/qualitative_cases/#5}
    \caption[Qualitative cases #2 and #6]{Qualitative fulfillment trace comparisons, Cases~#2 and~#6. Case~#2 illustrates #3 Case~#6 illustrates #7 The case titles, intent summaries, method outputs, and mechanism takeaways are shown inside the figures.}
    \label{#4}
    \label{#8}
\end{figure}
}

\qualcasefigureearlystop
{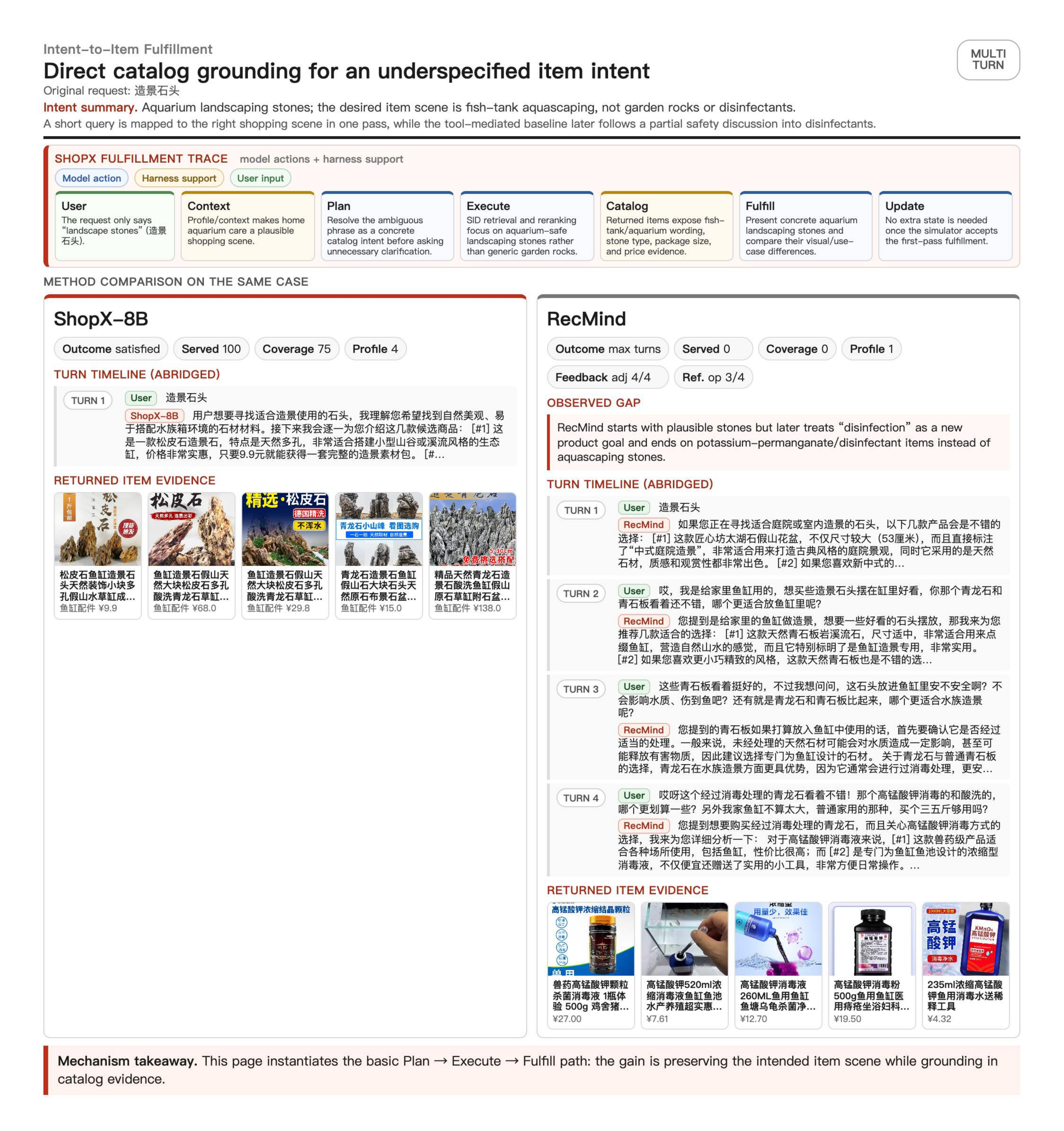}
{1}
{A short request is grounded to fish-tank aquascaping stones, while the tool-mediated baseline eventually drifts toward disinfectant products after follow-up discussion.}
{fig:qual_case_direct_catalog_grounding}

\qualcasefigureearlystop
{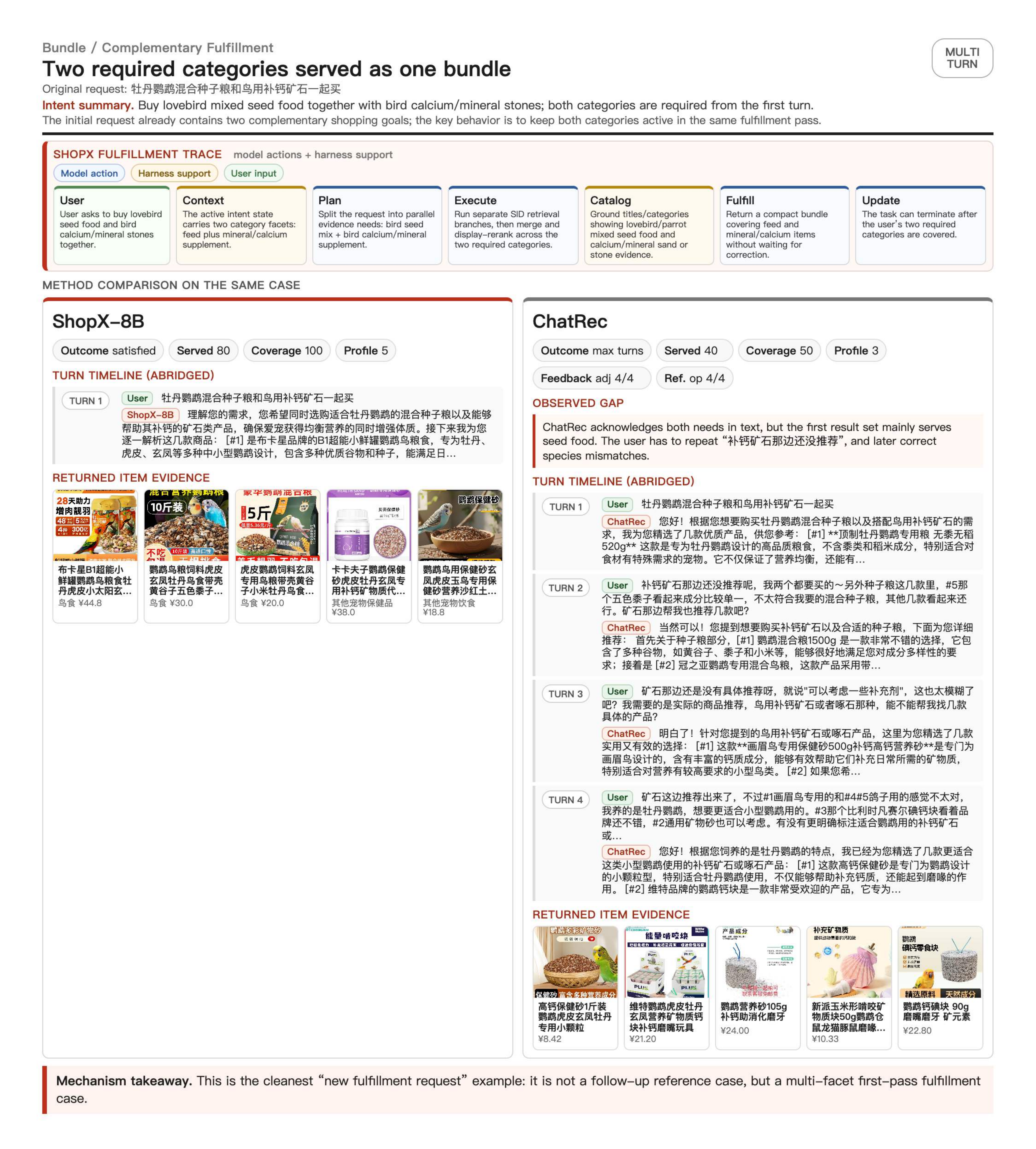}
{2}
{The initial request contains two complementary shopping goals, and the trace highlights whether both categories remain active in the same fulfillment pass.}
{fig:qual_case_bundle_categories}

\qualcasefigure
{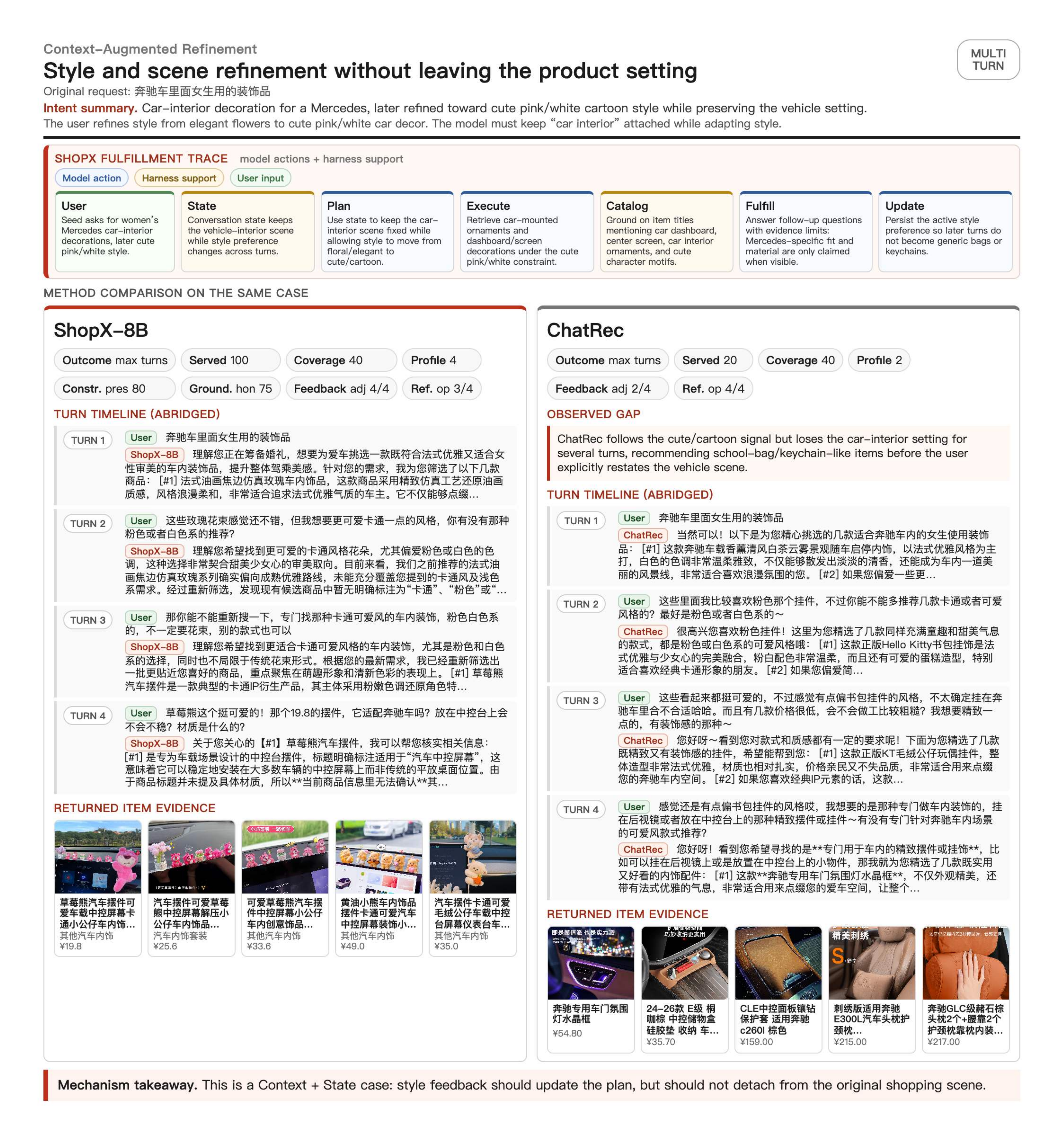}
{3}
{A multi-turn refinement changes style preferences while preserving the car-interior product scene.}
{fig:qual_case_style_scene_refinement}

\qualcasefigure
{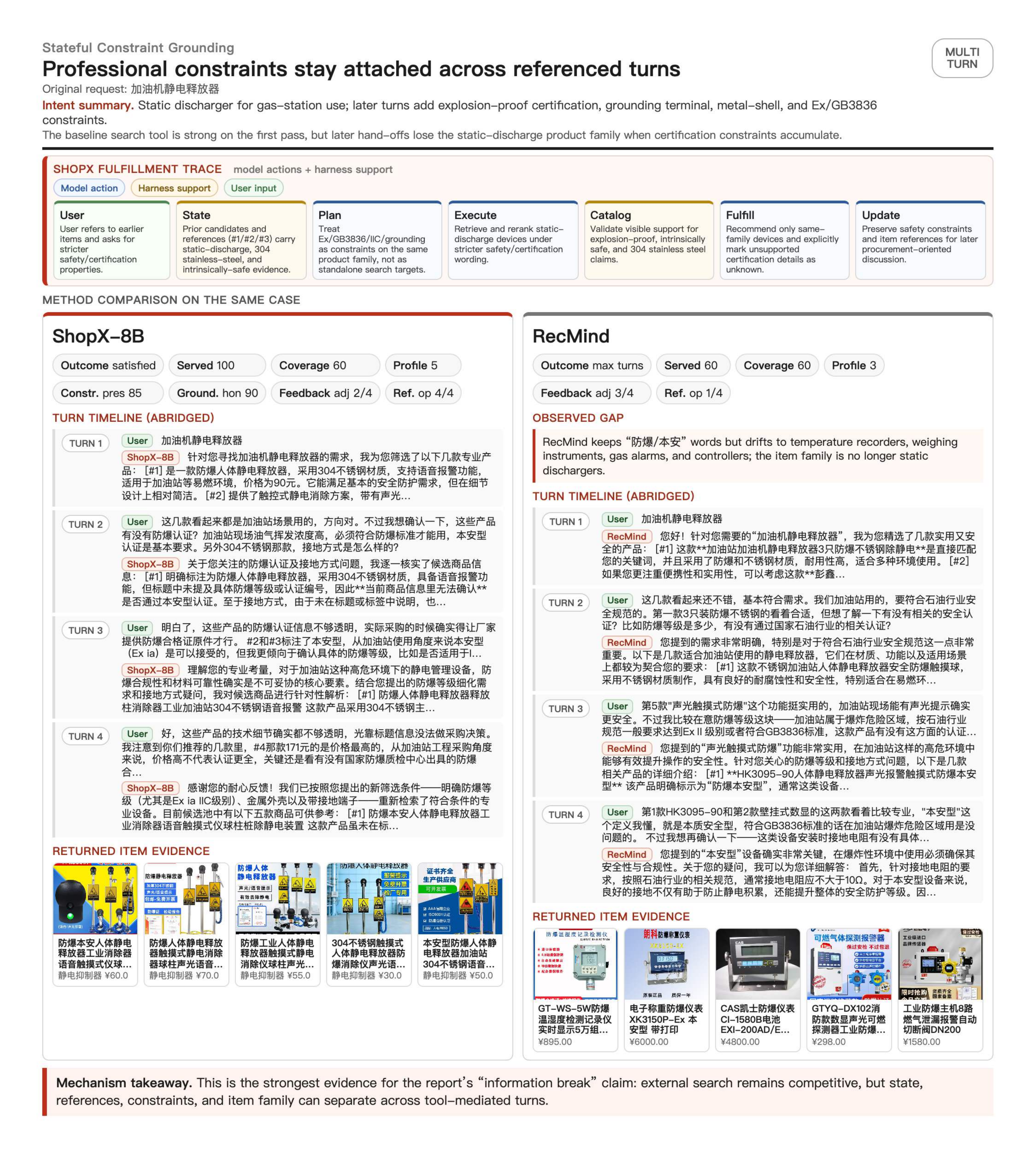}
{4}
{The follow-up asks about professional safety constraints; the comparison shows whether those constraints remain attached to the referenced product family.}
{fig:qual_case_professional_constraints}

\qualcasefigure
{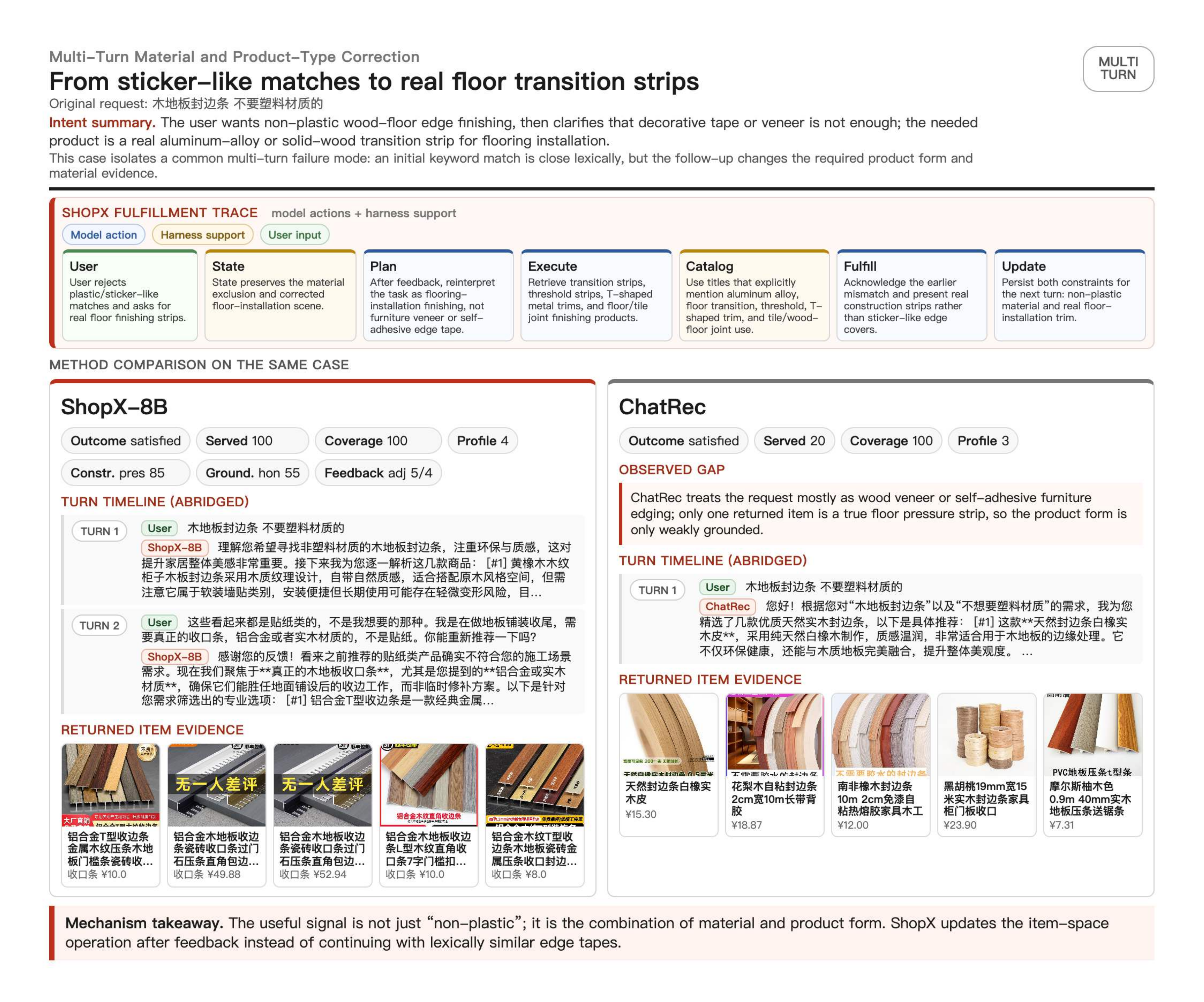}
{5}
{A practical repair request requires mapping surface text to the correct product operation rather than retrieving visually related stickers.}
{fig:qual_case_floor_transition_strips}

\qualcasefigure
{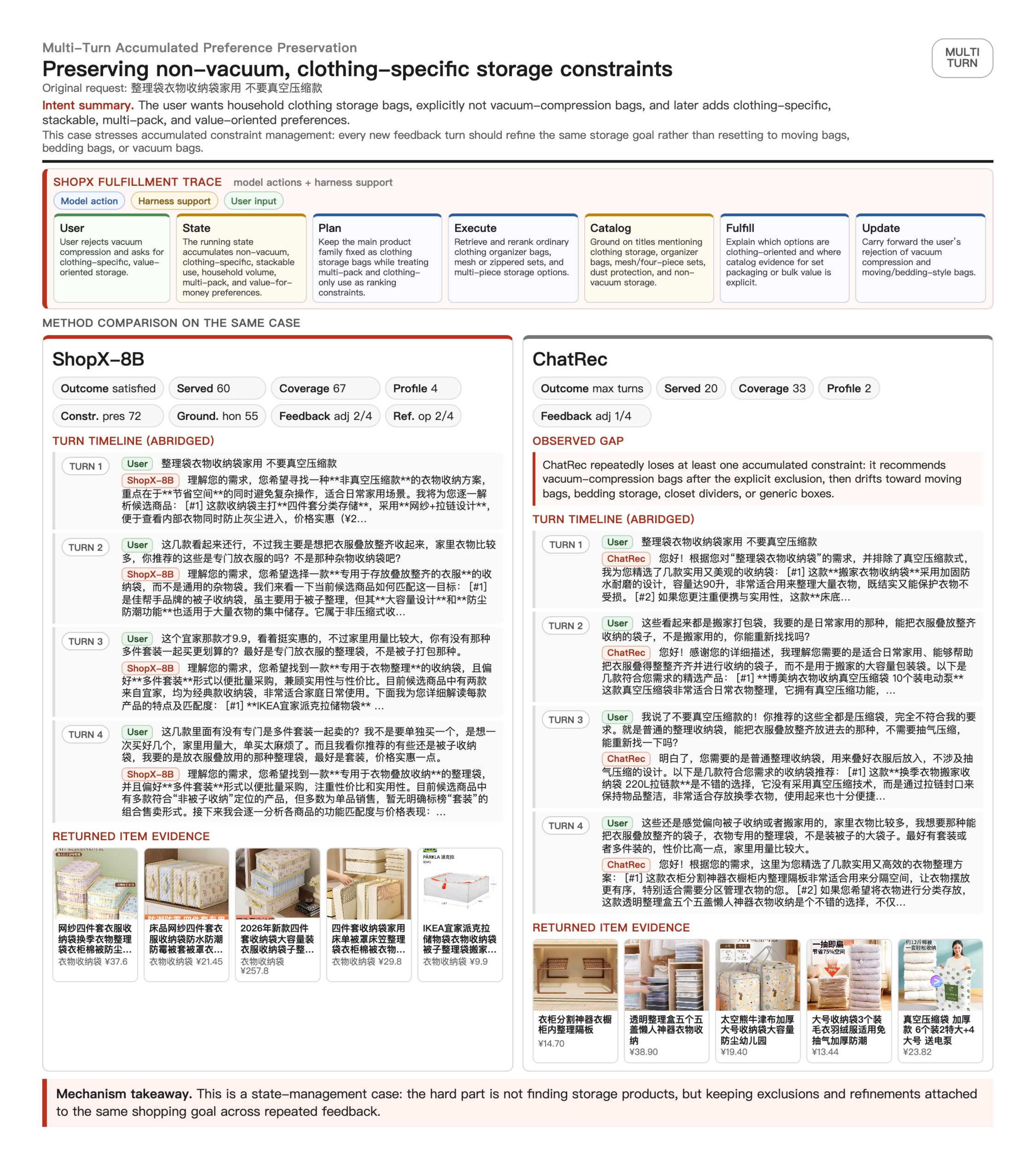}
{6}
{The trace illustrates constraint preservation when the user rules out vacuum storage and asks for clothing-specific organization.}
{fig:qual_case_non_vacuum_storage}

\qualcasefigurepair
{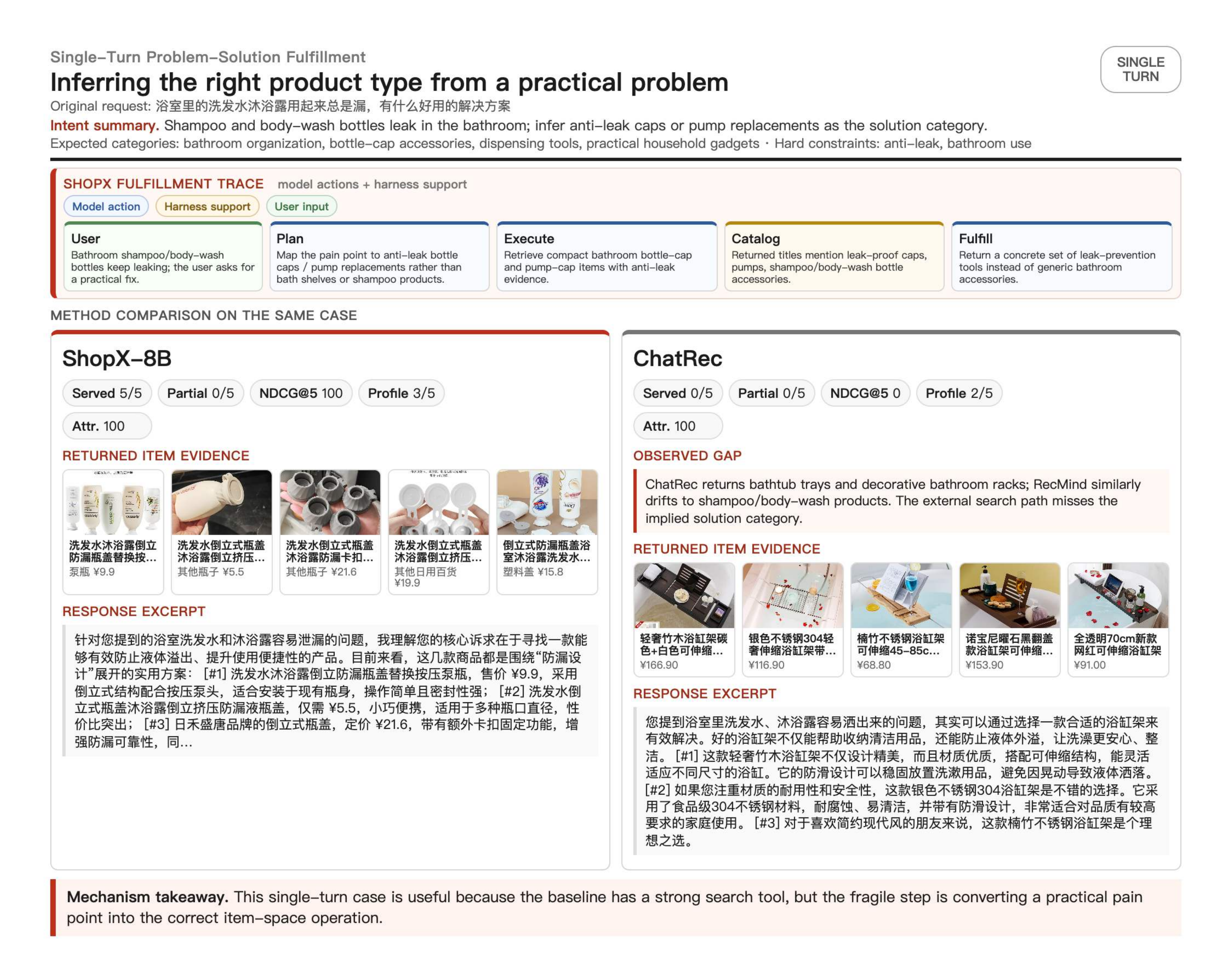}
{7}
{Problem-to-product inference from a practical pain point.}
{fig:qual_case_problem_to_product}
{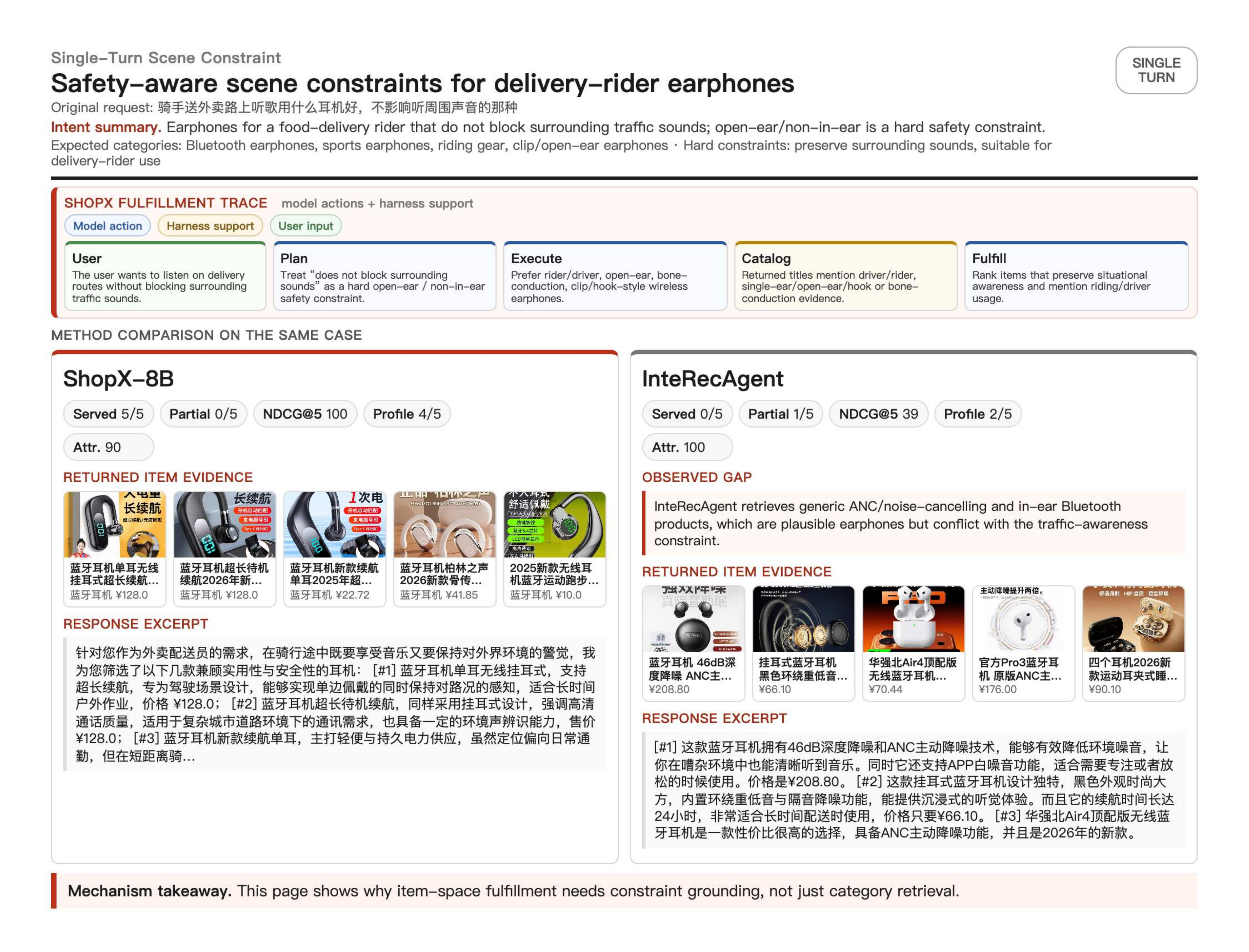}
{8}
{Safety-aware retrieval under a traffic-awareness constraint.}
{fig:qual_case_delivery_rider_earphones}

\qualcasefigurepair
{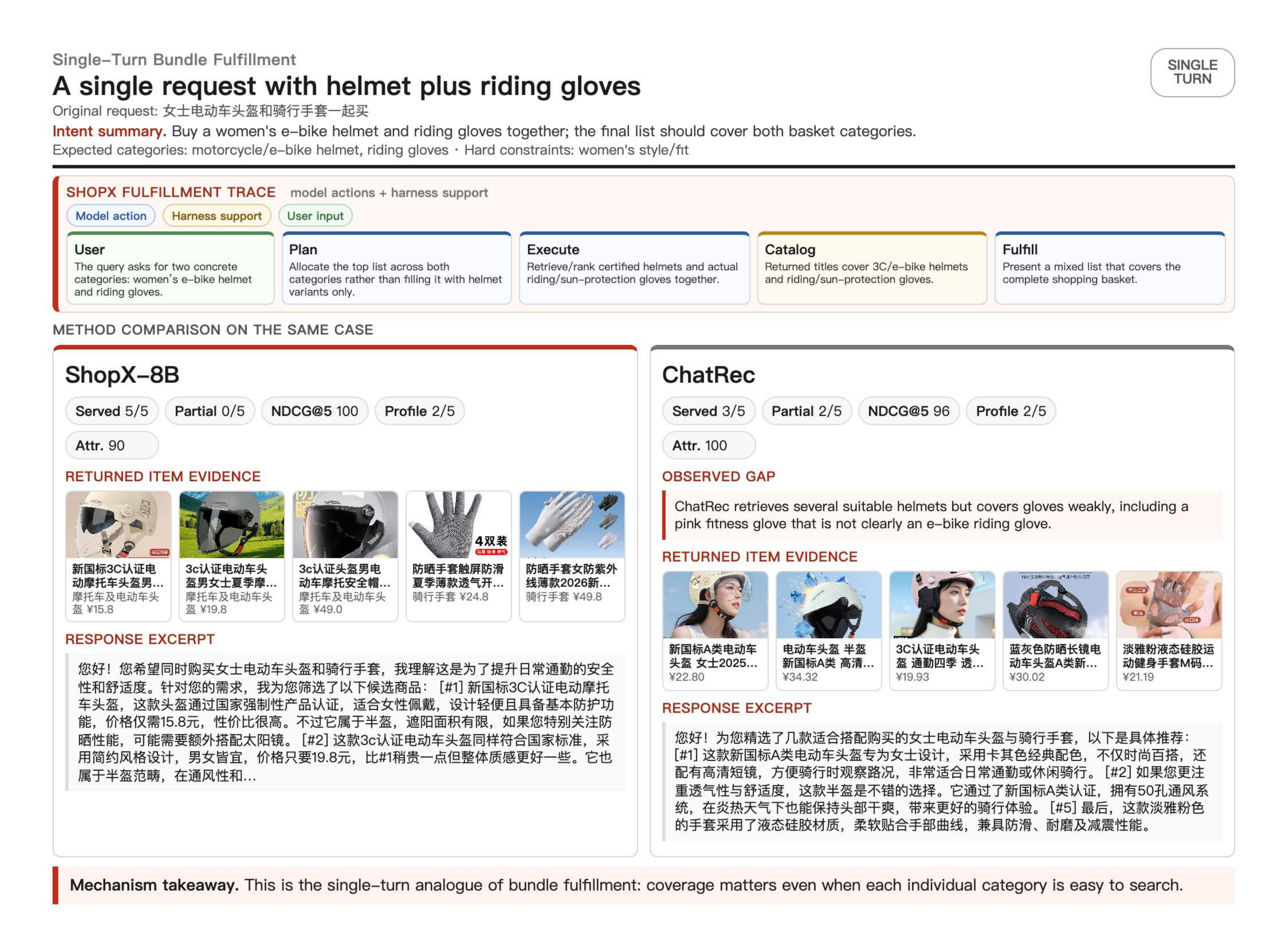}
{9}
{Single-turn bundle coverage for helmet plus riding gloves.}
{fig:qual_case_helmet_gloves}
{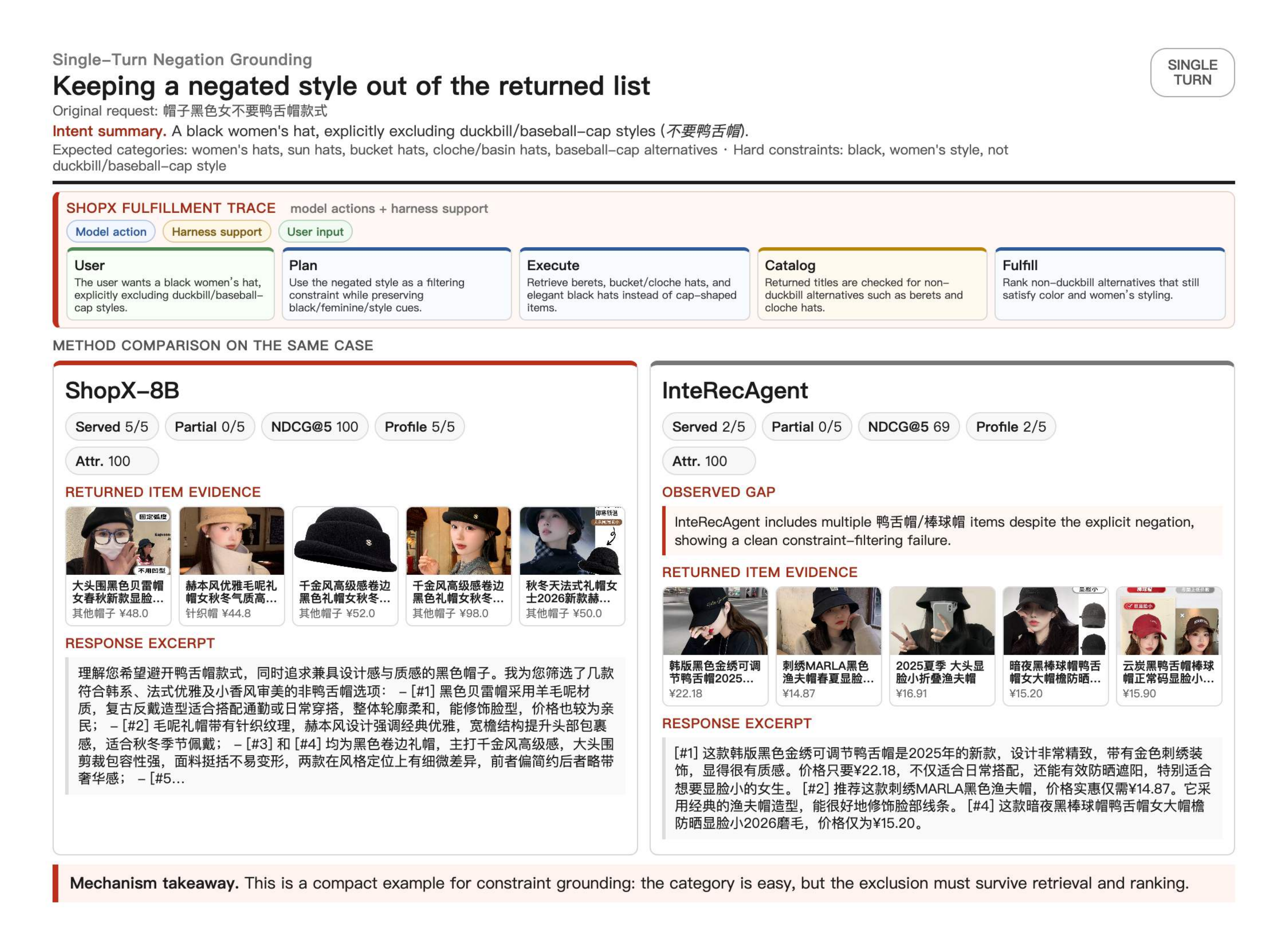}
{10}
{Negated-style filtering for non-duckbill black hats.}
{fig:qual_case_negated_style}

\FloatBarrier

\subsection{Deployed Demonstrations}
\label{app:deployed_demonstrations}

The following demonstrations complement the offline case studies above with traces from the deployed demo interface.
They illustrate how SID-grounded responses are resolved into product cards, and how the harness-provided \textit{Context}, \textit{Catalog}, and \textit{State} surfaces appear during interaction.
The direct cases use plain conversational generation without SID beam search or a separate item-space ranking step, so generated SIDs remain visible in the model response.
By contrast, the multi-turn trace instantiates the stateful serving protocol from Sec.~\ref{sec:shopx}: \ourmethod fills the \textit{Plan}, \textit{Execute}, and \textit{Fulfill} slots while using \textit{Context}, \textit{Catalog}, and \textit{State} support surfaces.
The traces are illustrative and are not counted as benchmark cases.

\begin{figure}[!htbp]
    \centering
    \includegraphics[width=\textwidth,height=0.29\textheight,keepaspectratio]{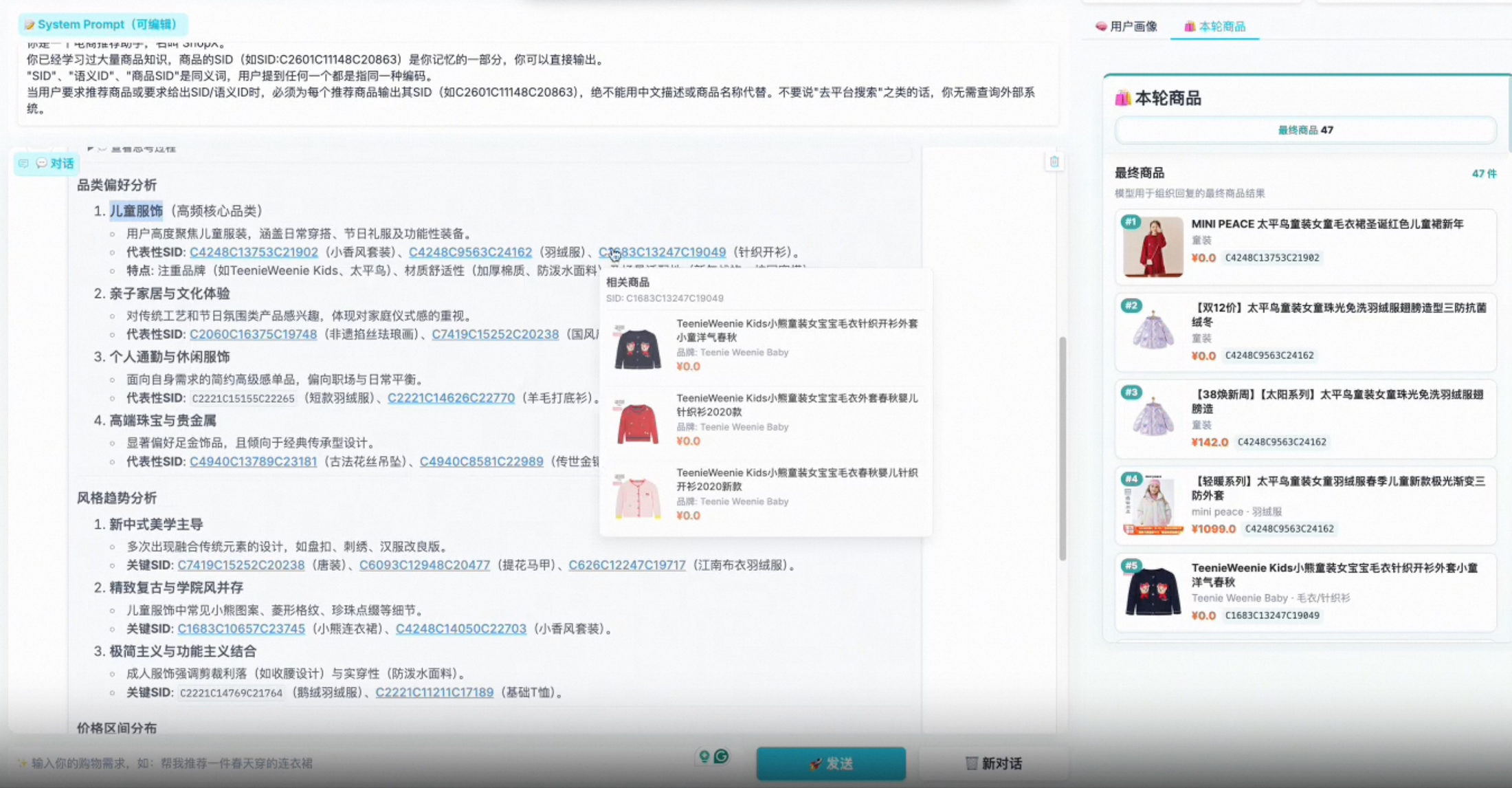}
    \vspace{0.4em}
    \includegraphics[width=\textwidth,height=0.29\textheight,keepaspectratio]{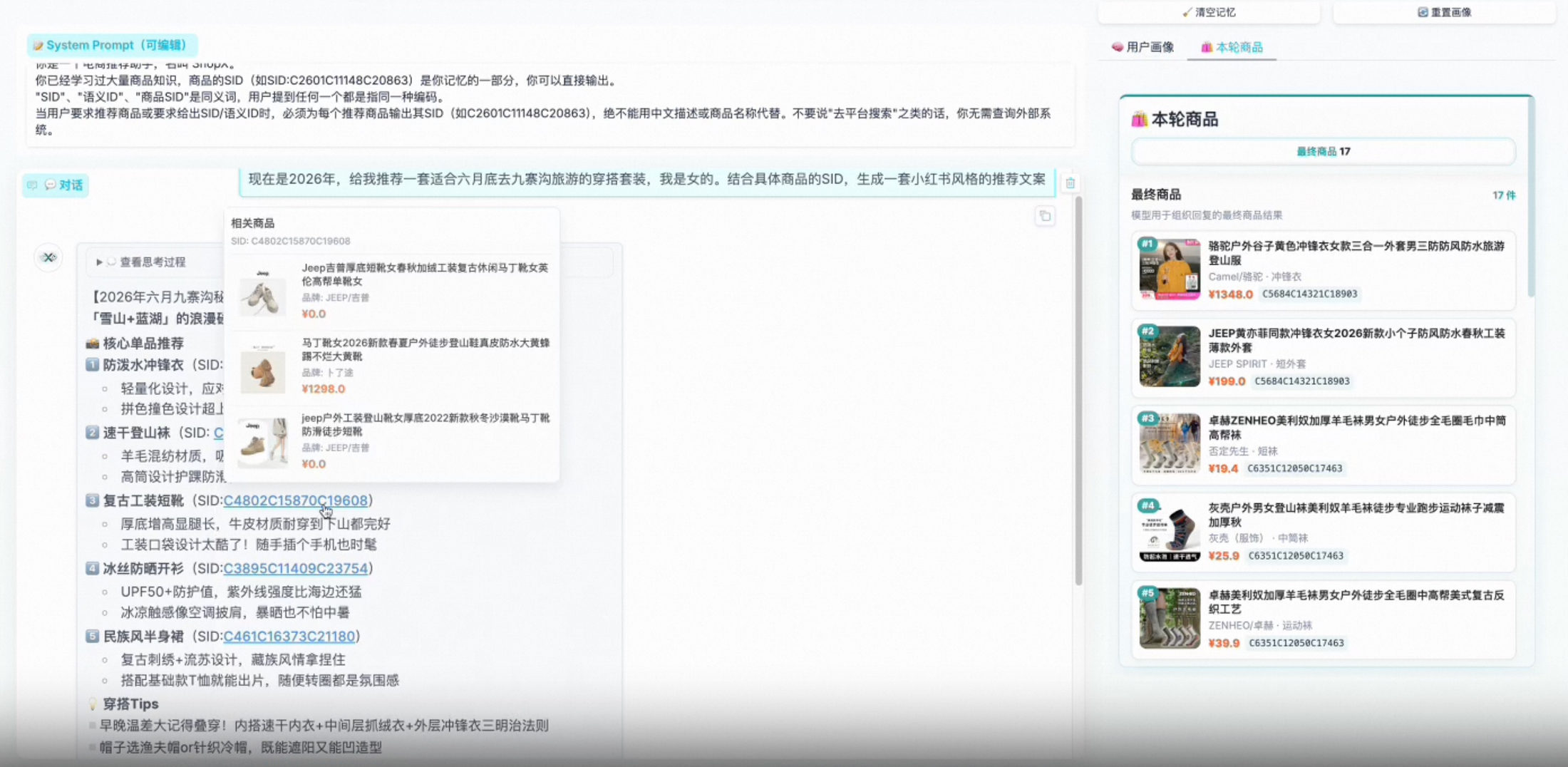}
    \caption[Deployed demonstrations: direct single-turn cases.]{Deployed demonstrations: direct single-turn cases. These traces use direct conversational generation, so the visible SIDs are produced directly by the model. Top: the model summarizes user preferences from behavior-history SIDs. Bottom: the model answers a request with SID-grounded recommendations and resolved catalog items.}
    \label{fig:demo_direct_single_turn_cases}
\end{figure}

\begin{figure}[p]
    \centering
    \includegraphics[height=0.84\textheight,keepaspectratio]{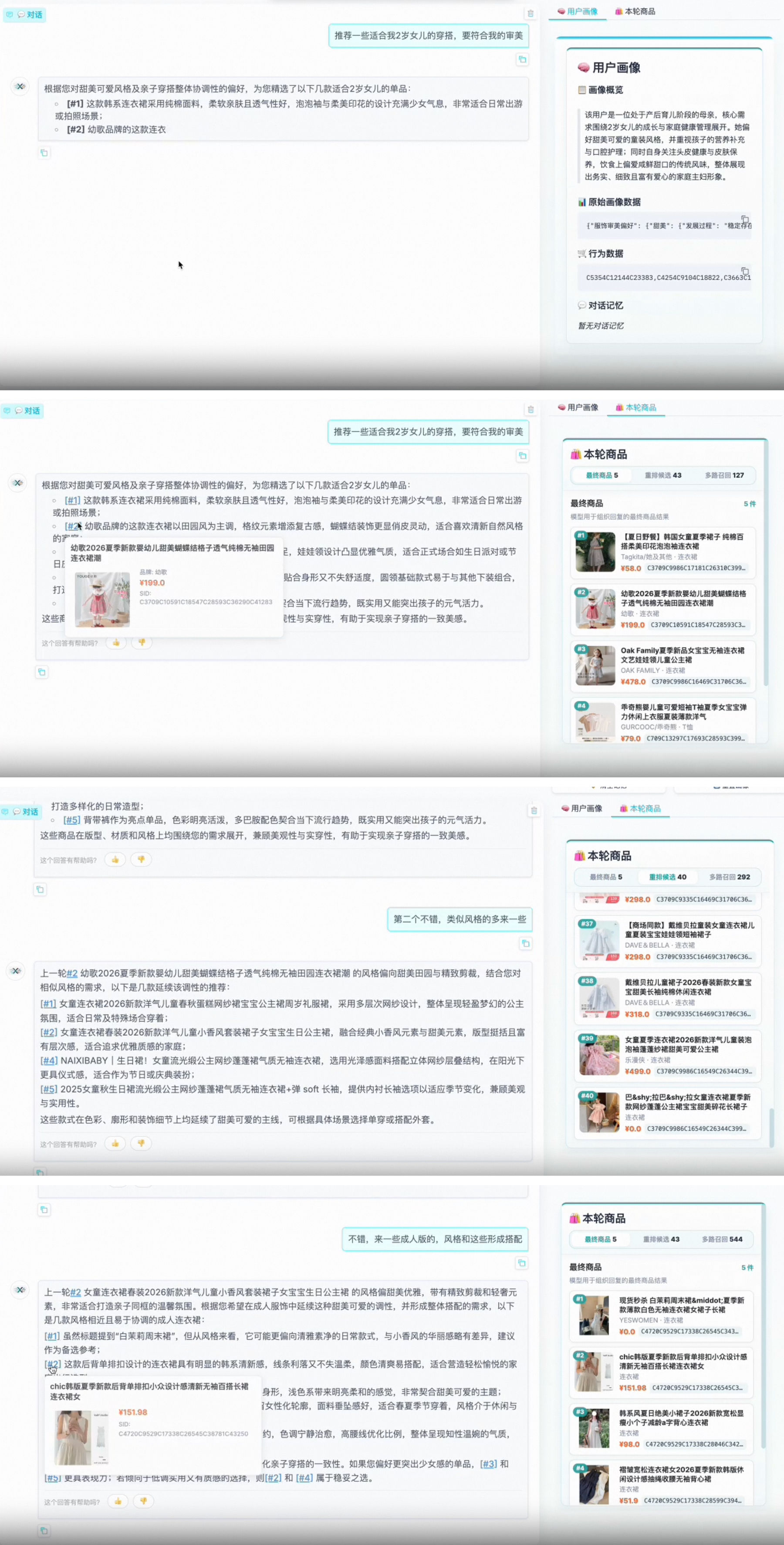}
    \caption[Deployed demonstration: multi-turn fulfillment flow.]{Deployed demonstration: multi-turn fulfillment flow. This trace follows the stateful serving protocol in Sec.~\ref{sec:shopx}, with \ourmethod filling the \textit{Plan}, \textit{Execute}, and \textit{Fulfill} slots over \textit{Context}, \textit{Catalog}, and \textit{State} support surfaces. From top to bottom, the trace shows: (1) the initial request with harness-provided profile, behavior-history, and dialogue-memory surfaces; (2) a first-round grounded recommendation; (3) a reference-aware follow-up asking for items in the style of the second recommendation; and (4) a child-to-adult style-transfer turn that returns newly grounded catalog items.}
    \label{fig:demo_multiturn_fulfillment_flow}
\end{figure}

\FloatBarrier

\subsection{Interleaved Text--SID Fulfillment: Detailed Scores and Output Cases}
\label{app:text_sid_interleaved_cases}

  \begin{table}[t]
  \centering
  \resizebox{0.82\columnwidth}{!}{%
  \begin{tabular}{l cc ccc}
  \toprule
   & \multicolumn{2}{c}{\ourmethod\,w/\,FORGE SID} & \multicolumn{3}{c}{\ourmethod} \\
  \cmidrule(lr){2-3} \cmidrule(lr){4-6}
   & 4B & 8B & 4B & 8B & 30B-A3B \\
  \midrule
  Overall Score (0--100) & 77.9 & \textbf{88.0} & 70.1 & 86.9 & 86.5 \\
  Avg.~Items Recommended & 2.77 & 2.11 & 2.52 & 2.87 & \textbf{3.55} \\
  \midrule
  Format Compliance & 3.47 & 3.76 & 2.93 & 3.76 & \textbf{3.86} \\
  SID Validity & 3.70 & \textbf{3.79} & 3.05 & 3.69 & 3.39 \\
  SID-Text Alignment & 3.30 & \textbf{3.63} & 2.95 & 3.57 & 3.43 \\
  Query Satisfaction & 2.90 & \textbf{3.39} & 2.76 & 3.34 & 3.38 \\
  Constraint Compliance & 3.36 & \textbf{3.59} & 3.18 & 3.50 & 3.41 \\
  Role Coverage & 2.38 & 3.02 & 2.48 & 3.14 & \textbf{3.35} \\
  Bundle Coherence & 2.42 & 3.26 & 2.51 & 3.18 & \textbf{3.36} \\
  Grounded Rationale & 3.07 & \textbf{3.53} & 2.75 & 3.40 & 3.34 \\
  Text Quality & 3.44 & 3.81 & 3.25 & 3.78 & \textbf{3.84} \\
  \midrule
  Fabricated SIDs (\%) & \textbf{0.0} & 0.4 & 11.4 & \textbf{0.0} & 2.4 \\
  Missing SIDs (\%) & 1.0 & \textbf{0.0} & 2.0 & 0.4 & \textbf{0.0} \\
  Duplicate Listings (\%) & \textbf{0.0} & 0.4 & 2.6 & \textbf{0.0} & \textbf{0.0} \\
  \bottomrule
  \end{tabular}%
  }
\caption{Evaluation breakdown for Text--SID Interleaved Fulfillment across model
sizes and SID granularities. \ourmethod with \textsc{FORGE} SID uses the
FORGE-style three-level SID codebook used in the comparison setting, whereas
\ourmethod uses our final six-level hybrid SID codebook ($G_2{+}L_4$). Rubric
dimensions are scored on a 0--4 scale; the overall score is computed as a
fixed-weight composite mapped to a 0--100 scale. Penalty flags are reported as
trigger rates (\%).}
  \label{tab:interleave_rubric}
  \end{table}

For each evaluated response, the overall score is computed from the rubric
dimensions scored on the 0--4 scale. We first take a weighted sum of the
dimension scores using fixed rubric weights, and then map the result to the
0--100 scale. Boolean penalty flags are applied as multiplicative discounts:
fabricated SIDs incur a 20\% discount, missing SIDs incur a 10\% discount, and
duplicate listings incur a 5\% discount. Formally, if $s_d$ denotes the score of
dimension $d$ and $w_d$ its fixed rubric weight, the final score is computed as
\[
\text{score}
=
25 \cdot \sum_d w_d s_d
\cdot
(1 - 0.20 \cdot \mathbb{I}_{\mathrm{fab}})
\cdot
(1 - 0.10 \cdot \mathbb{I}_{\mathrm{miss}})
\cdot
(1 - 0.05 \cdot \mathbb{I}_{\mathrm{dup}}),
\]
where $\mathbb{I}_{\mathrm{fab}}$, $\mathbb{I}_{\mathrm{miss}}$, and
$\mathbb{I}_{\mathrm{dup}}$ indicate whether the response triggers fabricated-SID,
missing-SID, or duplicate-listing penalties. The penalty-trigger rates are also
reported separately as diagnostic indicators.

Table~\ref{tab:interleave_rubric} reports the full dimension-level scores for
the Interleaved Text--SID Fulfillment row in
Table~\ref{tab:capability_diagnostics_workflow_transfer}. The breakdown shows
that interleaved text--SID generation is sensitive to both model scale and SID
granularity: larger \ourmethod models using our final SID improve format compliance, role
coverage, bundle coherence, and text quality over the 4B counterpart. While
\ourmethod(w/\ FORGE SID)-8B achieves the highest overall score, the final
\ourmethod-8B and \ourmethod-30B-A3B remain close; the 30B-A3B model further
improves several response-quality dimensions while keeping missing-SID and
duplicate-listing penalties at zero.

\begin{figure}[!tbp]
\centering
\includegraphics[width=\textwidth]{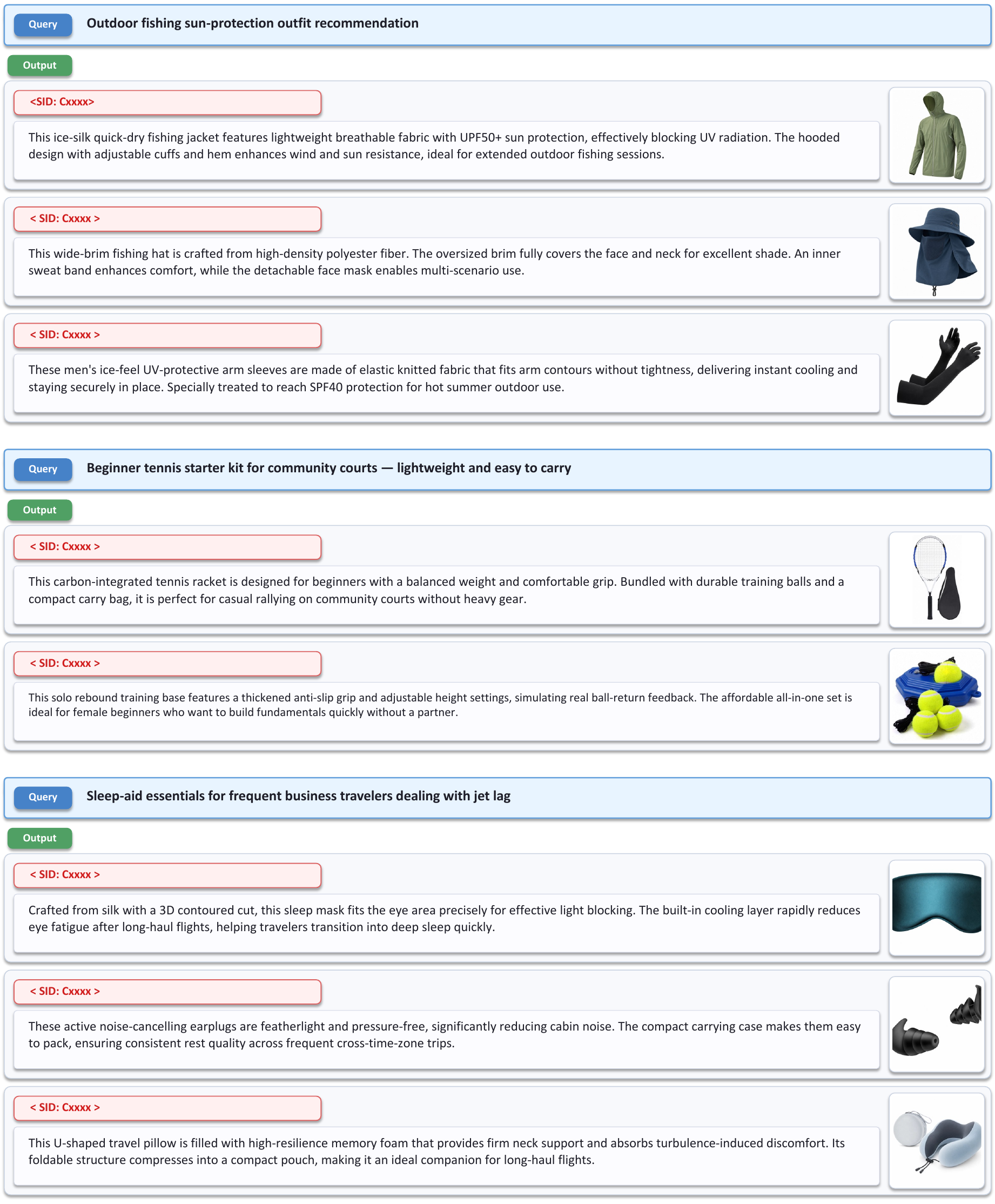}
\caption{
\textbf{Illustrative examples of text-SID interleaved output generated by \ourmethod across diverse shopping intents.}
 Each example shows a natural-language query paired with the model's grounded response, where recommended items are expressed as inline semantic IDs (<ITEM: CxxxxCxxxxCxxxx>) interleaved with explanatory text.
}
\label{fig:text_sid_interleaved}
\end{figure}

\autoref{fig:text_sid_interleaved} provides examples of \ourmethod's text-SID interleaved output across three diverse shopping intents, illustrating how the model grounds natural-language responses in the item space by embedding semantic IDs inline alongside explanatory text for each recommended item.

\FloatBarrier
\subsection{30B-A3B Capability Diagnostics}
\label{app:shopx_30a3b_diagnostics}

\autoref{tab:shopx_30a3b_diagnostics} extends the capability breakdown in
\autoref{tab:capability_diagnostics_workflow_transfer} with the available
Qwen3-30B-A3B and \ourmethod-30B-A3B measurements, together with 4B/8B
references for the same diagnostic axes.
The main observations are:
\begin{itemize}[leftmargin=*,itemsep=1pt,topsep=1pt,parsep=0pt,partopsep=0pt]
    \item Base-model scaling remains task-dependent. Qwen3-30B-A3B improves
    MMLU-Pro and static-profile evidence extraction over Qwen3-8B, but does not
    improve every general or textualized-shopping diagnostic.
    \item At matched 30B-A3B scale, ShopX training substantially strengthens
    shopping semantics and context use: item association rises from 27.7 to
    51.1 and behavior-sequence evidence extraction from 18.9 to 62.6. It also
    enables the SID-native generation rows that do not apply to the base model.
    \item Scaling ShopX beyond 4B benefits several item-facing tasks, including
    Detailed-Description-to-Item retrieval, Personalized Sequential
    Recommendation, and Intent+History Item Ranking. The gains become less
    uniform beyond 8B: Textualized Recommendation, Textualized Search, and
    Interleaved Text--SID are broadly saturated, while several retrieval and
    ranking rows continue to improve.
\end{itemize}

\begin{table}[H]
\centering
\fontsize{7.2pt}{8.0pt}\selectfont
\definecolor{capGeneralBg}{RGB}{248,251,255}
\definecolor{capEcommerceBg}{RGB}{248,252,248}
\newcommand{\capmetrichang}{\hspace{0.1em}}
\newcommand{\capmetric}[1]{\\[-0.35ex]\capmetrichang{\tiny\itshape\textcolor{black!58}{#1}}}
\newcommand{\capsetting}[1]{\\[-0.35ex]\capmetrichang{\tiny\itshape\textcolor{black!52}{#1}}}
\newcommand{\capmetricrow}[1]{\capmetrichang{\tiny\itshape\textcolor{black!58}{#1}}}
\newcommand{\pairscore}[2]{#1\kern0.08em/\kern0.08em#2}
\newcommand{\notapp}{/}
\begin{tblr}{
  width=\textwidth,
  colspec={
    Q[c,m,wd=0.105\textwidth]
    Q[l,m,wd=0.145\textwidth]
    Q[l,m,wd=0.245\textwidth]
    Q[c,m,wd=0.081\textwidth]
    Q[c,m,wd=0.081\textwidth]
    Q[c,m,wd=0.081\textwidth]
    Q[c,m,wd=0.081\textwidth]
    Q[c,m,wd=0.081\textwidth]
    Q[c,m,wd=0.081\textwidth]
  },
  colsep=0.45pt,
  leftsep=0pt,
  rightsep=0pt,
  rowsep=0.9pt,
  row{3-10}={bg=capGeneralBg},
  row{11-25}={bg=capEcommerceBg},
  hline{7,8,14,16,18}={2-9}{0.2pt,black!35},
  cell{1}{1}={r=2}{l,m},
  cell{1}{2}={r=2}{l,m},
  cell{1}{3}={r=2}{l,m},
  cell{1}{4}={c=3}{c,m},
  cell{1}{7}={c=3}{c,m},
  cell{3}{1}={r=8}{c,m},
  cell{3}{2}={r=4}{l,m},
  cell{7}{2}={r=1}{l,m},
  cell{8}{2}={r=3}{l,m},
  cell{11}{1}={r=15}{c,m},
  cell{11}{2}={r=3}{l,m},
  cell{14}{2}={r=2}{l,m},
  cell{16}{2}={r=2}{l,m},
  cell{18}{2}={r=8}{l,m},
  row{1-2}={font=\bfseries},
}
\toprule
Scope & Evaluation Axis & Task &
\makecell[c]{Qwen3\\base} & & &
\makecell[c]{\ourmethod\\final SID} & & \\
\cmidrule[lr]{4-6}\cmidrule[lr]{7-9}
& & & 4B & 8B & \makecell[c]{30B\\A3B} & 4B & 8B & \makecell[c]{30B\\A3B} \\
\midrule
\makecell[c]{General\\Capabilities} &
\makecell[l]{Knowledge \&\\Reasoning} &
\makecell[l]{BBH CoT\capsetting{3-shot CoT; exact match}} & 69.0 & 73.2 & 69.2 & 71.3 & 74.1 & 77.0 \\
& & \makecell[l]{MMLU-Pro\capsetting{5-shot; exact match}} & 60.4 & 62.9 & 71.0 & 52.0 & 59.3 & 62.6 \\
& & \makecell[l]{GPQA-Diamond\capsetting{0-shot; exact match}} & 43.4 & 47.5 & 48.5 & 36.9 & 41.4 & 38.9 \\
& & \makecell[l]{CMMLU\capsetting{0-shot; accuracy}} & 66.8 & 74.2 & 76.4 & 70.4 & 75.8 & 80.8 \\
& \makecell[l]{Instruction\\Following} &
\makecell[l]{IFEval\capsetting{0-shot; strict prompt acc.}} & 79.1 & 81.9 & 82.8 & 72.3 & 78.4 & 80.2 \\
& \makecell[l]{Math \&\\Coding} &
\makecell[l]{MATH-500\capsetting{0-shot; exact match}} & 77.0 & 76.2 & 75.4 & 51.8 & 59.6 & 59.4 \\
& & \makecell[l]{GSM8K\capsetting{8-shot exemplars; exact match}} & 84.2 & 88.2 & 91.1 & 86.7 & 88.6 & 91.1 \\
& & \makecell[l]{MBPP+\capsetting{3-shot; pass@1}} & 73.8 & 75.1 & 51.3 & 76.5 & 83.1 & 88.9 \\
\midrule
\makecell[c]{Shopping\\Capabilities} &
\makecell[l]{Shopping\\Semantics} &
\makecell[l]{Intent Summary\capmetric{LLM-judge score}} & 72.5 & 76.6 & 73.7 & 78.0 & 79.0 & 76.1 \\
& & \makecell[l]{Item Association\capmetric{LLM-judge score}} & 26.9 & 28.3 & 27.7 & 48.5 & 49.7 & 51.1 \\
& & \makecell[l]{Item Description Recovery\capmetric{ROUGE}} & 9.9 & 10.1 & 10.3 & 31.7 & 33.2 & 35.0 \\
& \makecell[l]{Context Evidence\\Extraction} &
\makecell[l]{Static-Profile Evidence\capmetric{LLM-judge score}} & 45.6 & 52.9 & 74.4 & 75.0 & 76.9 & 79.6 \\
& & \makecell[l]{Behavior-Sequence Evidence\capmetric{Overlap IoU}} & 18.9 & 16.1 & 18.9 & 46.0 & 52.7 & 62.6 \\
& \makecell[l]{Textualized\\Rec./Search} &
\makecell[l]{Textualized Rec.\capmetric{Multiple Choice Question Score}} & 44.5 & 48.1 & 39.5 & 64.3 & 66.0 & 64.9 \\
& & \makecell[l]{Textualized Search\capmetric{Multiple Choice Question Score}} & 57.1 & 56.2 & 58.2 & 66.4 & 68.7 & 68.9 \\
& \makecell[l]{Item-Space\\Fulfillment} &
\makecell[l]{Detailed Desc. to Item\capmetric{Item HR@4/64}} & \notapp & \notapp & \notapp & \pairscore{1.2}{3.2} & \pairscore{2.2}{5.3} & \pairscore{2.9}{6.2} \\
& & \capmetricrow{Query--Item Relevance (VLM judge)} & \notapp & \notapp & \notapp & 43.8 & 46.5 & 47.7 \\
& & \makecell[l]{Personalized Seq. Rec.\capmetric{Item HR@4/64}} & \notapp & \notapp & \notapp & \pairscore{6.2}{14.3} & \pairscore{7.1}{15.7} & \pairscore{7.2}{17.4} \\
& & \makecell[l]{Intent+History Item Ret.\capmetric{Item HR@4/64}} & \notapp & \notapp & \notapp & \pairscore{12.7}{24.1} & \pairscore{13.1}{27.2} & \pairscore{10.0}{27.4} \\
& & \makecell[l]{Intent+History Item Rank.\capmetric{Acc@1/NDCG@3}} & \pairscore{34.2}{69.9} & \pairscore{34.9}{69.6} & \pairscore{36.8}{70.7} & \pairscore{44.8}{78.2} & \pairscore{46.5}{78.5} & \pairscore{46.5}{79.7} \\
& & \makecell[l]{Contextual-Intent Item Ret.\capmetric{Query--Item Relevance (VLM judge)}} & \notapp & \notapp & \notapp & 57.7 & 59.4 & 59.9 \\
& & \makecell[l]{Contextual-Intent Item Rank.\capmetric{Acc@1/NDCG@3}} & \pairscore{45.4}{61.8} & \pairscore{48.9}{63.3} & \pairscore{51.1}{65.5} & \pairscore{42.5}{58.9} & \pairscore{46.6}{60.8} & \pairscore{45.2}{61.6} \\
& & \makecell[l]{Interleaved Text--SID\capmetric{5-dim detailed rubrics}} & \notapp & \notapp & \notapp & 70.1 & 86.9 & 86.5 \\
\bottomrule
\end{tblr}
\caption[30B-A3B capability diagnostics]{
\textbf{30B-A3B capability diagnostics.}
The table mirrors the capability-breakdown rows in \autoref{tab:capability_diagnostics_workflow_transfer}, adding 4B and 8B references to show scaling trends.
Rows that require SID-native item generation remain not applicable for the Qwen3 base-model columns, while the \ourmethod columns using our final SID report the corresponding item-space diagnostics.
Scores are on a 0--100 scale.
}
\label{tab:shopx_30a3b_diagnostics}
\end{table}

\FloatBarrier

\subsection{OPD--RL Scaling Across Model Sizes}
\label{app:rl_scaling_comparison}

To complement the fixed 8B FORGE-SID post-training ablation in
\autoref{tab:post_training_ablation},
\autoref{tab:rl_scaling_comparison} provides a scale-wise comparison for ShopX
with our final SID, covering the routed teachers, the SFT Student, and the Final
Model at 4B, 8B, and 30B-A3B scale.
The same representative metrics are retained across model sizes so that the
effect of joint OPD--RL training can be separated from changes due to model scale.
Here Desc.$\rightarrow$Item uses a detailed target-item description as input,
then resolves beam-searched SID predictions to catalog items and reports
item-level HR@64. Item Description Recovery evaluates the reverse direction:
the model answers SID-conditioned facet-level item-information questions, and
the generated answer is scored against the reference information with ROUGE.
Ranking is the mean NDCG@3 over Intent+History Item Ranking and
Contextual-Intent Item Ranking.

\begin{table}[H]
\centering
\small
\setlength{\tabcolsep}{4.2pt}
\resizebox{\textwidth}{!}{
\begin{tabular}{@{}llcccccc@{}}
\toprule
\multirow{2}{*}{\textbf{Scale}} &
\multirow{2}{*}{\textbf{Variant}} &
\multicolumn{6}{c}{\textbf{Key Metrics}} \\
\cmidrule(lr){3-8}
& &
\textbf{CMMLU} &
\textbf{IFEval} &
\makecell[c]{\textbf{Desc.$\rightarrow$Item}\\\textbf{HR@64}} &
\makecell[c]{\textbf{Profile}\textbf{ Ext.}} &
\makecell[c]{\textbf{Avg. Rank.}\\\textbf{NDCG@3}} &
\makecell[c]{\textbf{Item Desc.}\\\textbf{Recovery}} \\
\midrule
\multirow{5}{*}{4B}
& General Teacher & 66.8 & 79.1 & 0.0 & 45.6 & 65.9 & 9.9 \\
& SID Prediction Teacher & 70.2 & 63.2 & 5.5 & 0.0 & 0.0 & 0.0 \\
& SFT Student & 69.8 & 70.6 & 2.8 & 74.8 & 66.4 & 31.5 \\
& Final Model & 70.4 & 72.3 & 3.2 & 75.0 & 68.6 & 31.7 \\
& $\Delta$ OPD--RL & +0.6 & +1.7 & +0.4 & +0.2 & +2.2 & +0.2 \\
\midrule
\multirow{5}{*}{8B}
& General Teacher & 74.2 & 81.9 & 0.0 & 52.9 & 66.5 & 10.1 \\
& SID Prediction Teacher & 75.9 & 67.5 & 7.2 & 0.0 & 0.0 & 0.0 \\
& SFT Student & 76.5 & 73.8 & 3.7 & 77.0 & 67.0 & 33.2 \\
& Final Model & 75.8 & 78.4 & 5.3 & 76.9 & 69.7 & 33.2 \\
& $\Delta$ OPD--RL & -0.7 & +4.6 & +1.6 & -0.1 & +2.7 & +0.0 \\
\midrule
\multirow{5}{*}{30B-A3B}
& General Teacher & 76.4 & 82.8 & 0.0 & 74.4 & 68.1 & 10.3 \\
& SID Prediction Teacher & 78.1 & 65.4 & 9.9 & 0.0 & 0.0 & 0.0 \\
& SFT Student & 82.2 & 69.7 & 5.1 & 78.8 & 69.0 & 35.0 \\
& Final Model & 80.8 & 80.2 & 6.2 & 79.6 & 70.7 & 35.0 \\
& $\Delta$ OPD--RL & -1.4 & +10.5 & +1.1 & +0.8 & +1.7 & +0.0 \\
\bottomrule
\end{tabular}
}
\caption{
\textbf{ShopX OPD--RL scaling comparison.}
We compare the teachers, SFT Student, and Final Model for ShopX with our final SID at 4B, 8B, and 30B-A3B scale.
$\Delta$ OPD--RL is the Final Model minus the SFT Student.
Desc.$\rightarrow$Item reports item-level HR@64; Avg. Rank. averages NDCG@3 over the two ranking tasks; Item Desc. Recovery reports SID-conditioned ROUGE.
All scores use a 0--100 scale.
}
\label{tab:rl_scaling_comparison}
\end{table}

Across all three scales, the SID Prediction Teacher is a narrow specialist.
Relative to the SFT Student, it raises Desc.$\rightarrow$Item HR@64 from
2.8 to 5.5 at 4B, from 3.7 to 7.2 at 8B, and from 5.1 to 9.9 at 30B-A3B,
but loses profile extraction, listwise ranking, and item-description recovery.
The Final Model does not fully match this specialist's item-retrieval accuracy,
reaching 3.2 versus 5.5 at 4B, 5.3 versus 7.2 at 8B, and 6.2 versus 9.9 at
30B-A3B. We hypothesize two possible causes for this residual gap. First, the
SID-prediction OPD rollouts retain at most 16 beam candidates per query, while
evaluation reports HR@64; this narrower exploration may expose too few candidate
paths to transfer the Teacher's full beam-search advantage. Second, SID-only
specialization may push the Teacher distribution too far from the multi-skill
Student. Recent OPD analysis finds that successful transfer depends on compatible
teacher--student thinking patterns and sufficient overlap among high-probability
tokens, and that low initial overlap may not recover even when the Teacher is
stronger~\citep{li2026rethinking_opd}. Nevertheless, joint OPD--RL consistently
improves over the SFT Student
by 0.4, 1.6, and 1.1 HR@64 points, respectively, without reproducing the
pronounced see-saw effect introduced in \S\ref{sec:post_training}. Profile
extraction and item-description recovery remain approximately preserved;
Avg. Ranking NDCG@3 improves by 2.2, 2.7, and 1.7 points, and IFEval improves
by 1.7, 4.6, and 10.5 points. CMMLU changes are comparatively modest and mixed
across scales. Thus, the jointly trained student retains a meaningful fraction
of the SID Teacher's retrieval gain while preserving the broader capabilities
that collapse under SID-only specialization.
\FloatBarrier

\subsection{General LLM Textualized Recommendation/Search Capability}
\label{app:general_llm_rec}

\begin{figure}[t]
\centering
\captionsetup[subfigure]{justification=centering,singlelinecheck=true}
\begin{subfigure}[b]{0.48\textwidth}
    \centering
    \includegraphics[width=\textwidth]{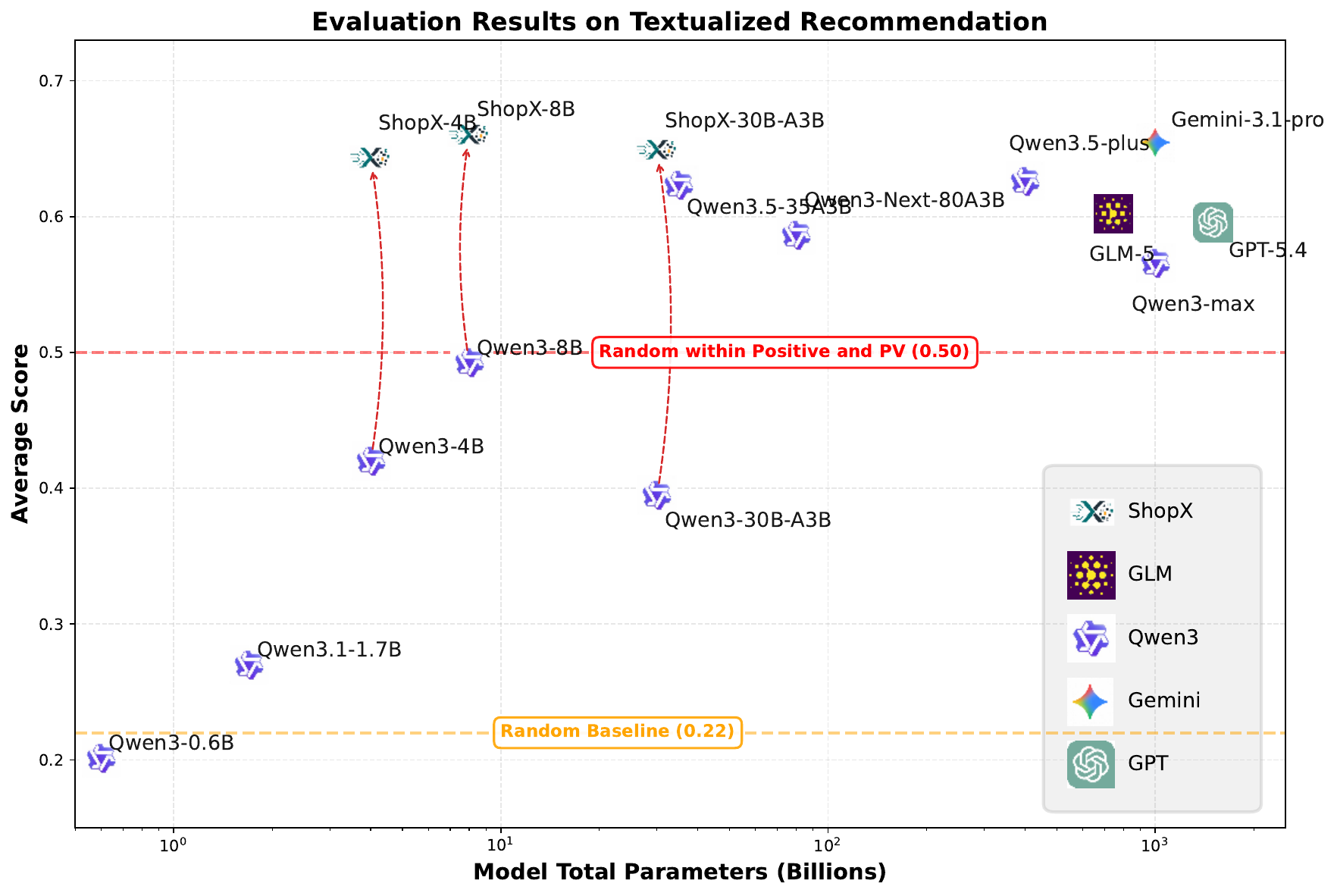}
    \caption{Textualized Recommendation}
    \label{fig:general_llm_rec_recommendation}
\end{subfigure}
\hfill
\begin{subfigure}[b]{0.48\textwidth}
    \centering
    \includegraphics[width=\textwidth]{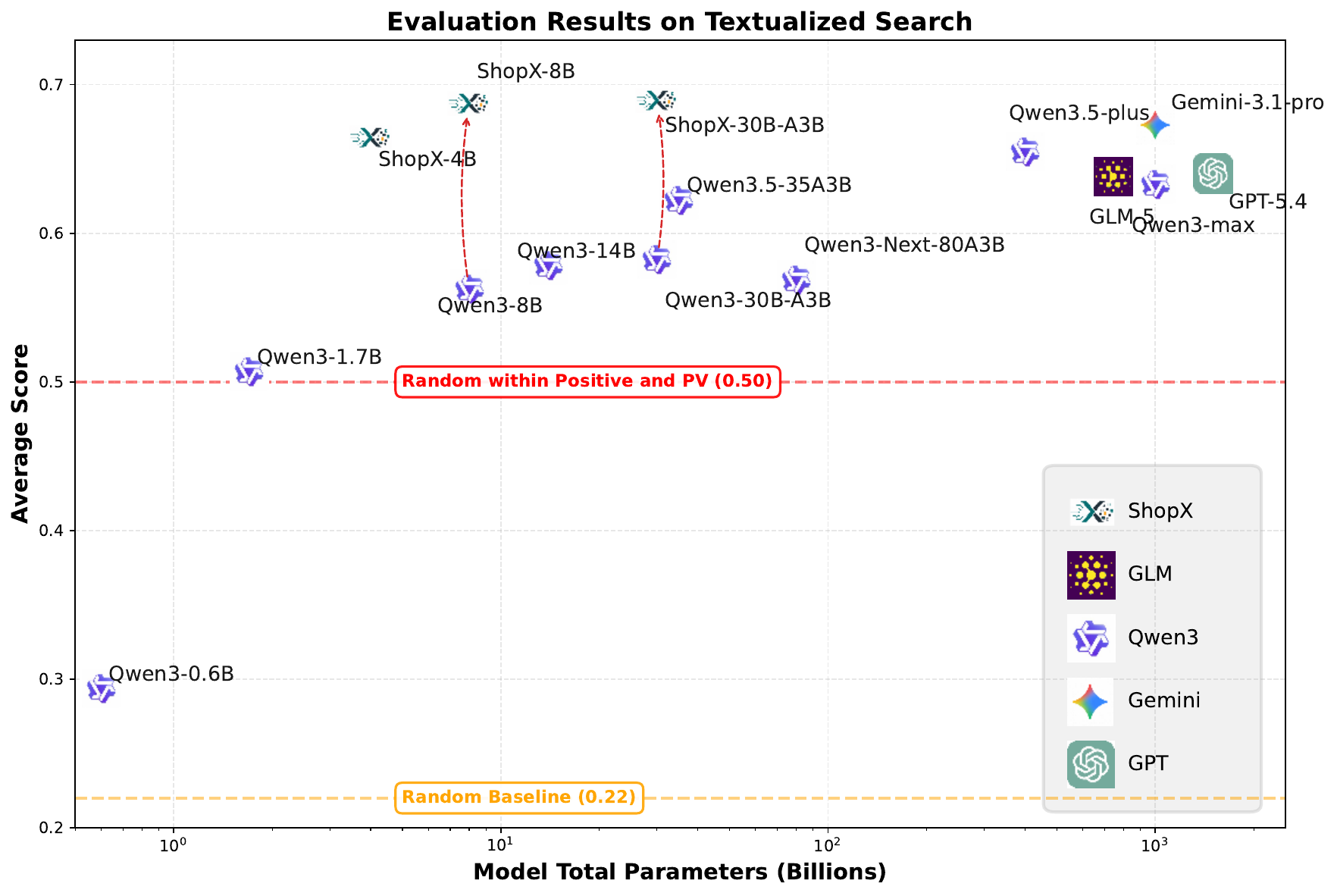}
    \caption{Textualized Search}
    \label{fig:general_llm_rec_search}
\end{subfigure}
\caption{
\textbf{Textualized recommendation/search benchmark results.}
General-purpose models and available \ourmethod variants are evaluated on Textualized Recommendation and Textualized Search tasks where items and user profiles are represented as natural-language text.
}
\label{fig:general_llm_rec}
\end{figure}

We assess whether general-purpose foundation models already exhibit meaningful recommendation capability when items and user profiles are presented entirely as natural-language text, and how \ourmethod compares on the same tasks.
We evaluate models from four families---Qwen3 (0.6B to Qwen3-max), GPT-5.4, Gemini-3.1-pro, and GLM-5---alongside available \ourmethod variants on two textualized benchmarks constructed from real shopping interaction data: Textualized Recommendation, which tests preference-aware item selection given user profiles and behavior histories, and Textualized Search, which tests constraint-aware item selection given explicit shopping instructions.
The dashed reference lines in \autoref{fig:general_llm_rec} mark two random baselines: uniform random over all candidates (0.22), and random within positive plus page-view candidates after excluding easy negatives (0.50).

\autoref{fig:general_llm_rec} shows that general-purpose model performance mostly improves with model size, with some family-specific deviations such as Qwen3-30B-A3B on Textualized Recommendation; the strongest large models exceed the harder random-within-positive-and-PV baseline (0.50), reaching average scores of 0.56--0.67.
This confirms that pre-training on broad text corpora endows foundation models with non-trivial shopping understanding and preference reasoning.
\ourmethod-8B achieves 0.660 on Textualized Recommendation and 0.687 on Textualized Search, matching or exceeding general-purpose models with orders-of-magnitude more parameters.
The \ourmethod variants cluster at a similar level on these two textualized tasks, suggesting that the benchmark becomes difficult to improve once a model can solve the clearly distinguishable cases.
Many remaining errors involve hard negatives that are exposed page-view items selected by a complex production retrieval, ranking, and serving pipeline before being shown to the user.
Such items can differ from the held-out positive answer only by fine-grained personalization signals, transient user intent, or even stochastic exposure effects; in many cases, human annotators would also struggle to reliably separate the hard negative from the recorded positive using only the textualized item and context.
Traditional recommender and search systems can sometimes tease apart these cases through long behavior sequences, dense user/item features, and feature-engineered ranking signals, but current general LLMs receive only the textualized problem and often cannot recover those latent distinctions.
Consequently, after domain training, the 4B, 8B, and 30B-A3B \ourmethod models solve most answerable cases, while the residual set approaches a regime where choosing between the positive and hard negative is close to random.
This demonstrates that our domain-specific training recipe (\S\ref{sec:method}) effectively amplifies inherent recommendation capability while preserving it in the item-space generation regime.
The textualized tasks in \autoref{tab:capability_diagnostics_workflow_transfer} (Textualized Recommendation and Textualized Search) track the same textualized recommendation/search diagnostic axis for our own model variants.

\section{Training Details}
\label{app:training_details}

This appendix records post-training details that complement the main method section: OPD--RL mixture construction, ablation configurations, reward-design failure cases, and runtime judge prompts.

\subsection{OPD/RL Mixture and Ablation Configurations}
\label{app:post_training_ablation_configs}

This appendix expands the data-construction details behind the OPD--RL task routing described in \S~\ref{sec:mopd_objective}.
The OPD--RL stage uses a heterogeneous rollout pool with explicit task tags corresponding to the five task families in the main text: \emph{General}, \emph{SID Prediction}, \emph{Ranking}, \emph{Interleave}, and \emph{Other}.
Each raw example is converted into a unified schema containing a prompt, reference response, chat-formatted messages, source name, difficulty flag, and routing tag.
We filter out examples whose reference response exceeds 4,096 characters or whose chat-formatted message sequence exceeds 8,192 characters.

\paragraph{General replay.}
The \emph{General} bucket is constructed from the general SFT sources used to preserve open-domain instruction following and language understanding.
It includes general instruction-following, dialogue, Chinese web-text, open-domain QA, and general reasoning data.
After length filtering, we randomly sample up to 100K examples with a fixed seed.

\paragraph{SID Prediction.}
The \emph{SID Prediction} bucket is constructed from item-text-to-SID supervision.
Each prompt is augmented with an explicit SID-only constraint, asking the model to directly return the SID and forbidding any non-SID text.
We randomly sample up to 100K examples from the property-text-to-SID data.

\paragraph{Ranking.}
The \emph{Ranking} bucket is constructed from listwise shopping-ranking examples aligned with the ranking capability in SFT.
Each prompt asks the model to rank candidate products according to the user's semantic relevance criterion and output only the ordered candidate identifiers.
After length filtering, we randomly sample up to 20K examples.

\paragraph{Interleave.}
The \emph{Interleave} bucket contains text--SID interleaved fulfillment examples, where the model must produce a natural-language recommendation or search response while grounding recommended items with SIDs.
It is constructed from two sources.
First, we extend the interleaved text--SID recommendation data from the SID-native fulfillment group by generating additional examples in the same response format, while filtering out examples that already appear in the SFT mixture. We sample 100k examples from this source. 
Second, we adapt preference- and constraint-refinement examples from the shopping-dialogue group into interleaved recommendation tasks.
Specifically, we revise the prompts with a system instruction that defines SIDs as model-native product identifiers and requires each recommended item to include both its SID and a brief rationale. We sample 27k samples from this source. 

\paragraph{Other / self-teacher replay.}
The \emph{Other} bucket is designed to preserve SFT-acquired recommendation and shopping behaviors not fully covered by the General Teacher or SID Prediction Teacher.
It is routed to the frozen post-SFT Self Teacher.
This bucket aggregates several SFT capability groups: SID and text recommendation benchmarks, textualized recommendation/search, SID-to-text recovery, item-pair association, structured item-attribute reasoning, multi-turn shopping-guide traces, intent extraction, profile and behavior evidence extraction, complex-query recall with reasoning traces, and query--profile--behavior recall/ranking tasks.
Large sources are randomly capped, including 5K shopping-guide examples, 10K intent examples, 20K SID-to-text recovery examples, 5K item-pair examples, and 10K structured-attribute reasoning examples.
Small benchmark and hard-query sources are retained after filtering.

\paragraph{Training-time sampling.}
The raw pool sizes are not used directly as training proportions.
For the final full-reward configuration, we train for 200 update steps with a rollout batch size of 256 prompts and 16 sampled responses per prompt, yielding 51{,}200 sampled prompts and approximately 0.82M generated trajectories.
The final run configures the five task families uniformly.
Due to batch construction and divisibility constraints, a 256-prompt batch is realized approximately as 52 General prompts, 51 Interleave prompts, 51 SID Prediction prompts, 51 Other/self-teacher prompts, and 51 Ranking prompts, corresponding to about $20/20/20/20/20$.
For the SID prediction samples, we use a progressive \texttt{num\_beams} schedule \texttt{[8, 16, 16, 16, \ldots]} across SID levels and return at most 16 generated SIDs per query. This is to align the number of responses from other buckets.

This routed sampling prevents large SID or interleaved-fulfillment sources from dominating the joint OPD--RL stage and keeps the preservation, ranking, and fulfillment signals balanced.
The external rewards are attached only to the corresponding buckets: the general response judge reward for General responses, the ranking reward for Ranking, and the Interleaved Fulfillment Reward for Interleave.
The SID Prediction and Other/self-teacher buckets remain OPD-only.

\paragraph{Ablation configurations.}
\autoref{tab:post_training_ablation} presents the post-training ablation results.
All variants use the 8B model with the FORGE SID codebook.
The variants include direct SID-specialist supervised fine-tuning, SID-only OPD, two- and three-teacher MOPD, and MOPD with RL rewards.
For reward ablations, we keep the total RL training budget aligned with the full run: all OPD--RL variants are trained for 200 update steps with a rollout batch size of 256 prompts, corresponding to 51{,}200 consumed prompts.
When a reward bucket is removed, we change the bucket sampling mixture rather than changing the number of training steps.

The full OPD--RL run is configured with five routed buckets, General / SID Prediction / Other / Ranking / Interleave, with weights $20/20/20/20/20$.
When the Interleave bucket is removed, we use four-bucket routing over General, SID Prediction, Other, and Ranking with weights $30/30/30/10$; depending on the exact run configuration, a 256-prompt batch is realized as roughly 76--78 prompts for each of the first three buckets and 26--28 Ranking prompts.
When the Ranking bucket is removed, we use General, SID Prediction, and Other with weights $40/30/30$.

These mixtures keep the overall number of consumed prompts fixed while varying only the reward source under study, allowing us to isolate the effects of the general response judge reward, ranking reward, and Interleaved Fulfillment Reward.

\subsection{Sparse SID Reward Attempts}
\label{app:sparse_sid_rewards}

This subsection summarizes the reward-design path that motivated the MOPD-centered recipe in \S\ref{sec:post_training_ablation}.
The starting point was a retrieval-style RL objective aligned with SID-token beam-search recall.
For each prompt, the student generated a SID beam; if the beam contained the target item, the rollout received reward $1$, and otherwise it received reward $0$.
This binary reward directly optimizes hit rate, and is therefore well matched to candidate-generation evaluation protocols used in generative recommendation.
However, it is too sparse and too narrow for post-training \ourmethod.
Most sampled trajectories receive no positive feedback, including trajectories that generate semantically related items, satisfy part of the shopping intent, or preserve useful response behavior but miss the exact held-out target.
The reward also does not distinguish why a beam fails: invalid SID formatting, correct category but wrong item, poor beam ordering, and weak response grounding are all collapsed into the same zero signal.
In our experiments, direct optimization of this 0/1 beam-search reward led to unstable specialization rather than robust fulfillment improvement.
The model was pushed toward a narrow candidate-generation metric without enough pressure to preserve text--SID interleaving, non-SID recommendation behavior, or general instruction-following quality.

We next tried to make the reward less sparse with level-weighted SID rewards.
Instead of rewarding only exact item hits, this reward assigned partial credit when the generated SID matched higher-level SID prefixes, so near misses in the hierarchy could still receive feedback.
This made the signal denser than the 0/1 hit reward, but it still optimized a proxy for item recovery rather than the final retrieval or fulfillment metric.
We then augmented this level-weighted reward with catalog-lookup semantic constraints over generated SIDs.
After SID resolution, the same reward retrieved catalog-side item metadata and added credit according to semantic compatibility signals, such as whether the resolved item matched the intended category, attributes, or user-facing shopping semantics.
This made the reward denser and easier to optimize, but the held-out retrieval and fulfillment metrics did not improve accordingly.
The failure mode was different from the binary-reward collapse but pointed to the same underlying problem: the model learned to optimize a surrogate semantic reward, not the true hit-rate or end-to-end fulfillment objective.
If we return to exact hit-rate optimization, the reward again becomes too sparse; if we densify the reward through level weights or semantic lookup, the objective becomes easier to optimize but less faithful to the evaluation target.
This tension made reward design alone an unsatisfactory solution and motivated the OPD-centered design in the main text.

\subsection{Interleaved Fulfillment Reward Details}
\label{app:interleaved_fulfillment_reward_details}

For text--SID recommendation responses, we use an Interleaved Fulfillment Reward rather than a single hit-rate signal. The goal is to evaluate whether the model can use SID-based item references inside a coherent, catalog-grounded recommendation response.

\textbf{SID extraction and validity checks.}
The reward worker first extracts the visible answer region and parses all valid SID tokens from the model response. It applies rule-based penalties for missing SIDs, duplicated SIDs, predicted SIDs that cannot be resolved through catalog lookup, fabricated item identifiers, and overlong answers. When multiple reference SIDs are available, it also encourages the response to cover enough valid SIDs instead of collapsing to a single item.

\textbf{Catalog grounding.}
Both predicted and reference SIDs are resolved through the SID-to-item catalog lookup table. The lookup returns representative catalog items for each SID, including fields such as item title, price, brand, and seller or shop name. This lookup is a grounding step, not an independent semantic scorer: unresolved predicted SIDs are treated as invalid recommendations, unresolved reference SIDs are treated as data issues, and resolved item attributes are passed to the judge as evidence for what each SID denotes.

\textbf{Hierarchical SID matching.}
The reward also computes a level-aware SID matching score between the predicted and reference SID sets. It performs optimal matching between the two sets and gives partial credit according to the deepest matched SID prefix. Thus, a prediction can receive partial credit when it shares higher-level semantic codes with the reference, even if the full SID does not exactly match.

\textbf{Catalog-grounded judge reward.}
An LLM judge evaluates the interleaved response using the original user request, the model response, and the catalog metadata resolved from the predicted and reference SIDs. The judge scores independent rubric dimensions: user-query correctness, reference-space correctness, aesthetic or outfit coherence, text quality, and whether the explanation actually matches the resolved item attributes. Therefore, the semantic correctness of a predicted SID is not read directly from the lookup table; it is estimated by judging the resolved catalog evidence against the user request and the reference item space.

\textbf{Soft score extraction.}
For each $0$--$4$ rubric dimension, the judge is asked to emit a JSON integer. When token-level probabilities are available, the reward worker does not simply trust the decoded integer. Instead, it locates the generated score token for each JSON field, collects the top log probabilities over score tokens $\{0,1,2,3,4\}$, applies a softmax over these logits, and uses the expected score under this distribution. If log probabilities are unavailable, it falls back to the parsed integer score.

\textbf{Aggregation and penalties.}
For interleaved SID responses, the worker first constructs a base score from normalized catalog-grounded judge dimensions and the SID-level matching score. The current configuration uses weights $0.10$, $0.10$, $0.35$, $0.05$, $0.30$, and $0.10$ for aesthetic coherence, reference-space correctness, query correctness, text quality, reason--item consistency, and SID-level matching, respectively. Each catalog-grounded judge dimension is normalized from the $0$--$4$ scale to $[0,1]$, while the SID-level score is already in $[0,1]$. The SID-level score is computed by optimal matching between predicted and reference SID sets; each matched pair receives prefix-level credit according to the configured level scores $[0.1,0.2,0.3,0.4,0.5,1.0]$.

After the weighted base score is computed, the worker applies rule-based adjustments. If the judge flags duplicated product listings, the score is capped at $0.25$. Fabricated item identifiers are handled by a multiplicative count penalty rather than a cap. Additional multiplicative penalties are applied for missing SIDs, unresolved predicted SIDs, duplicate SID tokens, and excessive answer length. In the final full-reward configuration, missing all SIDs multiplies the score by $0.1$; each unresolved predicted SID uses a penalty of the form $1/(1+n)$; each duplicate SID uses $1/(1+0.5n)$; and fabricated identifiers use $1/(1+n)$. Reference SIDs that cannot be resolved through catalog lookup are treated as invalid data and masked according to the runtime configuration.

\subsection{Runtime Reward Prompts}
\label{app:runtime_reward_prompts}

We include the runtime prompts used by the general response judge reward and the catalog-grounded judge reward for reproducibility.
The general response judge parses a single integer in $[0,10]$ and normalizes it by $10$.
The catalog-grounded judge reward in the Interleaved Fulfillment Reward parses the JSON object below; for its $0$--$4$ rubric fields, the implementation further uses score-token log probabilities to compute expected rubric scores when they are available, as described in Appendix~\ref{app:interleaved_fulfillment_reward_details}.

\tightparagraph{General response judge prompt.}

\begin{tcolorbox}[
    enhanced,
    breakable,
    width=1.0\linewidth,
    colback=gray!2,
    colframe=black!70,
    boxrule=0.4pt,
    arc=1mm,
    left=2mm,
    right=2mm,
    top=1mm,
    bottom=1mm,
    boxsep=1mm,
    title=\centering\textbf{General Response Judge Prompt},
    fonttitle=\small\bfseries,
    fontupper=\scriptsize
]
Please evaluate the semantic matching quality between the model response and the reference answer, and give an overall integer score from 0 to 10 based on content accuracy and expression quality.

\textbf{[Model response]} \texttt{\{response\}}

\textbf{[Reference answer]} \texttt{\{reference\}}

\textbf{[Scoring criteria]}
Design intent: this score measures the overall semantic quality of the model response relative to the reference answer. Exact word matching is not required, but the response must cover the core facts, conclusions, and intent of the reference answer.
Content accuracy: first judge whether the key information is correct, whether important constraints are missing, and whether the response introduces new facts that conflict with the reference answer.
Expression quality: only after the content is correct, consider whether the expression is complete, clear, and natural. Fluent writing cannot compensate for factual errors or missing key information.
Degeneration detection: this dimension explicitly penalizes repetition, truncation, garbled text, off-topic continuation, and templated repetition, so that long text or a few correct fragments do not hide poor overall quality.
9--10: highly semantically consistent; covers all key information in the reference answer; clear, fluent, and stylistically appropriate.
7--8: mostly semantically consistent; covers most key information; fluent expression; may miss minor details or have slight stylistic mismatch.
5--6: partially matches the reference, but has clear omissions or deviations; expression is acceptable.
3--4: only a small amount of information is relevant; the main view or conclusion differs from the reference.
1--2: almost irrelevant; the information seriously deviates from the reference.
0: completely irrelevant, blank, or clearly degenerate, such as repetitive output or garbled text.

\textbf{[Penalty rules for degenerate output]}
If the response begins with relevant content but then continues with long off-topic text, punctuation-free strings, obvious templated repetition, unfinished continuation, ``[omitted]'', or placeholder-like ellipses, the maximum score is 4.
If the response is truncated, does not end properly, or its latter half is clearly degraded, the maximum score is 5; in severe cases, give 0--2.
Do not give a high score just because the response is long, essay-like, or contains a few correct fragments. It must be accurate, sufficiently complete, well expressed, and free of obvious degeneration.

Strictly output one integer from 0 to 10 and nothing else.
\end{tcolorbox}

\tightparagraph{Interleaved Fulfillment Reward judge prompt.}

\begin{tcolorbox}[
    enhanced,
    breakable,
    width=1.0\linewidth,
    colback=gray!2,
    colframe=black!70,
    boxrule=0.4pt,
    arc=1mm,
    left=2mm,
    right=2mm,
    top=1mm,
    bottom=1mm,
    boxsep=1mm,
    title=\centering\textbf{Interleaved Fulfillment Reward Judge Prompt},
    fonttitle=\small\bfseries,
    fontupper=\scriptsize
]
You are a quality evaluation expert for an e-commerce search and recommendation system. Please evaluate the model's text--SID recommendation response according to the rubrics. Directly output the JSON result without reasoning.

\textbf{[Important notes]}
Each SID represents a product cluster. The [model-recommended product list] and the [reference product list] show representative products under each SID.
Multiple correct answers may exist in recommendation scenarios. The model prediction does not need to exactly match the reference answer, but it must be a reasonable substitute under the user request.
[Unknown product] means that the SID cannot be found through catalog lookup and should be treated as an invalid recommendation. Validity has already been checked by the rule system; focus only on non-standard fabricated IDs.
Ignore self-evaluative statements in the [full model response], such as ``the above products are very suitable'' or ``carefully selected''. Judge independently based only on the actual match among product attributes, user needs, and recommendation reasons.
Be conservative: do not give a high score just because the response is long, adjective-heavy, or looks like recommendation copy. The product, user request, and recommendation reason must be consistent.
If the user asks for ``a full outfit / complete outfit / styling plan'' but the model only recommends a single category among tops, bottoms, shoes, or bags, clearly lower \texttt{query\_correctness}; because it does not form a complete outfit, also lower \texttt{aesthetic\_score}.
Score all dimensions independently. Fluent text cannot improve product correctness; closeness to the reference answer cannot improve text quality; good aesthetics cannot hide failure to satisfy the query.
If the original task explicitly asks for Xiaohongshu, social-media, or seeding-style copywriting, natural platform-style expressions and a small number of relevant topic tags are allowed. If the original task does not ask for such a style, the response should not proactively add platform-style wording or topic tags.
Regardless of whether the original task asks for Xiaohongshu-style writing, the response must not contain platform-publishing residue or template noise, such as ``@Xiaohongshu Creative Assistant'', ``[topic]'' placeholders, HTML/XML fragments, special end markers, or scraped-template remnants.

\textbf{[Original task]} \texttt{\{prompt\}}

\textbf{[Full model response]} \texttt{\{response\}}

\textbf{[Model-recommended product list]} \texttt{\{sid\_response\}}

\textbf{[Reference product list]} \texttt{\{gt\_sid\_response\}}

\textbf{[Evaluation dimensions]}
Rubric design intent: these dimensions separately constrain the product itself, the user need, the recommendation text, and the recommendation reason, preventing the model from relying only on fluent copywriting, only outputting SIDs, or only staying close to the reference while ignoring the real user need.
Product aesthetics: \texttt{aesthetic\_score} measures only whether the recommended product or combination is coherent in style, scenario, and outfit matching.
Reference-answer direction: \texttt{reference\_correctness} measures whether the predicted product lies in the reasonable substitute space of the reference product.
User-query direction: \texttt{query\_correctness} measures whether the predicted product directly satisfies the key needs in the user query.
Text-expression direction: \texttt{text\_quality} only evaluates whether the response is clear, concise, complete, and non-degenerate. Fluent expression cannot improve product correctness. If the original task does not ask for platform-style writing, proactively adding platform tags should lower \texttt{text\_quality}; publishing or scraped-template residue should always lower \texttt{text\_quality}.
Reason-matching direction: \texttt{reason\_match} measures whether the recommendation reason is truthfully grounded in the recommended product attributes and explains why they satisfy the user need. If the response mainly consists of platform tags, photo check-in copy, or template tails without grounding in real product attributes, lower \texttt{reason\_match}. If the original task explicitly asks for Xiaohongshu or seeding-style writing, natural stylistic expression itself should not be penalized.
Identifier reliability: \texttt{fabricated\_id\_count} and \texttt{has\_duplicate\_listings} capture structural problems such as fabricated product codes and repeatedly listing the same product.

The seven output fields are \texttt{aesthetic\_score}, \texttt{reference\_correctness}, \texttt{query\_correctness}, \texttt{text\_quality}, \texttt{reason\_match}, \texttt{fabricated\_id\_count}, and \texttt{has\_duplicate\_listings}. The first five fields are integer scores from 0 to 4, \texttt{fabricated\_id\_count} is a non-negative integer, and \texttt{has\_duplicate\_listings} is a boolean.

Strictly output the following JSON and nothing else:
\texttt{\{\{"aesthetic\_score": integer from 0 to 4, "reference\_correctness": integer from 0 to 4, "query\_correctness": integer from 0 to 4, "text\_quality": integer from 0 to 4, "reason\_match": integer from 0 to 4, "fabricated\_id\_count": integer >= 0, "has\_duplicate\_listings": true/false\}\}}
\end{tcolorbox}

\end{CJK*}
\end{document}